\documentclass[11pt,a4paper]{article}

\usepackage{jheppub}
\usepackage{amsmath}
\usepackage{multicol,bbm}
\usepackage{enumerate}
\usepackage{bibentry}

\usepackage{amssymb}
\usepackage{bm}
\usepackage{multirow}
\usepackage{caption}
\usepackage{tabularx}
\usepackage{enumitem}
\usepackage{extarrows}
\usepackage{tikz}
\usepackage{verbatim} 
\usepackage{float} 

\graphicspath{{figures/}}  

\def\be{\begin{equation}}
\def\ee{\end{equation}}
\def\ba{\begin{eqnarray}}
\def\ea{\end{eqnarray}}

\def\Li{\textrm{Li}}

\def\CP1{\mathbb{CP}^1}
\def\SL2C{\mathrm{SL}(2,\mathbb{C})}

\def\Z2{\mathbb{Z}_2}

\def\su2{{SU(2)}}

\def\[{\left[}
\def\]{\right]}

\def\({\left(}
\def\){\right)}
\def\[{\left[}
\def\]{\right]}

\def\<{\langle}
\def\>{\rangle}

\def\i2{\frac{i}{2}}

\def\2F1{\,_2{\rm F}_1}

\usepackage{float}
\usepackage{tikz}
\usepackage{mathtools}
\usetikzlibrary{shapes.geometric,calc}
\usetikzlibrary{arrows.meta}
\usetikzlibrary{decorations.pathmorphing}
\usepackage{CJKutf8}
\usepackage{nicematrix}
\usepackage{physics2}
\usephysicsmodule{ab,ab.braket}
\usepackage{amssymb}
\usetikzlibrary{arrows.meta}

\usepackage{caption}
\usepackage[labelformat=simple]{subcaption}

\usepackage[textsize=footnotesize]{todonotes}

\newcommand{\dif}{\mathrm{d}} 

\definecolor{nicered}{rgb}{0.7,0.1,0.1}
\definecolor{nicegreen}{rgb}{0.1,0.5,0.1}
\definecolor{kblue}{RGB}{0, 47, 167}
\definecolor{ultramarine}{RGB}{16,9,143}
\definecolor{winered}{RGB}{127,23,52}
\definecolor{nicecyn}{RGB}{0,99,107}
\definecolor{mathblue}{rgb}{0.368417, 0.506779, 0.709798}
\definecolor{mathyellow}{rgb}{0.880722, 0.611041, 0.142051}
\hypersetup{colorlinks,citecolor= kblue,linkcolor= kblue,urlcolor=nicecyn}

\newcommand{\Ep}[2]{\mathrm{E}(#1;#2)}
\newcommand{\Ephat}[2]{\hat{\mathrm{E}}(#1;#2)}
\newcommand{\Ecut}[2]{\mathrm{E}_{\text{cut}}(#1;#2)}

\newcommand{\bcfwhex}{
    \begin{tikzpicture}[baseline={([yshift=-0.8ex]current bounding box.center)}]
    \coordinate (O1) at (-0.5,0);
    \coordinate (O2) at (0.5,0);
    \draw[line width=0.6mm] (O1)--(O2);
    \draw[line width=0.2mm] (O1) -- +(90:0.6)node[above,scale=0.8]{$\widehat{1}$};
    \draw[line width=0.2mm] (O1) -- +(180:0.6)node[left,scale=0.8]{$6$};
    \draw[line width=0.2mm] (O1) -- +(270:0.6)node[below,scale=0.8]{$5$};
     \draw[line width=0.2mm] (O2) -- +(90:0.6)node[above,scale=0.8]{$\widehat{2}$};
    \draw[line width=0.2mm] (O2) -- +(0:0.6)node[right,scale=0.8]{$3$};
    \draw[line width=0.2mm] (O2) -- +(-90:0.6)node[below,scale=0.8]{$4$};
    \node[circle,fill=gray!70,draw,minimum size=0.3cm,inner sep=0pt] at (O1) {};
    \node[circle,fill=gray!70,draw,minimum size=0.3cm,inner sep=0pt] at (O2) {};
\end{tikzpicture}
}

\newcommand{\osdhex}{
    \begin{tikzpicture}[baseline={([yshift=-0.8ex]current bounding box.center)}]
     \coordinate (O) at (-0.5,0);
     \foreach \i in {0,1,2} {
        \path (O)+(90 + \i*120 : 0.4) coordinate (V\i) ;       
    }
     \draw[line width=0.2mm] (V0)--(V1)--(V2)--cycle;
      \draw[line width=0.2mm] (V0)+(120 :0.5)node[above,scale=0.8,xshift=-2]{$1$}--(V0)--+(60 : 0.5)node[above,xshift=2,scale=0.8]{$2$};
      \draw[line width=0.2mm] (V1)+(240 :0.5)node[below,xshift=-1,scale=0.8]{$5$}--(V1)--+(180 : 0.5)node[left,xshift=2,scale=0.8]{$6$};
    \draw[line width=0.2mm] (V2)+(0:0.5)node[right,xshift=-1,scale=0.8]{$3$}--(V2)--+(300 : 0.5)node[below,xshift=2,scale=0.8]{$4$};
     \foreach \i in {0,1,2}{
        \node[circle,fill=gray!70,draw,minimum size=0.3cm,inner sep=0pt] at (V\i) {};
     }
\end{tikzpicture}
}

\newcommand{\bcfwoctl}{
    \begin{tikzpicture}[baseline={([yshift=-0.8ex]current bounding box.center)}]
    \coordinate (O1) at (-0.5,0);
    \coordinate (O2) at (0.5,0);
    \draw[line width=0.6mm] (O1)--(O2);
    \draw[line width=0.2mm] (O1) -- +(90:0.6)node[above,yshift=-0.2ex,scale=0.8]{$\widehat{1}$};
    \draw[line width=0.2mm] (O1) -- +(135:0.65)node[above left,xshift=0.5ex,yshift=-0.5ex,scale=0.8]{$8$};
    \draw[line width=0.2mm] (O1) -- +(180:0.6)node[left,xshift=0.4ex,scale=0.8]{$7$};
      \draw[line width=0.2mm] (O1) -- +(-135:0.65)node[below left,xshift=0.5ex,yshift=0.5ex,scale=0.8]{$6$};
    \draw[line width=0.2mm] (O1) -- +(270:0.6)node[below,yshift=0.2ex,scale=0.8]{$5$};
     \draw[line width=0.2mm] (O2) -- +(90:0.6)node[above,yshift=-0.2ex,scale=0.8]{$\widehat{2}$};
    \draw[line width=0.2mm] (O2) -- +(0:0.6)node[right,xshift=-0.4ex,scale=0.8]{$3$};
    \draw[line width=0.2mm] (O2) -- +(-90:0.6)node[below,yshift=0.2ex,scale=0.8]{$4$};
    \node[circle,fill=gray!70,draw,minimum size=0.3cm,inner sep=0pt] at (O1) {};
    \node[circle,fill=gray!70,draw,minimum size=0.3cm,inner sep=0pt] at (O2) {};
\end{tikzpicture}
}

\newcommand{\bcfwoctr}{
     \begin{tikzpicture}[baseline={([yshift=-0.8ex]current bounding box.center)}]
    \coordinate (O1) at (-0.5,0);
    \coordinate (O2) at (0.5,0);
    \draw[line width=0.6mm] (O1)--(O2);
    \draw[line width=0.2mm] (O1) -- +(90:0.6)node[above,yshift=-0.2ex,scale=0.8]{$\widehat{1}$};
    \draw[line width=0.2mm] (O1) -- +(180:0.6)node[left,xshift=0.4ex,scale=0.8]{$8$};
    \draw[line width=0.2mm] (O1) -- +(270:0.6)node[below,yshift=0.2ex,scale=0.8]{$7$};
    \draw[line width=0.2mm] (O2) -- +(90:0.6)node[above,yshift=-0.2ex,scale=0.8]{$\widehat{2}$};
    \draw[line width=0.2mm] (O2) -- +(45:0.65)node[above right,xshift=-0.5ex,yshift=-0.5ex,scale=0.8]{$3$};
    \draw[line width=0.2mm] (O2) -- +(0:0.6)node[right,xshift=-0.4ex,scale=0.8]{$4$};
    \draw[line width=0.2mm] (O2) -- +(-45:0.65)node[below right,xshift=-0.5ex,yshift=0.5ex,scale=0.8]{$5$};
    \draw[line width=0.2mm] (O2) -- +(-90:0.6)node[below,yshift=0.2ex,scale=0.8]{$6$};
    \node[circle,fill=gray!70,draw,minimum size=0.3cm,inner sep=0pt] at (O1) {};
    \node[circle,fill=gray!70,draw,minimum size=0.3cm,inner sep=0pt] at (O2) {};
\end{tikzpicture}
}

\newcommand{\osdoctl}{
    \begin{tikzpicture}[baseline={([yshift=-0.6ex]current bounding box.center)}]
    \coordinate (O) at (0,0);
    \path (O)+(60:0.6) coordinate(P1);
    \path (O)+(0:0.6) coordinate(P2);
    \path (O)+(180:0.6) coordinate(Q1);
    \path (O)+(-120:0.6) coordinate(Q2);
    \draw[line width=0.2mm] (O) -- (P1) -- (P2)--cycle;
    \draw[line width=0.2mm] (O)--(Q1)--(Q2)--cycle;
    \draw[line width=0.2mm]  (P1)+(120:0.4)node[above,xshift=-0.4ex,yshift=-0.2ex,scale=0.8]{1} -- (P1)--
     +(60:0.4)node[above,xshift=0.4ex,yshift=-0.2ex,scale=0.8]{2};
     \draw[line width=0.2mm]  (P2)+(0:0.4)node[right,xshift=-0.4ex,scale=0.8]{3} -- (P2)--
     +(-60:0.4)node[below,xshift=0.2ex,yshift=0.4ex,scale=0.8]{4};
     \draw[line width=0.2mm]  (Q1)+(120:0.4)node[above,xshift=-0.2ex,yshift=-0.4ex,scale=0.8]{8} -- (Q1)--
     +(-180:0.4)node[left,xshift=0.4ex,scale=0.8]{7};
      \draw[line width=0.2mm]  (Q2)+(-60:0.4)node[below,xshift=0.4ex,yshift=0.2ex,scale=0.8]{5} -- (Q2)--
     +(-120:0.4)node[below,xshift=-0.4ex,yshift=0.2ex,scale=0.8]{6};
     \node[circle,fill=gray!70,draw,minimum size=0.3cm,inner sep=0pt] at (P1) {};
    \node[circle,fill=gray!70,draw,minimum size=0.3cm,inner sep=0pt] at (P2) {};
         \node[circle,fill=gray!70,draw,minimum size=0.3cm,inner sep=0pt] at (Q1) {};
    \node[circle,fill=gray!70,draw,minimum size=0.3cm,inner sep=0pt] at (Q2) {};
       \node[circle,fill=gray!70,draw,minimum size=0.3cm,inner sep=0pt] at (O) {};
\end{tikzpicture} 
}

\newcommand{\osdoctr}{
    \begin{tikzpicture}[baseline={([yshift=-0.6ex]current bounding box.center)}]
    \coordinate (O) at (0,0);
    \path (O)+(120:0.6) coordinate(P1);
    \path (O)+(180:0.6) coordinate(P2);
    \path (O)+(0:0.6) coordinate(Q1);
    \path (O)+(-60:0.6) coordinate(Q2);
    \draw[line width=0.2mm] (O) -- (P1) -- (P2)--cycle;
    \draw[line width=0.2mm] (O)--(Q1)--(Q2)--cycle;
    \draw[line width=0.2mm]  (P1)+(120:0.4)node[above,xshift=-0.4ex,yshift=-0.2ex,scale=0.8]{1} -- (P1)--
     +(60:0.4)node[above,xshift=0.4ex,yshift=-0.2ex,scale=0.8]{2};
     \draw[line width=0.2mm]  (P2)+(180:0.4)node[left,xshift=0.4ex,scale=0.8]{8} -- (P2)--
     +(-120:0.4)node[below,xshift=-0.2ex,yshift=0.4ex,scale=0.8]{7};
     \draw[line width=0.2mm]  (Q1)+(60:0.4)node[above,xshift=0.2ex,yshift=-0.4ex,scale=0.8]{3} -- (Q1)--
     +(0:0.4)node[right,xshift=-0.4ex,scale=0.8]{4};
      \draw[line width=0.2mm]  (Q2)+(-60:0.4)node[below,xshift=0.4ex,yshift=0.2ex,scale=0.8]{5} -- (Q2)--
     +(-120:0.4)node[below,xshift=-0.4ex,yshift=0.2ex,scale=0.8]{6};
     \node[circle,fill=gray!70,draw,minimum size=0.3cm,inner sep=0pt] at (P1) {};
    \node[circle,fill=gray!70,draw,minimum size=0.3cm,inner sep=0pt] at (P2) {};
         \node[circle,fill=gray!70,draw,minimum size=0.3cm,inner sep=0pt] at (Q1) {};
    \node[circle,fill=gray!70,draw,minimum size=0.3cm,inner sep=0pt] at (Q2) {};
       \node[circle,fill=gray!70,draw,minimum size=0.3cm,inner sep=0pt] at (O) {};
\end{tikzpicture} 
}

\begin{document}


\title{BDS ansatz in ABJM via scaffolding triangulations}

\date{\today}

\author[a,b,c]{Yu-tin Huang,}
\author[d]{Chia-Kai Kuo}
\author[e]{and Chi Zhang \begin{CJK*}{UTF8}{gbsn}(张驰)\end{CJK*}}


\affiliation[a]{Department of Physics and Center for Theoretical Physics, National Taiwan University, Taipei 10617, Taiwan}
\affiliation[b]{Physics Division, National Center for Theoretical Sciences, Taipei 10617, Taiwan}
\affiliation[c]{Max Planck{-}IAS{-}NTU Center for Particle Physics, Cosmology and Geometry, Taipei 10617, Taiwan}
\affiliation[d]{Max-Planck-Institut für Physik, Werner-Heisenberg-Institut, D-85748 Garching bei München, Germany
}
\affiliation[e]{Bethe Center for Theoretical Physics, Universität Bonn, 53115 Bonn, Germany}

\emailAdd{yutinyt@gmail.com}
\emailAdd{chia-kai.kuo@mpp.mpg.de}
\emailAdd{czhang@uni-bonn.de}

\preprint{ \begin{flushright} BONN-TH-2025-31 \\ 
MPP-2025-142\end{flushright}}

\abstract{In this work, we analyze the infrared divergence of two-loop amplitudes at arbitrary multiplicity in three-dimensional $\mathcal{N}=6$ Chern-Simons matter theory. We introduce the Bern-Dixon-Smirnov (BDS) integrand, which captures the full infrared structure while remaining free of unphysical cuts. We show that these local integrands, together with their kinematic prefactors, are naturally organized by the scaffolding triangulations of $n=2k$-gon, with distinct triangulations yielding different local representations. Remarkably, this triangulation structure also persists at the level of the integrated functions. This observation provides a graphical proof of both the cancellation of elliptic cuts and the triangulation independence of the integrated result. As a direct consequence, we obtain a simple proof that the integrated BDS integrand coincides with the one-loop maximally-helicity-violating (MHV) amplitude (the BDS ansatz) of $\mathcal{N}=4$ super Yang-Mills theory for all $n=2k$. 
}


\maketitle

\section{Introduction and summary of results}
The study of planar amplitudes of $\mathcal{N}=6$ supersymmetric three-dimensional Chern-Simons matter theory~\cite{Hosomichi:2008jb, Aharony:2008ug}, commonly referred to as ABJM theory, has exhibit strong similarity with its four-dimensional counterpart $\mathcal{N}=4$ super Yang-Mills (sYM) theory. Both enjoy enhanced dual superconformal symmetry~\cite{Bargheer:2010hn, Huang:2010qy}, whose building block can be identified as positroid cells in (orthogonal) positive grassmanian manifolds~\cite{Arkani-Hamed:2012zlh, Huang:2013owa,  Huang:2014xza,Lee:2010du}. Furthermore, it's all loop-integrand admits an alternative formulation in terms of canonical forms of positive geometry~\cite{Arkani-Hamed:2013jha,  Arkani-Hamed:2017vfh,Damgaard:2019ztj, Huang:2021jlh, He:2021llb, He:2022cup, He:2023rou,He:2023exb,He:2025zen}. 
While these are tantalizing similarities, they mainly address properties of rational functions, such as tree-amplitudes and loop integrands. Naively, one does not expect such close resemblance at the integrated level. After all, the analytic behaviour of a three-dimensional theory is drastically different from a four-dimensional one. 

However, early results for two-loop four- and six-point amplitudes~\cite{Chen:2011vv, Caron-Huot:2012sos} indicate that the integrated amplitudes of the two theories share the same infrared (IR) structure. In particular, the leading-order infrared  divergence in the coupling for ABJM theory, which occurs at two loops, is identical to the one-loop divergence of $\mathcal{N}=4$ sYM whose exponentiation is known as the Bern-Dixon-Smirnov (BDS) ansatz~\cite{Bern:2005iz}. Similar results was noticed for four-~\cite{Henn:2010ps} and general $n$-point bosonic Wilson-loops~\cite{Wiegandt:2011uu}. In the previous two-loop eight-point result which includes the two of the authors~\cite{He:2022lfz}, a ``BDS function" was defined that captures the infrared divergence of the 3D ABJM theory, denoted as $\text{BDS}_{n}^{\rm 3D}$ with $n=2k$. It is related to that of $\mathcal{N}=4$ sYM ($\text{BDS}_n^{\rm 4D}$) through
\begin{equation}\label{eq: BDS3}
\text{BDS}_n^{\rm 3D} := \frac{1}{2}\operatorname{BDS}^{\text{4D}}_n -\frac{(n-3)\pi^2}{6},
\end{equation}
where BDS$^{\rm 4D}_n$ is the one-loop $n$-point maximally-helicity-violating (MHV) amplitude of $\mathcal{N}=4$ sYM~\cite{Bern:1994zx}, and reads in mass regularization (also known as Higgs regularization~\cite{Alday:2009zm,Henn:2010ir}) as\footnote{The general pattern breaks at 4-points, which is given a  $\text{BDS}^{\rm 4D}_4=\frac{\pi^2}{2}- \log \frac{4 \mu_{\text{IR}}^2  }{ x_{1,3}^2} \log \frac{4 \mu_{\text{IR}}^2  }{ x_{2,4}^2}$. To ensure consistency with all even $n$, we normalize it by a factor of 2 $i.e.$ $\text{BDS}^{\rm 4D}_4 \rightarrow 2\, \text{BDS}^{\rm 4D}_4$.}
\begin{align}\label{eq:BDS}
       &\text{BDS}^\text{4D}_n= \frac{n}{6}\pi^2{+}\frac{1}{2} \sum_{i=1}^{n}  \log^2 \bigg(\frac{x_{i,i{+}3}^2}{x_{i{+}1,i{+}3}^2} \bigg){-}\log^2  \left( \frac{4\mu^2_{\text{IR}} x_{i,i{+}3}^2}{x_{i,i{+}2}^2 x_{i{+}1,i{+}3}^2} \right) \nonumber \\
        &{-}\sum_{i=1}^n \sum_{j=0}^{\frac{n}{2}{-}4}   \log  \left( \frac{x_{i,i+j+2}^2  x_{i-1,i+j+3}^2}{x_{i,i+j+3}^2  x_{i-1,i+j+2}^2}\right) \log\left( \frac{4 \mu^2_{\text{IR}}}{x_{i{-}1,i{+}j{+}2}^2} \right) {+} \text{Li}_2 \left(1{-}\frac{x_{i,i+j+2}^2  x_{i-1,i+j+3}^2}{x_{i,i+j+3}^2  x_{i-1,i+j+2}^2}\right)  \nonumber  \\
        &{-}\sum_{i=1}^{\frac{n}{2}}  
\log\left(\frac{x_{i,i+\frac{n}{2}-1}^2  x_{i-1,i+\frac{n}{2}}^2}{x_{i,i+\frac{n}{2}}^2  x_{i-1,i+\frac{n}{2}-1}^2}\right) \log\left( \frac{4 \mu^2_{\text{IR}}}{x_{i{-}1,i{+}\frac{n}{2}{-}1}^2} \right) {+} \text{Li}_2 \left(1{-}\frac{x_{i,i+\frac{n}{2}-1}^2  x_{i-1,i+\frac{n}{2}}^2}{x_{i,i+\frac{n}{2}}^2  x_{i-1,i+\frac{n}{2}-1}^2}\right)  \:.
\end{align}
Here, we introduce the dual coordinate notation where $x^2_{i,j}\equiv (p_i{+}p_{i{+}1}{+}\cdots p_{j{-}1})^2$.

In this paper, we determine the BDS integrand for all multiplicities and verify that it indeed yields $\text{BDS}_n^{\rm 3D}$, establishing this remarkable 3D/4D correspondence. Importantly, we find that when the amplitude is expanded in the basis of leading singularities, each leading singularity is accompanied by a distinct set of local integrals that all evaluate to the same function, $\text{BDS}_n^{\rm 3D}$. These distinct sets are in one-to-one correspondence with the triangulations of an $n$-gon, strongly suggesting a combinatoric geometry behind $\text{BDS}_n^{\rm 3D}$.

\subsection{A sketch of IR divergences in ABJM theory} \label{sec: subIntro}

To analyze the IR divergence of amplitudes in ABJM theory, we start with an expansion of the two-loop amplitudes in a basis of local integrals, using the general unitarity method~\cite{Bern:1994cg,Bourjaily:2017wjl}, which gives
\begin{equation} \label{eq:amplitude_expansion}
     \begin{tikzpicture}[baseline={([yshift=-0.7ex]current bounding box.center)}]
           \foreach \i in {-1,...,2}{
               \draw[line width=0.2mm] (0,0) -- (170+\i*20:0.8);
           }
              \foreach \i in {-1,...,1}{
               \draw[line width=0.2mm] (0,0) -- (0+\i*30:0.8);
           }
               \foreach \i in {-1,...,1} {
                \fill[black] (0,0)+(-15+\i *8: 0.70) circle (0.14ex);
            }
            \draw[line width=0.2mm](0,0) -- (15:0.8);
            \node[ellipse,fill=gray!70,draw,minimum width=1.0cm,minimum height=0.8cm,inner sep=0pt] at (0,0) {};
            \fill[white] (-0.23,0) circle (1ex);
             \fill[white] (0.23,0) circle (1ex);
     \end{tikzpicture} \bigg\vert_{\scalebox{0.8}{$\substack{\text{planar}\\ \text{ABJM}}$}} = 
     \sum a_{i}\,
     \begin{tikzpicture}[baseline={([yshift=-0.55ex]current bounding box.center)}]
          \path (-30:0.8) -- +(30:0.8) coordinate(P1);
          \draw[line width=0.2mm] (0,0) -- (30:0.8) -- (P1)--(-30:0.8)--cycle;
          \draw[line width=0.2mm] (-30:0.8) -- (30:0.8);
          \draw[line width=0.2mm]  (150:0.4)--(0,0)--(-150:0.4);
                 \foreach \i in {-1,...,1} {
                \fill[black] (0,0)+(-180+\i *15: 0.3) circle (0.12ex);
            }
          \draw[line width=0.2mm]  (P1)+(30:0.4)--(P1)--+(-30:0.4);
           \foreach \i in {-1,...,1} {
                \fill[black] (P1)+(0+\i *15: 0.3) circle (0.12ex);
            }
        \draw[line width=0.2mm,dash=on 0.4ex off 0.2ex phase 0pt] (30:0.8)--+(60:0.4);
        \draw[line width=0.2mm,dash=on 0.4ex off 0.2ex phase 0pt] (30:0.8)--+(120:0.4);
        \draw[line width=0.2mm,dash=on 0.4ex off 0.2ex phase 0pt] (-30:0.8)--+(-60:0.4);
        \draw[line width=0.2mm,dash=on 0.4ex off 0.2ex phase 0pt] (-30:0.8)--+(-120:0.4);
         \foreach \i in {-1,...,0} {
                \fill[black] (30:0.8)+(102+\i *24: 0.3) circle (0.12ex);
            }
          \foreach \i in {-1,...,0} {
                \fill[black] (-30:0.8)+(-102-\i *24: 0.3) circle (0.12ex);
            }
     \end{tikzpicture}
     +\sum b_{i} \,
     \begin{tikzpicture}[baseline={([yshift=-0.55ex]current bounding box.center)}]
          \path (-30:0.8) -- +(30:0.8) coordinate(R);
          \path (30:0.8) -- +(180:0.8) coordinate(L1);
           \path (-30:0.8) --+(180:0.8) coordinate(L2);
          \draw[line width=0.2mm] (L1) -- (30:0.8) -- (R)--(-30:0.8)--(L2)--cycle;
          \draw[line width=0.2mm] (30:0.8) --(-30:0.8);
          \draw[line width=0.2mm,dash=on 0.4ex off 0.2ex phase 0pt]  (L1) -- +(135-30:0.4);
          \draw[line width=0.2mm]  (L1) -- +(135+30:0.4);
          \draw[line width=0.2mm,dash=on 0.4ex off 0.2ex phase 0pt]  (L2) -- +(-135-30:0.4);
          \draw[line width=0.2mm]  (L2) -- +(-135+30:0.4);
          \foreach \i in {-1,...,1} {
                \fill[black] (L1)+(135+\i *15: 0.3) circle (0.12ex);
            }
           \foreach \i in {-1,...,1} {
                \fill[black] (L2)+(-135+\i *15: 0.3) circle (0.12ex);
            }
          \draw[line width=0.2mm]  (30:0.8) -- +(75+30:0.4);
          \draw[line width=0.2mm,dash=on 0.4ex off 0.2ex phase 0pt]  (30:0.8) -- +(75-30:0.4);
          \draw[line width=0.2mm]  (-30:0.8) -- +(-75+30:0.4);
          \draw[line width=0.2mm,dash=on 0.4ex off 0.2ex phase 0pt]  (-30:0.8) -- +(-75-30:0.4);
             \foreach \i in {-1,...,1} {
                \fill[black] (-30:0.8)+(-75-\i *15: 0.3) circle (0.12ex);
            }
            \foreach \i in {-1,...,1} {
                \fill[black] (30:0.8)+(75+\i *15: 0.3) circle (0.12ex);
            }
                  \draw[line width=0.2mm]  (P1)+(30:0.4)--(P1)--+(-30:0.4);
           \foreach \i in {-1,...,1} {
                \fill[black] (P1)+(0+\i *15: 0.3) circle (0.12ex);
            }
     \end{tikzpicture}
     +\sum c_{i}\,
     \begin{tikzpicture}[baseline={([yshift=-0.55ex]current bounding box.center)}]
          \path (30:0.8) -- +(0:0.8) coordinate(R1);
            \path (-30:0.8) -- +(0:0.8) coordinate(R2);
          \path (30:0.8) -- +(180:0.8) coordinate(L1);
           \path (-30:0.8) --+(180:0.8) coordinate(L2);
          \draw[line width=0.2mm] (L1) -- (30:0.8) -- (R1) -- (R2)--(-30:0.8)--(L2)--cycle;
                    \draw[line width=0.2mm] (30:0.8) --(-30:0.8);
          \draw[line width=0.2mm,dash=on 0.4ex off 0.2ex phase 0pt]  (L1) -- +(135-30:0.4);
          \draw[line width=0.2mm]  (L1) -- +(135+30:0.4);
          \draw[line width=0.2mm,dash=on 0.4ex off 0.2ex phase 0pt]  (L2) -- +(-135-30:0.4);
          \draw[line width=0.2mm]  (L2) -- +(-135+30:0.4);
          \foreach \i in {-1,...,1} {
                \fill[black] (L1)+(135+\i *15: 0.3) circle (0.12ex);
            }
           \foreach \i in {-1,...,1} {
                \fill[black] (L2)+(-135+\i *15: 0.3) circle (0.12ex);
            }
          \draw[line width=0.2mm,dash=on 0.4ex off 0.2ex phase 0pt]  (R1) -- +(45-30:0.4);
          \draw[line width=0.2mm]  (R1) -- +(45+30:0.4);
          \draw[line width=0.2mm,dash=on 0.4ex off 0.2ex phase 0pt]  (R2) -- +(-45-30:0.4);
          \draw[line width=0.2mm]  (R2) -- +(-45+30:0.4);
          \foreach \i in {-1,...,1} {
                \fill[black] (R1)+(45+\i *15: 0.3) circle (0.12ex);
            }
           \foreach \i in {-1,...,1} {
                \fill[black] (R2)+(-45+\i *15: 0.3) circle (0.12ex);
            }
            \draw[line width=0.2mm,dash=on 0.4ex off 0.2ex phase 0pt]  (30:0.8) -- +(90-30:0.4);
           \draw[line width=0.2mm,dash=on 0.4ex off 0.2ex phase 0pt]  (30:0.8) -- +(90+30:0.4);
                       \draw[line width=0.2mm,dash=on 0.4ex off 0.2ex phase 0pt]  (-30:0.8) -- +(-90-30:0.4);
           \draw[line width=0.2mm,dash=on 0.4ex off 0.2ex phase 0pt]  (-30:0.8) --+(-90+30:0.4);
             \foreach \i in {-1,...,0} {
                \fill[black] (30:0.8)+(102+\i *24: 0.3) circle (0.12ex);
            }
              \foreach \i in {-1,...,0} {
                \fill[black] (-30:0.8)+(-102-\i *24: 0.3) circle (0.12ex);
            }
     \end{tikzpicture} \:,
\end{equation}
where the coefficients are given by leading singularities (LSs) --- residues of amplitudes on the corresponding maximal cuts of the local integrals. Naively, each type of integrals can be divergent, such as the double-triangle integrals with a massless corner. However, huge cancellations are required by the absence of odd-point amplitudes in 3D~\cite{Chen:2011vv, Caron-Huot:2012sos,He:2022lfz}. As a consequence, the net IR divergence arises only from the \emph{factorizable} divergence~\cite{Caron-Huot:2012sos} of double-box integrals with two consecutive massless corners,
\begin{equation}\label{eq:divergent_db}
\begin{tikzpicture}[baseline=1.1em]
  \draw[line width=0.2mm] (0,0)--(0,1) -- (1,1) -- (1,0) --(0,0);
  \draw[line width=0.2mm] (0,1) -- +(135:0.5);
  \draw[line width=0.2mm] (0,0) -- +(-135:0.5);
  \draw[line width=0.2mm] (1,1) -- (2,1) -- (2,0) -- (1,0);
  \draw[line width=0.2mm] (2,1) --  +(75:0.45);
  \draw[line width=0.2mm,dash=on 0.4ex off 0.2ex phase 0pt] (2,1) --  +(15:0.45);
  \draw[line width=0.2mm,dash=on 0.4ex off 0.2ex phase 0pt] (2,0) --  +(-75:0.45);
  \draw[line width=0.2mm] (2,0) --  +(-15 :0.45);
    \foreach \i in {-1,...,1} {
                \fill[black] (2,1)+(45+\i *16: 0.35) circle (0.14ex);
            }
    \foreach \i in {-1,...,1} {
                \fill[black] (2,0)+(-45+\i *16: 0.35) circle (0.14ex);
            }
\end{tikzpicture} \sim \log^2 \mu_{\text{IR}}^2\:, \qquad
\begin{tikzpicture}[baseline=1.1em]
  \draw[line width=0.2mm] (0,0)--(0,1) -- (1,1) -- (1,0) --(0,0);
  \draw[line width=0.2mm] (0,1) -- +(135:0.5);
  \draw[line width=0.2mm] (0,0) -- +(-135:0.5);
  \draw[line width=0.2mm] (1,1) -- (2,1) -- (2,0) -- (1,0);
   \draw[line width=0.2mm] (2,1) --  +(75:0.45);
  \draw[line width=0.2mm,dash=on 0.4ex off 0.2ex phase 0pt] (2,1) --  +(15:0.45);
  \draw[line width=0.2mm,dash=on 0.4ex off 0.2ex phase 0pt] (2,0) --  +(-75:0.45);
  \draw[line width=0.2mm] (2,0) --  +(-15 :0.45);
    \foreach \i in {-1,...,1} {
                \fill[black] (2,1)+(45+\i *16: 0.35) circle (0.14ex);
            }
    \foreach \i in {-1,...,1} {
                \fill[black] (2,0)+(-45+\i *16: 0.35) circle (0.14ex);
            }
  \draw[line width=0.2mm] (1,0) --  +(-60 :0.45);
  \draw[line width=0.2mm,dash=on 0.4ex off 0.2ex phase 0pt]   (1,0)-- +(-120 :0.45);
   \foreach \i in {-1,...,1} {
                \fill[black] (1,0)+(-90+\i *16: 0.35) circle (0.14ex);
            }
\end{tikzpicture} \sim \log \mu_{\text{IR}}^2 \:.
\end{equation}

For this cancellation to occur, there are numerous linear relations among these coefficients $a_{i},b_{i},c_{i}$. 
Therefore, it is more convenient to collect local integrals according to their coefficients.
Since the IR divergence of two-loop amplitudes are proportional to the corresponding tree amplitudes~\cite{Britto:2004ap},
one should organize these local integrals as follows:
\emph{\begin{enumerate}
    \item Choose a basis (not necessarily linearly independent) for these coefficients such that the associated local integrals combine into the same infrared divergent function, namely $\mathrm{BDS}_n^{\rm 3D}$. \label{guide_1}
    \item The sum of the elements in this basis gives the tree amplitude. \label{guide_2}
\end{enumerate}}

\begin{figure}
    \begin{center}
        \begin{tikzpicture}[baseline={([yshift=-0.6ex]current bounding box.center)}]
    \coordinate (O) at (0,0);
    \path (O)+(60:0.6) coordinate(P1);
    \path (O)+(0:0.6) coordinate(P2);
    \path (O)+(180:0.6) coordinate(Q1);
    \path (O)+(-120:0.6) coordinate(Q2);
    \draw[line width=0.2mm] (O) -- (P1) -- (P2)--cycle;
    \draw[line width=0.2mm] (O)--(Q1)--(Q2)--cycle;
    \draw[line width=0.2mm]  (P1)+(120:0.4)node[above,xshift=-0.4ex,yshift=-0.2ex,scale=0.8]{5} -- (P1)--
     +(60:0.4)node[above,xshift=0.4ex,yshift=-0.2ex,scale=0.8]{6};
     \draw[line width=0.2mm]  (P2)+(0:0.4)node[right,xshift=-0.4ex,scale=0.8]{7} -- (P2)--
     +(-60:0.4)node[below,xshift=0.2ex,yshift=0.4ex,scale=0.8]{8};
     \draw[line width=0.2mm]  (Q1)+(120:0.4)node[above,xshift=-0.2ex,yshift=-0.4ex,scale=0.8]{4} -- (Q1)--
     +(-180:0.4)node[left,xshift=0.4ex,scale=0.8]{3};
      \draw[line width=0.2mm]  (Q2)+(-60:0.4)node[below,xshift=0.4ex,yshift=0.2ex,scale=0.8]{1} -- (Q2)--
     +(-120:0.4)node[below,xshift=-0.4ex,yshift=0.2ex,scale=0.8]{2};
     \node[circle,fill=gray!70,draw,minimum size=0.3cm,inner sep=0pt] at (P1) {};
    \node[circle,fill=gray!70,draw,minimum size=0.3cm,inner sep=0pt] at (P2) {};
         \node[circle,fill=gray!70,draw,minimum size=0.3cm,inner sep=0pt] at (Q1) {};
    \node[circle,fill=gray!70,draw,minimum size=0.3cm,inner sep=0pt] at (Q2) {};
       \node[circle,fill=gray!70,draw,minimum size=0.3cm,inner sep=0pt] at (O) {};
\end{tikzpicture} $\Leftrightarrow$  \begin{tikzpicture}[baseline={([yshift=-0.7ex]current bounding box.center)}]
        \foreach \i  in {1,...,8} {
        \coordinate (P\i) at (-45-45*\i:0.7);
        \node[scale=0.7] at (P\i) [shift=(-45-45*\i:0.25)] {\i};
    }
    \draw[line width=0.15mm] (P1) \foreach \i in {3,5,7} { -- (P\i) } -- cycle;
    \draw[line width=0.15mm] (P1)--(P5);
    \draw[line width=0.3mm] (P1) \foreach \i in {2,...,8} { -- (P\i) } -- cycle;
     \end{tikzpicture}
    \end{center}
     \caption{A leading singularity at eight points, the dual graph for the on-shell diagram is a scaffolding triangulation of an octagon.
     }
     \label{fig: onshell}
\end{figure}
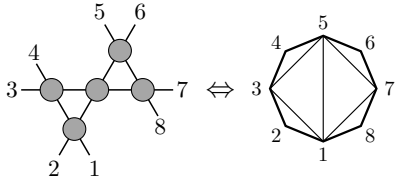

With the guiding principle~\ref{guide_2}, a natural choice for this basis is a special class of \emph{on-shell diagrams (OSDs)}, which appear in the BCFW recursion relation of tree amplitudes~\cite{Gang:2010gy}. These on-shell diagrams are built through the gluing of quartic vertices as usual, but are constrained to form $k-2$ successive triangles~\cite{Huang:2013owa},
see for example the left in figure~\ref{fig: onshell}. The same information can be encoded in the \emph{scaffolding triangulation}\footnote{The similar combinatorial structure also appears in~\cite{Arkani-Hamed:2023swr,Arkani-Hamed:2023jry} as a way representing the $n$-gluon scattering in terms of $2n$ scalars. Therefore, we adopt the same jargon.} of a $2k$-gon, where the dissection of the polygon is realized by chords connecting either only odd vertices or only even vertices, see for example the right in figure~\ref{fig: onshell}. For more details, see section~\ref{sec:preliminary}.

Next, given an arbitrary OSD described above, or equivalently a scaffolding triangulation, we need to identify the local integrals whose on-shell data (i.e., their coefficients in the expansion~\eqref{eq:amplitude_expansion}) can contribute to that OSD. It turns out that the main contribution arises from the \emph{double-boxes}, with the double-triangles serving as their counterterms. Therefore, let us briefly recall several fundamental facts about double-boxes, especially their associated on-shell data.

In terms of dual coordinates, the double-box integrals in our basis are defined as
\begin{equation}\label{eq:db_def}
        I^{\text{db}}(i,j,k;r,s,t):= \begin{tikzpicture}[baseline={([yshift=-0.55ex]current bounding box.center)}]
          \path (30:0.8) -- +(0:0.8) coordinate(R1);
            \path (-30:0.8) -- +(0:0.8) coordinate(R2);
          \path (30:0.8) -- +(180:0.8) coordinate(L1);
           \path (-30:0.8) --+(180:0.8) coordinate(L2);
           \path(L1)+(0.4,-0.4) coordinate(Xa);
           \path(R1)+(-0.4,-0.4) coordinate(Xb);
          \draw[line width=0.2mm] (L1) -- (30:0.8) -- (R1) -- (R2)--(-30:0.8)--(L2)--cycle;
                    \draw[line width=0.2mm] (30:0.8) --(-30:0.8);
          \draw[line width=0.2mm,dash=on 0.4ex off 0.2ex phase 0pt]  (L1) -- +(135-30:0.4);
          \draw[line width=0.2mm]  (L1) -- +(135+30:0.4);
          \draw[line width=0.2mm,dash=on 0.4ex off 0.2ex phase 0pt]  (L2) -- +(-135-30:0.4);
          \draw[line width=0.2mm]  (L2) -- +(-135+30:0.4);
          \foreach \i in {-1,...,1} {
                \fill[black] (L1)+(135+\i *15: 0.3) circle (0.12ex);
            }
           \foreach \i in {-1,...,1} {
                \fill[black] (L2)+(-135+\i *15: 0.3) circle (0.12ex);
            }
          \draw[line width=0.2mm,dash=on 0.4ex off 0.2ex phase 0pt]  (R1) -- +(45-30:0.4);
          \draw[line width=0.2mm]  (R1) -- +(45+30:0.4);
          \draw[line width=0.2mm,dash=on 0.4ex off 0.2ex phase 0pt]  (R2) -- +(-45-30:0.4);
          \draw[line width=0.2mm]  (R2) -- +(-45+30:0.4);
          \foreach \i in {-1,...,1} {
                \fill[black] (R1)+(45+\i *15: 0.3) circle (0.12ex);
            }
           \foreach \i in {-1,...,1} {
                \fill[black] (R2)+(-45+\i *15: 0.3) circle (0.12ex);
            }
            \draw[line width=0.2mm,dash=on 0.4ex off 0.2ex phase 0pt]  (30:0.8) -- +(90-30:0.4);
           \draw[line width=0.2mm,dash=on 0.4ex off 0.2ex phase 0pt]  (30:0.8) -- +(90+30:0.4);
                       \draw[line width=0.2mm,dash=on 0.4ex off 0.2ex phase 0pt]  (-30:0.8) -- +(-90-30:0.4);
           \draw[line width=0.2mm,dash=on 0.4ex off 0.2ex phase 0pt]  (-30:0.8) --+(-90+30:0.4);
             \foreach \i in {-1,...,0} {
                \fill[black] (30:0.8)+(102+\i *24: 0.3) circle (0.12ex);
            }
              \foreach \i in {-1,...,0} {
                \fill[black] (-30:0.8)+(-102-\i *24: 0.3) circle (0.12ex);
            }
            \fill[mathblue] (Xa) circle (0.3ex);
            \fill[mathblue] (Xb) circle (0.3ex);
            \draw[line width=0.4mm,color=mathblue] (Xa)node[scale=0.8,below,xshift=-0.6em,color=black]{$x_{a}$}--(Xb)node[scale=0.8,below,xshift=0.7em,color=black]{$x_{b}$};
            \draw[line width=0.4mm,color=mathblue] (Xa)--+(90:0.6)node[scale=0.8,above,yshift=-0.5ex,color=black]{$x_{k}$};
            \draw[line width=0.4mm,color=mathblue] (Xa)--+(180:0.6)node[scale=0.8,left,xshift=0.5ex,color=black]{$x_{j}$};
            \draw[line width=0.4mm,color=mathblue] (Xa)--+(-90:0.6)node[scale=0.8,below,yshift=0.5ex,color=black]{$x_{i}$};
            \draw[line width=0.4mm,color=mathblue] (Xb)--+(90:0.6)node[scale=0.8,above,yshift=-0.5ex,color=black]{$x_{r}$};
            \draw[line width=0.4mm,color=mathblue] (Xb)--+(0:0.6)node[scale=0.8,right,xshift=-0.5ex,color=black]{$x_{s}$};
            \draw[line width=0.4mm,color=mathblue] (Xb)--+(-90:0.6)node[scale=0.8,below,yshift=0.5ex,color=black]{$x_{t}$};
     \end{tikzpicture}=\int \frac{\mathcal{G}_{\{a,i,j,k\}}^{\{b,r,s,t\}}\:\dif^3 x_{a} \dif^{3}x_{b}}{x_{a,i}^2 x_{a,j}^2 x_{a,k}^2 x_{a,b}^2 x_{b,r}^2 x_{b,s}^2 x_{b,t}^2}\,,
\end{equation}
where $i<j<k\leq r<s<t\leq i$ is understood cyclically, as indicated by the diagram. Here we introduce Gram determinants in the embedding space\footnote{The embedding-space formalism~\cite{Dirac:1936fq} is usually adopted when there are conformal symmetries. In the embedding space, a point $x\in\mathbb{R}^{d}$ is represented by a projective light ray $X^{M}:=[1:x^{\mu}:x^2]\in\mathbb{R}^{d+1,1}$ (in light-cone coordinates), such that $x_{i,j}^2=-2X_{i}\cdot X_{j}=:X_{ij}=(X_{i},X_{j})$. } as
\begin{equation}\label{eq:Gram_def}
    \mathcal{G}_{B}^{A}=\det(x_{a,b}^2)\vert_{a\in A,b\in B} \quad \text{and} \quad \mathcal{G}_{A}:=\mathcal{G}_{A}^{B}\:.
\end{equation}
The numerator is designed such that the (sum of) residues of $I^{\text{db}}$ is unity on the kissing-triangle cut defined by $x^2_{a,i}=x^2_{a,j}=x^2_{a,k}=x^2_{b,r}=x^2_{b,s}=x^2_{b,t}=0$, but vanishes on any box-triangle cut such as 
$x^2_{a,i}=x^2_{a,j}=x^2_{a,b}=x^2_{b,r}=x^2_{b,s}=x^2_{b,t}=0$. As such, the on-shell data associated with this double-box integral is simply the residue of the two-loop amplitude on the corresponding kissing-triangle cut, which is a sum of the OSDs described above. 

As will be explained in detail in section~\ref{sec:preliminary} and \ref{sec: bds_3d}, the precise relation can be conveniently expressed through a graphical rule using the scaffolding triangulation: each double-box integral can be represented by a $2k$-gon with two shaded triangles (2ST), and the associated on-shell data is then a sum of the on-shell diagrams with a compatible scaffolding triangulation. For instance,
\begin{equation}
    \begin{tikzpicture}[baseline={([yshift=-0.7ex]current bounding box.center)}]
          \path (30:0.8) -- +(0:0.8) coordinate(R1);
            \path (-30:0.8) -- +(0:0.8) coordinate(R2);
          \path (30:0.8) -- +(180:0.8) coordinate(L1);
           \path (-30:0.8) --+(180:0.8) coordinate(L2);
           \path (L1)--+(0:0.4) coordinate (Lum);
              \path (L1)--+(-90:0.4) coordinate (Lm);
            \path (R1)--+(-90:0.4) coordinate (Rm);
           \path (L2)--+(0:0.4) coordinate (Ldm);
           \path (R1)--+(180:0.4) coordinate (Rum);
            \path (R2)--+(180:0.4) coordinate (Rdm);
          \draw[line width=0.2mm] (L1) -- (30:0.8) -- (R1) -- (R2)--(-30:0.8)--(L2)--cycle;
                    \draw[line width=0.2mm] (30:0.8) --(-30:0.8);
          \draw[line width=0.2mm]  (L1) -- +(135-30:0.3);
          \draw[line width=0.2mm]  (L1) -- +(135+30:0.3);
          \draw[line width=0.2mm]  (L2) -- +(-135-30:0.3);
          \draw[line width=0.2mm]  (L2) -- +(-135+30:0.3);
          \draw[line width=0.2mm]  (R1) -- +(45-30:0.3);
          \draw[line width=0.2mm]  (R1) -- +(45+30:0.3);
          \draw[line width=0.2mm]  (R2) -- +(-45-30:0.3);
          \draw[line width=0.2mm]  (R2) -- +(-45+30:0.3);
          \foreach \i in {-2,...,1} {
           \draw[line width=0.2mm]  (30:0.8) -- +(102+24*\i:0.3);
            }
        \node[left,scale=0.8,xshift=-8pt] at (0,0) {$x_3$};
        \node[below, scale=0.8] at (-30:0.8) {$x_{1}$};
        \node[right,scale=0.8,xshift=4pt] at (Rm) {$x_{11}$};
        \node[above,scale=0.8,xshift=-0.55cm,yshift=0.1cm] at (30:0.8) {$x_5$};
        \node[above,scale=0.8,xshift=0.55cm,yshift=0.1cm] at (30:0.8) {$x_9$};
     \end{tikzpicture} =
     \begin{tikzpicture}[baseline={([yshift=-0.7ex]current bounding box.center)}]
        \foreach \i  in {1,...,12} {
        \coordinate (P\i) at (-60-30*\i:0.7);
    }
    \foreach \i in {1,3,5,7,9,11} \node[scale=0.7] at (P\i) [shift=(-60-30*\i:0.25)] {\i};
    \fill[color=gray!70] (P1)--(P3)--(P5)--cycle;
    \fill[color=gray!70] (P1)--(P9)--(P11)--cycle;
     \foreach \i in {1,5} \draw[line width=0.15mm] (P3)--(P\i);
     \foreach \i in {1,9} \draw[line width=0.15mm] (P11)--(P\i);
     \foreach \i in {5,9} \draw[line width=0.15mm] (P1)--(P\i);
         \draw[line width=0.3mm] (P1) \foreach \i in {2,...,12} { -- (P\i) } -- cycle;
     \end{tikzpicture}
     \Rightarrow
        \begin{tikzpicture}[baseline={([yshift=-0.7ex]current bounding box.center)}]
        \foreach \i  in {1,...,12} {
        \coordinate (P\i) at (-60-30*\i:0.7);
    }
    \foreach \i in {1,3,5,7,9,11} \node[scale=0.7] at (P\i) [shift=(-60-30*\i:0.25)] {\i};
    \draw[line width=0.15mm] (P1) \foreach \i in {3,5,7,9,11} { -- (P\i) } -- cycle;
    \draw[line width=0.15mm] (P9)--(P1)--(P5);
    \draw[line width=0.15mm] (P9)--(P5);
         \draw[line width=0.3mm] (P1) \foreach \i in {2,...,12} { -- (P\i) } -- cycle;
     \end{tikzpicture}
     +   \begin{tikzpicture}[baseline={([yshift=-0.7ex]current bounding box.center)}]
        \foreach \i  in {1,...,12} {
        \coordinate (P\i) at (-60-30*\i:0.7);
    }
     \foreach \i in {1,3,5,7,9,11} \node[scale=0.7] at (P\i) [shift=(-60-30*\i:0.25)] {\i};
    \draw[line width=0.15mm] (P1) \foreach \i in {3,5,7,9,11} { -- (P\i) } -- cycle;
    \draw[line width=0.15mm] (P9)--(P1)--(P5);
    \draw[line width=0.15mm] (P1)--(P7);
         \draw[line width=0.3mm] (P1) \foreach \i in {2,...,12} { -- (P\i) } -- cycle;
     \end{tikzpicture} \:.
\end{equation}

With the graphical rule described above, it is easy to identify local integrals that contribute to the BDS integrand:
for any OSD that admits a scaffolding triangulation, the BDS integrand is given by a sum of all possible integrals whose 2ST representation is compatible with that triangulation. For example, the eight-point BDS integrand, up to an overall constant, can be expressed as a sum of the 2STs compatible with the eight-point OSD in figure~\ref{fig: onshell}:
\begin{equation} \label{eq:BDSoct_TSTrep}
   \begin{tikzpicture}[baseline={([yshift=-0.7ex]current bounding box.center)}]
        \foreach \i  in {1,...,8} {
        \coordinate (P\i) at (-45-45*\i:0.7);
    }
    \draw[line width=0.15mm] (P1) \foreach \i in {3,5,7} { -- (P\i) } -- cycle;
    \draw[line width=0.15mm] (P1)--(P5);
        \fill[color=gray!70] (P1)--(P2)--(P3)--cycle;
     \fill[color=gray!70] (P3)--(P4)--(P5)--cycle;
    \draw[line width=0.3mm] (P1) \foreach \i in {2,...,8} { -- (P\i) } -- cycle;
     \end{tikzpicture}
     +
      \begin{tikzpicture}[baseline={([yshift=-0.7ex]current bounding box.center)}]
        \foreach \i  in {1,...,8} {
        \coordinate (P\i) at (-45-45*\i:0.7);
    }
    \draw[line width=0.15mm] (P1) \foreach \i in {3,5,7} { -- (P\i) } -- cycle;
    \draw[line width=0.15mm] (P1)--(P5);
        \fill[color=gray!70] (P1)--(P2)--(P3)--cycle;
     \fill[color=gray!70] (P5)--(P6)--(P7)--cycle;
    \draw[line width=0.3mm] (P1) \foreach \i in {2,...,8} { -- (P\i) } -- cycle;
     \end{tikzpicture}
     + \begin{tikzpicture}[baseline={([yshift=-0.7ex]current bounding box.center)}]
        \foreach \i  in {1,...,8} {
        \coordinate (P\i) at (-45-45*\i:0.7);
    }
    \fill[color=gray!70] (P1)--(P2)--(P3)--cycle;
     \fill[color=gray!70] (P1)--(P3)--(P5)--cycle;
    \draw[line width=0.15mm] (P1) \foreach \i in {3,5,7} { -- (P\i) } -- cycle;
    \draw[line width=0.15mm] (P1)--(P5);
    \draw[line width=0.3mm] (P1) \foreach \i in {2,...,8} { -- (P\i) } -- cycle;
     \end{tikzpicture}
     +
     \begin{tikzpicture}[baseline={([yshift=-0.7ex]current bounding box.center)}]
        \foreach \i  in {1,...,8} {
        \coordinate (P\i) at (-45-45*\i:0.7);
    }
    \fill[color=gray!70] (P1)--(P2)--(P3)--cycle;
     \fill[color=gray!70] (P1)--(P5)--(P7)--cycle;
    \draw[line width=0.15mm] (P1) \foreach \i in {3,5,7} { -- (P\i) } -- cycle;
    \draw[line width=0.15mm] (P1)--(P5);
    \draw[line width=0.3mm] (P1) \foreach \i in {2,...,8} { -- (P\i) } -- cycle;
     \end{tikzpicture}
      +
     \begin{tikzpicture}[baseline={([yshift=-0.7ex]current bounding box.center)}]
        \foreach \i  in {1,...,8} {
        \coordinate (P\i) at (-45-45*\i:0.7);
    }
     \fill[color=gray!70] (P1)--(P3)--(P5)--(P7)--cycle;
    \draw[line width=0.15mm] (P1) \foreach \i in {3,5,7} { -- (P\i) } -- cycle;
    \draw[line width=0.15mm] (P1)--(P5);
    \draw[line width=0.3mm] (P1) \foreach \i in {2,...,8} { -- (P\i) } -- cycle;
     \end{tikzpicture}
     +\text{10 more}\:.
\end{equation}
Two caveats are in order:
\begin{itemize}[topsep=2pt,partopsep=2pt,itemsep=-2pt]
    \item This expression includes not only the divergent double-boxes, 
          but also finite ones, such as the double-box $I^{\text{db}}(1,3,5;5,7,1)$ 
          corresponding to the fifth term in eq.~\eqref{eq:BDSoct_TSTrep}. 
          These finite double-boxes are necessary for the cancellation of 
          unphysical elliptic cuts.
    \item The degenerate double-box integrals with $k=r$ and/or $i=t$ should be 
          understood as accompanied by some double-triangle 
          integrals, for the cancellation of unphysical three-point cuts 
          introduced by the internal 3-vertices. Viewing these double-triangle integrals as counterterms, 
          the degenerate double-box integrals are modified and hence denoted as $\hat{I}^{\text{db}}$.
\end{itemize}
We will address these two statements more carefully in section \ref{sec: bds_3d}.

In this way, one can easily build quite distinct-looking BDS integrands from different scaffolding triangulations. 
Remarkably, all these integrands give the same result, i.e. eq.~\eqref{eq: BDS3}, upon integration!
To see this, it is convenient to use a one-fold integral representation for the double-box integral, which is an analogue 
of the case in $\mathcal{N}=4$ sYM~\cite{Spiering:2024sea}:
\begin{align}\label{eq: IntroEdge}
        I^{\text{db}}(i,j,k;r,s,t)
        &=   \Ep{i,j}{r,s} - \Ep{i,j}{r,t} + \Ep{i,j}{s,t} - \Ep{i,k}{r,s} +\Ep{i,k}{r,t} \nonumber \\
        &\quad -\Ep{i,k}{s,t}+\Ep{j,k}{r,s} -\Ep{j,k}{r,t}+\Ep{j,k}{s,t}  \:.
\end{align}
Here for each pairing of edges, one from each of the two shaded triangles, we introduce an \emph{edge-pairing} function $\Ep{i,j}{k,l}$, which is a one-fold integral of the four-dimensional one-loop box and hence depends on $x_{i},x_{j},x_{k},x_{l}$ and 
potentially the regulator $\mu_{\text{IR}}$. Again, the edge-pairing functions such as $\Ep{i,j}{j,k}$ arising from the degenerate double-boxes will be modified by the double-triangles, and the modified version is denoted as $\Ephat{i,j}{j,k}$. The derivation of eq.~\eqref{eq: IntroEdge} and explicit expressions of (modified) edge-pairing functions will 
be given in section~\ref{sec: tri_func}.

\begin{figure}
    \centering
    \begin{tikzpicture}[baseline={([yshift=-0.7ex]current bounding box.center)}]
        \begin{scope}[baseline={([yshift=-0.7ex]current bounding box.center)}]
        \foreach \i  in {1,...,8} {
        \coordinate (P\i) at (-45-45*\i:0.7);
    }
    \foreach \i in {1,3,5,7} \node[scale=0.7] at (P\i) [shift=(-45-45*\i:0.25)] {\i};
    \foreach \i in {5,7} {\draw[line width=0.15mm] (P1) -- (P\i);}
    \fill[color=gray!50] (P1)--(P2)--(P3)--cycle;
    \fill[color=gray!50] (P5)--(P6)--(P7)--cycle;
    \draw[line width=0.15mm] (P3)--(P5);
    \draw[line width=0.3mm,color=red!75!black] (P1) -- (P3);
    \draw[line width=0.3mm,color=blue!75!black] (P5) -- (P7);
    \draw[line width=0.15mm] (P1)--(P5);
    \draw[line width=0.3mm] (P1) \foreach \i in {2,...,8} { -- (P\i) } -- cycle;
     \end{scope}
     \draw [stealth-stealth,thick] (1,0)--(3,0);
    \draw [stealth-stealth,thick] (0,-1)--(0,-2);
    \draw [stealth-stealth,thick] (1,-3)--(3,-3);
    \draw [stealth-stealth,thick] (4,-1)--(4,-2);
    \begin{scope}[baseline={([yshift=-0.7ex]current bounding box.center)},shift={(4,0)}]
        \foreach \i  in {1,...,8} {
        \coordinate (P\i) at (-45-45*\i:0.7);
    }
        \foreach \i in {1,3,5,7} \node[scale=0.7] at (P\i) [shift=(-45-45*\i:0.25)] {\i};
    \foreach \i in {5,7} {\draw[line width=0.15mm] (P1) -- (P\i);}
    \fill[color=gray!50] (P1)--(P3)--(P5)--cycle;
    \fill[color=gray!50] (P5)--(P6)--(P7)--cycle;
    \draw[line width=0.15mm] (P3)--(P5);
    \draw[line width=0.3mm,color=red!75!black] (P1) -- (P3);
    \draw[line width=0.3mm,color=blue!75!black] (P5) -- (P7);
    \draw[line width=0.15mm] (P1)--(P5);
    \draw[line width=0.3mm] (P1) \foreach \i in {2,...,8} { -- (P\i) } -- cycle;
     \end{scope}
       \begin{scope}[baseline={([yshift=-0.7ex]current bounding box.center)},shift={(4,-3)}]
        \foreach \i  in {1,...,8} {
        \coordinate (P\i) at (-45-45*\i:0.7);
    }
        \foreach \i in {1,3,5,7} \node[scale=0.7] at (P\i) [shift=(-45-45*\i:0.25)] {\i};
    \foreach \i in {5,7} {\draw[line width=0.15mm] (P1) -- (P\i);}
    \fill[color=gray!50] (P1)--(P3)--(P5)--cycle;
    \fill[color=gray!50] (P1)--(P5)--(P7)--cycle;
    \draw[line width=0.15mm] (P3)--(P5);
    \draw[line width=0.3mm,color=red!75!black] (P1) -- (P3);
    \draw[line width=0.3mm,color=blue!75!black] (P5) -- (P7);
    \draw[line width=0.15mm] (P1)--(P5);
    \draw[line width=0.3mm] (P1) \foreach \i in {2,...,8} { -- (P\i) } -- cycle;
     \end{scope}
       \begin{scope}[baseline={([yshift=-0.7ex]current bounding box.center)},shift={(0,-3)}]
        \foreach \i  in {1,...,8} {
        \coordinate (P\i) at (-45-45*\i:0.7);
    }
        \foreach \i in {1,3,5,7} \node[scale=0.7] at (P\i) [shift=(-45-45*\i:0.25)] {\i};
    \foreach \i in {5,7} {\draw[line width=0.15mm] (P1) -- (P\i);}
    \fill[color=gray!50] (P1)--(P2)--(P3)--cycle;
    \fill[color=gray!50] (P1)--(P5)--(P7)--cycle;
    \draw[line width=0.15mm] (P3)--(P5);
    \draw[line width=0.3mm,color=red!75!black] (P1) -- (P3);
    \draw[line width=0.3mm,color=blue!75!black] (P5) -- (P7);
    \draw[line width=0.15mm] (P1)--(P5);
    \draw[line width=0.3mm] (P1) \foreach \i in {2,...,8} { -- (P\i) } -- cycle;
     \end{scope}
    \end{tikzpicture}
    \caption{Four double-boxes (represented by 2STs) whose one-fold integral representations contain $\Ep{1,3}{5,7}$. The double arrows indicate that the signs of $\Ep{1,3}{5,7}$ in pairs of double-boxes are opposite.} \label{fig: cancellation_E1357}
\end{figure}
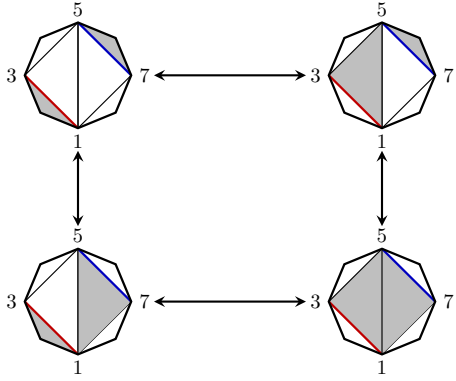

Without further knowledge of these edge-pairing functions, we can already anticipate that massive cancellations occurs among these functions in the BDS integrand due to their alternating signs in the expression~\eqref{eq: IntroEdge}. For example, the edge-pairing function $\Ep{1,3}{5,7}$ completely cancels out in the eight-point BDS function due to its opposite 
signs in 4 different double-boxes, see figure~\ref{fig: cancellation_E1357}. A more detailed study presented in section~\ref{sec: tri_func} indicates that for any representation of the BDS integrand, or equivalently, given any scaffolding triangulation, only 3 types of edge-pairing functions survive: (i) $\Ep{i,i+1}{j,j+1}$, (ii) $\Ephat{i,j}{j,k}$ and (iii) $\Ephat{i,j}{j,i}$.

To go further, we also need to know whether there are relations among the remaining edge-pairing functions. As will be shown in section~\ref{sec: tri_func}, there are two relations to remove the triangulation dependence: one for internal triangles and one for external triangles. With these two relations, the edge-pairing functions that contribute to the BDS integrand for an arbitrary triangulation sum to
\begin{equation}
  \underbrace{ \sum_{2 \leq j-i\leq n-2} \Ep{i,i{+}1}{j,j{+}1} + \sum_{i=1}^{n}\Ep{i{-}1,i}{i,i{+}1}}_{=\frac{1}{2}\text{BDS}_{n}^{\text{4D}}-n\pi^2/12} - \frac{\pi^{2}}{6}\ab(\frac{n}{2}-3)\:,
\end{equation}
and this leads to our final conclusion: the BDS integrands constructed from all scaffolding triangulations yield the same 
function $\text{BDS}_{n}^{\text{3D}}$ defined in eq.~\eqref{eq: BDS3}!

\section{On-shell diagrams and maximal cuts of ABJM}  \label{sec:preliminary}

In this section, we shall briefly review several basic concepts in ABJM theory, mainly focusing on the on‑shell data related to tree‑level amplitudes, to which the IR divergences are proportional. For more details, we refer to \cite{Elvang:2013cua} and references therein.

Unlike $\mathcal{N}=4$ sYM theory, there are two on-shell supermultiplets in ABJM theory: the fermionic $\bar{\Psi}$ and the bosonic $\Phi$. The amplitudes are therefore \emph{not} cyclically invariant. Unless otherwise stated, the 
$n$-point amplitudes throughout this paper are taken to be $A_{n}\coloneq A(\bar{\Psi}_{1},\Phi_{2},\ldots,\bar{\Psi}_{n-1},\Phi_{n})$.

\subsection{Tree-amplitudes and on-shell diagrams}

The connection between tree amplitudes and on-shell data in ABJM theory is established through the BCFW recursion relation~\cite{Gang:2010gy}: choosing a pair of legs and then summing over all possible factorizations where the chosen legs are separated across the factorization channel. Then, each term in the BCFW expansion can be represented by a connected triangle on-shell diagram~\cite{Huang:2013owa}. For example,
\begin{align}
  A_{6}^{\text{tree}}=  \bcfwhex 
&=
\osdhex  \label{eq: BCFW6pt}\\
A^{\text{tree}}_{8}=\bcfwoctl +  \bcfwoctr
&=
-\osdoctl - \osdoctr \label{eq: BCFW8pt}\:,
\end{align}
where each blob represents a tree amplitude, with heavy lines indicating off-shell, otherwise on-shell. The diagrammatic representation here is somewhat sloppy, since we also need to specify the particle ordering for each tree amplitude due to the absence of cyclicity. As such details are of no importance to the main topic of the paper, we therefore adopt this simplified version to avoid overcomplicating the exposition. The ensuing sign ambiguity is fixed by expressing on-shell diagrams in terms of (residues of) orthogonal Graßmannian integrals as in~\cite{Huang:2013owa,Huang:2014xza}, where the Yangian symmetries of the theory become manifest.

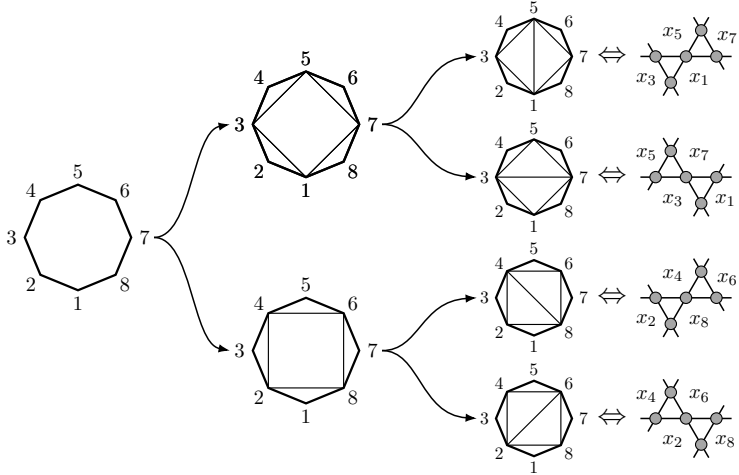
\begin{figure}
    \centering
    \begin{tikzpicture}
        \begin{scope}[baseline={([yshift=-0.7ex]current bounding box.center)}]
            \foreach \i  in {1,...,8} {
                \coordinate (P\i) at (-45-45*\i:0.7);
                \node[scale=0.7] at (P\i) [shift=(-45-45*\i:0.25)] {\i};
                }
             \draw[line width=0.3mm] (P1) \foreach \i in {2,...,8} { -- (P\i) } -- cycle;
        \end{scope}
        \draw[-{Latex[length=1.5mm]},line width=0.2mm] (1,0) ..controls (1.5,0) and (1.5,1.4).. (2,1.5);
        \draw[-{Latex[length=1.5mm]},line width=0.2mm] (1,0) ..controls (1.5,0) and (1.5,-1.4).. (2,-1.5);
       \begin{scope}[baseline={([yshift=-0.7ex]current bounding box.center)},shift={(3,1.5)}]
            \foreach \i  in {1,...,8} {
                \coordinate (P\i) at (-45-45*\i:0.7);
                \node[scale=0.7] at (P\i) [shift=(-45-45*\i:0.25)] {\i};
                }
         \draw[line width=0.15mm] (P1) \foreach \i in {3,5,7} { -- (P\i) } -- cycle;
         \draw[line width=0.3mm] (P1) \foreach \i in {2,...,8} { -- (P\i) } -- cycle;
        \end{scope}
    \begin{scope}[baseline={([yshift=-0.7ex]current bounding box.center)},shift={(3,1.5)}]
        \foreach \i  in {1,...,8} {
        \coordinate (P\i) at (-45-45*\i:0.7);
        \node[scale=0.7] at (P\i) [shift=(-45-45*\i:0.25)] {\i};
             }
        \draw[line width=0.15mm] (P1) \foreach \i in {3,5,7} { -- (P\i) } -- cycle;
        \draw[line width=0.3mm] (P1) \foreach \i in {2,...,8} { -- (P\i) } -- cycle;
    \end{scope}
     \draw[-{Latex[length=1.5mm]},line width=0.2mm] (4,1.5) ..controls (4.5,1.5) and (4.5,2.35).. (5.25,2.4);
     \draw[-{Latex[length=1.5mm]},line width=0.2mm] (4,1.5) ..controls (4.5,1.5) and (4.5,0.85).. (5.25,0.8);
    \begin{scope}[baseline={([yshift=-0.7ex]current bounding box.center)},shift={(3,-1.5)}]
        \foreach \i  in {1,...,8} {
        \coordinate (P\i) at (-45-45*\i:0.7);
        \node[scale=0.7] at (P\i) [shift=(-45-45*\i:0.25)] {\i};
            }
        \draw[line width=0.15mm] (P2) \foreach \i in {4,6,8} { -- (P\i) } -- cycle;
        \draw[line width=0.3mm] (P1) \foreach \i in {2,...,8} { -- (P\i) } -- cycle;
    \end{scope}
    \draw[-{Latex[length=1.5mm]},line width=0.2mm] (4,-1.5) ..controls (4.5,-1.5) and (4.5,-2.35).. (5.25,-2.4);
     \draw[-{Latex[length=1.5mm]},line width=0.2mm] (4,-1.5) ..controls (4.5,-1.5) and (4.5,-0.85).. (5.25,-0.8);
    \begin{scope}[baseline={([yshift=-0.7ex]current bounding box.center)},shift={(6,2.4)}]
        \foreach \i  in {1,...,8} {
        \coordinate (P\i) at (-45-45*\i:0.5);
        \node[scale=0.6] at (P\i) [shift=(-45-45*\i:0.25)] {\i};
            }
        \draw[line width=0.15mm] (P1) \foreach \i in {3,5,7} { -- (P\i) } -- cycle;
        \draw[line width=0.15mm] (P1)--(P5);
        \draw[line width=0.3mm] (P1) \foreach \i in {2,...,8} { -- (P\i) } -- cycle;
    \end{scope}
    \node at (7.03,2.4) {$\Leftrightarrow$};
    \begin{scope}[baseline={([yshift=-0.6ex]current bounding box.center)},shift={(8,2.4)}]
        \coordinate (O) at (0,0);
        \path (O)+(60:0.4) coordinate(P1);
        \path (O)+(0:0.4) coordinate(P2);
        \path (O)+(180:0.4) coordinate(Q1);
        \path (O)+(-120:0.4) coordinate(Q2);
        \node[scale=0.7] at (120:0.35) {$x_{5}$};
        \node[scale=0.7] at (-60:0.35) {$x_{1}$};
        \node[scale=0.7] at (25:0.60) {$x_{7}$};
        \node[scale=0.7] at (30-180:0.60) {$x_{3}$};
        \draw[line width=0.2mm] (O) -- (P1) -- (P2)--cycle;
        \draw[line width=0.2mm] (O)--(Q1)--(Q2)--cycle;
        \draw[line width=0.2mm]  (P1)+(120:0.2) -- (P1)-- +(60:0.2);
        \draw[line width=0.2mm]  (P2)+(0:0.2) -- (P2)--  +(-60:0.2);
        \draw[line width=0.2mm]  (Q1)+(120:0.2)-- (Q1)-- +(-180:0.2);
        \draw[line width=0.2mm]  (Q2)+(-60:0.2) -- (Q2)-- +(-120:0.2);
        \node[circle,fill=gray!70,draw,minimum size=0.15cm,inner sep=0pt] at (P1) {};
        \node[circle,fill=gray!70,draw,minimum size=0.15cm,inner sep=0pt] at (P2) {};
        \node[circle,fill=gray!70,draw,minimum size=0.15cm,inner sep=0pt] at (Q1) {};
        \node[circle,fill=gray!70,draw,minimum size=0.15cm,inner sep=0pt] at (Q2) {};
        \node[circle,fill=gray!70,draw,minimum size=0.15cm,inner sep=0pt] at (O) {};
    \end{scope} 
    \begin{scope}[baseline={([yshift=-0.7ex]current bounding box.center)},shift={(6,0.8)}]
        \foreach \i  in {1,...,8} {
        \coordinate (P\i) at (-45-45*\i:0.5);
        \node[scale=0.6] at (P\i) [shift=(-45-45*\i:0.25)] {\i};
            }
        \draw[line width=0.15mm] (P1) \foreach \i in {3,5,7} { -- (P\i) } -- cycle;
        \draw[line width=0.15mm] (P3)--(P7);
        \draw[line width=0.3mm] (P1) \foreach \i in {2,...,8} { -- (P\i) } -- cycle;
    \end{scope}
    \node at (7.03,0.8) {$\Leftrightarrow$};
    \begin{scope}[baseline={([yshift=-0.6ex]current bounding box.center)},shift={(8,0.8)}]
        \coordinate (O) at (0,0);
        \path (O)+(120:0.4) coordinate(P1);
        \path (O)+(180:0.4) coordinate(P2);
        \path (O)+(0:0.4) coordinate(Q1);
        \path (O)+(-60:0.4) coordinate(Q2);
        \node[scale=0.7] at (150:0.60) {$x_{5}$};
        \node[scale=0.7] at (-30:0.60) {$x_{1}$};
        \node[scale=0.7] at (60:0.35) {$x_{7}$};
        \node[scale=0.7] at (60-180:0.35) {$x_{3}$};
        \draw[line width=0.2mm] (O) -- (P1) -- (P2)--cycle;
        \draw[line width=0.2mm] (O)--(Q1)--(Q2)--cycle;
        \draw[line width=0.2mm] (P1)+(120:0.2) -- (P1) -- +(60:0.2);
        \draw[line width=0.2mm] (P2)+(180:0.2) -- (P2) -- +(-120:0.2);
        \draw[line width=0.2mm] (Q1)+(60:0.2) -- (Q1) -- +(0:0.2);
        \draw[line width=0.2mm] (Q2)+(-60:0.2) -- (Q2) -- +(-120:0.2);
        \node[circle,fill=gray!70,draw,minimum size=0.15cm,inner sep=0pt] at (P1) {};
        \node[circle,fill=gray!70,draw,minimum size=0.15cm,inner sep=0pt] at (P2) {};
        \node[circle,fill=gray!70,draw,minimum size=0.15cm,inner sep=0pt] at (Q1) {};
        \node[circle,fill=gray!70,draw,minimum size=0.15cm,inner sep=0pt] at (Q2) {};
        \node[circle,fill=gray!70,draw,minimum size=0.15cm,inner sep=0pt] at (O) {};
    \end{scope} 
    \begin{scope}[baseline={([yshift=-0.7ex]current bounding box.center)},shift={(6,-0.8)}]
        \foreach \i  in {1,...,8} {
        \coordinate (P\i) at (-45-45*\i:0.5);
        \node[scale=0.6] at (P\i) [shift=(-45-45*\i:0.25)] {\i};
            }
        \draw[line width=0.15mm] (P2) \foreach \i in {4,6,8} { -- (P\i) } -- cycle;
        \draw[line width=0.15mm] (P4)--(P8);
        \draw[line width=0.3mm] (P1) \foreach \i in {2,...,8} { -- (P\i) } -- cycle;
    \end{scope}
    \node at (7.03,-0.8) {$\Leftrightarrow$};
    \begin{scope}[baseline={([yshift=-0.6ex]current bounding box.center)},shift={(8,-0.8)}]
        \coordinate (O) at (0,0);
        \path (O)+(60:0.4) coordinate(P1);
        \path (O)+(0:0.4) coordinate(P2);
        \path (O)+(180:0.4) coordinate(Q1);
        \path (O)+(-120:0.4) coordinate(Q2);
        \node[scale=0.7] at (120:0.35) {$x_{4}$};
        \node[scale=0.7] at (-60:0.35) {$x_{8}$};
        \node[scale=0.7] at (25:0.60) {$x_{6}$};
        \node[scale=0.7] at (30-180:0.60) {$x_{2}$};
        \draw[line width=0.2mm] (O) -- (P1) -- (P2)--cycle;
        \draw[line width=0.2mm] (O)--(Q1)--(Q2)--cycle;
        \draw[line width=0.2mm]  (P1)+(120:0.2) -- (P1)-- +(60:0.2);
        \draw[line width=0.2mm]  (P2)+(0:0.2) -- (P2)--  +(-60:0.2);
        \draw[line width=0.2mm]  (Q1)+(120:0.2)-- (Q1)-- +(-180:0.2);
        \draw[line width=0.2mm]  (Q2)+(-60:0.2) -- (Q2)-- +(-120:0.2);
        \node[circle,fill=gray!70,draw,minimum size=0.15cm,inner sep=0pt] at (P1) {};
        \node[circle,fill=gray!70,draw,minimum size=0.15cm,inner sep=0pt] at (P2) {};
        \node[circle,fill=gray!70,draw,minimum size=0.15cm,inner sep=0pt] at (Q1) {};
        \node[circle,fill=gray!70,draw,minimum size=0.15cm,inner sep=0pt] at (Q2) {};
        \node[circle,fill=gray!70,draw,minimum size=0.15cm,inner sep=0pt] at (O) {};
    \end{scope} 
    \begin{scope}[baseline={([yshift=-0.7ex]current bounding box.center)},shift={(6,-2.4)}]
        \foreach \i  in {1,...,8} {
        \coordinate (P\i) at (-45-45*\i:0.5);
        \node[scale=0.6] at (P\i) [shift=(-45-45*\i:0.25)] {\i};
            }
        \draw[line width=0.15mm] (P2) \foreach \i in {4,6,8} { -- (P\i) } -- cycle;
        \draw[line width=0.15mm] (P2)--(P6);
        \draw[line width=0.3mm] (P1) \foreach \i in {2,...,8} { -- (P\i) } -- cycle;
    \end{scope}
    \node at (7.03,-2.4) {$\Leftrightarrow$};
    \begin{scope}[baseline={([yshift=-0.6ex]current bounding box.center)},shift={(8,-2.4)}]
        \coordinate (O) at (0,0);
        \path (O)+(120:0.4) coordinate(P1);
        \path (O)+(180:0.4) coordinate(P2);
        \path (O)+(0:0.4) coordinate(Q1);
        \path (O)+(-60:0.4) coordinate(Q2);
        \node[scale=0.7] at (150:0.60) {$x_{4}$};
        \node[scale=0.7] at (-30:0.60) {$x_{8}$};
        \node[scale=0.7] at (60:0.35) {$x_{6}$};
        \node[scale=0.7] at (60-180:0.35) {$x_{2}$};
        \draw[line width=0.2mm] (O) -- (P1) -- (P2)--cycle;
        \draw[line width=0.2mm] (O)--(Q1)--(Q2)--cycle;
        \draw[line width=0.2mm] (P1)+(120:0.2) -- (P1) -- +(60:0.2);
        \draw[line width=0.2mm] (P2)+(180:0.2) -- (P2) -- +(-120:0.2);
        \draw[line width=0.2mm] (Q1)+(60:0.2) -- (Q1) -- +(0:0.2);
        \draw[line width=0.2mm] (Q2)+(-60:0.2) -- (Q2) -- +(-120:0.2);
        \node[circle,fill=gray!70,draw,minimum size=0.15cm,inner sep=0pt] at (P1) {};
        \node[circle,fill=gray!70,draw,minimum size=0.15cm,inner sep=0pt] at (P2) {};
        \node[circle,fill=gray!70,draw,minimum size=0.15cm,inner sep=0pt] at (Q1) {};
        \node[circle,fill=gray!70,draw,minimum size=0.15cm,inner sep=0pt] at (Q2) {};
        \node[circle,fill=gray!70,draw,minimum size=0.15cm,inner sep=0pt] at (O) {};
    \end{scope} 
\end{tikzpicture}
\caption{An illustration of the leading singularities and their graphical representation at the case of $n=8$.}
\label{fig:2n to n}
\end{figure}

Since each on-shell diagram in the BCFW expansion of $2k$-point amplitudes consists of $(k-2)$ successive triangles, we can label it using a sequence of triplets, each of which consists of dual points surrounding the corresponding triangle. 
In this way, the two on-shell diagrams in eq.~\eqref{eq: BCFW8pt} are denoted as $\operatorname{OSD}_{\{\{1,3,5\},\{5,7,1\}\}}$ and $\operatorname{OSD}_{\{\{1,3,7\},\{3,5,7\}\}}$, respectively. Importantly, Such a labeling also links these on-shell diagrams to the so-called \emph{scaffolding triangulation} for a $2k$-gon. More precisely, the scaffolding triangulation starts by connecting either consecutive even vertices or consecutive odd vertices, which reduces the $2k$-gon to a $k$-gon, and then the triangulation of the reduced $k$-gon completes the dissection. The case of $n=8$ is illustrated in figure~\ref{fig:2n to n}. 
The tree-amplitude is then given by either the sum over even scaffolding triangulations or the negative sum over odd scaffolding triangulations, for example,
\begin{equation}\label{eq:8pt relation} 
     A_8^{\rm tree}=
     \begin{tikzpicture}[baseline={([yshift=-0.7ex]current bounding box.center)},shift={(6,-0.8)}]
        \foreach \i  in {1,...,8} {
        \coordinate (P\i) at (-45-45*\i:0.5);
        \node[scale=0.6] at (P\i) [shift=(-45-45*\i:0.25)] {\i};
            }
        \draw[line width=0.15mm] (P2) \foreach \i in {4,6,8} { -- (P\i) } -- cycle;
        \draw[line width=0.15mm] (P4)--(P8);
        \draw[line width=0.3mm] (P1) \foreach \i in {2,...,8} { -- (P\i) } -- cycle;
    \end{tikzpicture} + 
    \begin{tikzpicture}[baseline={([yshift=-0.7ex]current bounding box.center)},shift={(6,-2.4)}]
        \foreach \i  in {1,...,8} {
        \coordinate (P\i) at (-45-45*\i:0.5);
        \node[scale=0.6] at (P\i) [shift=(-45-45*\i:0.25)] {\i};
            }
        \draw[line width=0.15mm] (P2) \foreach \i in {4,6,8} { -- (P\i) } -- cycle;
        \draw[line width=0.15mm] (P2)--(P6);
        \draw[line width=0.3mm] (P1) \foreach \i in {2,...,8} { -- (P\i) } -- cycle;
    \end{tikzpicture} = - \ab( 
    \begin{tikzpicture}[baseline={([yshift=-0.7ex]current bounding box.center)},shift={(6,2.4)}]
        \foreach \i  in {1,...,8} {
        \coordinate (P\i) at (-45-45*\i:0.5);
        \node[scale=0.6] at (P\i) [shift=(-45-45*\i:0.25)] {\i};
            }
        \draw[line width=0.15mm] (P1) \foreach \i in {3,5,7} { -- (P\i) } -- cycle;
        \draw[line width=0.15mm] (P1)--(P5);
        \draw[line width=0.3mm] (P1) \foreach \i in {2,...,8} { -- (P\i) } -- cycle;
    \end{tikzpicture}+
    \begin{tikzpicture}[baseline={([yshift=-0.7ex]current bounding box.center)},shift={(6,0.8)}]
        \foreach \i  in {1,...,8} {
        \coordinate (P\i) at (-45-45*\i:0.5);
        \node[scale=0.6] at (P\i) [shift=(-45-45*\i:0.25)] {\i};
            }
        \draw[line width=0.15mm] (P1) \foreach \i in {3,5,7} { -- (P\i) } -- cycle;
        \draw[line width=0.15mm] (P3)--(P7);
        \draw[line width=0.3mm] (P1) \foreach \i in {2,...,8} { -- (P\i) } -- cycle;
    \end{tikzpicture}
    )\:.
\end{equation}
From the scaffolding triangulation representation for tree amplitudes, it follows immediately that the number of on-shell diagrams in the BCFW expansion for the $2k$-point tree amplitude is the Catalan number $C_{k-2}$ --- the number of triangulations of a $k$-gon.

\subsection{The maximal cuts of one- and two-loop amplitudes}

As mentioned in section \ref{sec: subIntro}, the on-shell data emerge when taking various maximal cuts of loop amplitudes. 
However, there is a subtlety compared to the tree-level case: the general maximal cut equations at $L$ loops yield $2^{L}$ solutions, thus defining $2^{L}$ \emph{branches} of leading singularities. These branches appear in particular linear combinations when they serve as the coefficients in the expansion of the amplitudes in terms of local integrals, 
while the on-shell diagrams above are merely the sum of all branches.

For example, at one loop, the maximal cut equations read $x^2_{a,i}=x^2_{a,j}=x^2_{a,k}=0$, which give two on-shell solutions $\ell^{\pm}_{\ast}$ for the loop momentum. The two branches of the leading singularity of one-loop amplitudes are then simply products of three tree-amplitudes, denoted as $\mathcal{C}_{i,j,k}^\pm$, each weighted by its respective Jacobian factor $\pm\sqrt{\mathcal{G}_{\{i,j,k\}}}$, see figure~\ref{fig: MaxCut}. To isolate individual branches, 
we can take the basis for local integrals to be \emph{chiral boxes}~\cite{He:2022lfz},
\begin{equation} \label{eq:chiral-tri}
    I^{\pm}(i,j,k):= \frac{1}{2} \int \ab(\frac{\sqrt{\mathcal{G}_{i,j,k}}}{x_{a,i}^2x_{a,j}^2x_{a,k}^2} \mp 
    \frac{ \epsilon(x_{a},x_{i},x_{j},x_{k},x_{\rho})}{x_{a,i}^2x_{a,j}^2x_{a,k}^2x_{a,\rho}^2})\,\dif^{3}x_{a}\:,
\end{equation}
where $x_{\rho}$ is some reference dual point and $\epsilon(x_{a},x_{i},x_{j},x_{k},x_{\rho}):=\epsilon_{M_{1}\cdots M_{5}}X_{a}^{M_{1}}\cdots X_{\rho}^{M_{5}}$ is the totally antisymmetric product of dual points in embedding space. 
This set of integrals is designed such that the maximal cuts of $I^{\pm}(i,j,k)$ at $\ell^{\mp}_{\ast}$ vanish respectively, so that their coefficients in the expansion of one-loop amplitudes are simply $\mathcal{C}_{i,j,k}^{\pm}/\sqrt{\mathcal{G}_{\{i,j,k\}}}$. It will be useful to define 
\begin{equation}\label{eq: 1LD}
    \mathcal{D}_{i,j,k}\equiv \frac{\mathcal{C}_{i,j,k}^{+}-\mathcal{C}_{i,j,k}^{-}}{\sqrt{2\mathcal{G}_{\{i,j,k\}}}} \:.
\end{equation}
The BCFW recursion relation for the $2k$-point tree-amplitude then can be expressed as 
\begin{equation}\label{eq: DandA}
        A^{\text{tree}}_n= (-1)^i \sum_{j=1}^{k{-}2}  \mathcal{D}_{i,i+2,i+2+2j}\,.
\end{equation}

\begin{figure}
    \centering
\begin{tikzpicture}
        \begin{scope}[baseline={([yshift=-0.7ex]current bounding box.center)}]
        \foreach \i  in {1,...,3} {
        \coordinate (P\i) at (-30-120*\i:0.8);
            }
        \draw[line width=0.2mm] (P1) \foreach \i in {2,3} { -- (P\i) } -- cycle;
        \draw[line width=0.2mm] (P1)+(-180:0.5) -- (P1) -- +(-120:0.5);
        \draw[line width=0.2mm] (P2)+(120:0.5) -- (P2) -- +(60:0.5);
        \draw[line width=0.2mm] (P3)+(0:0.5) -- (P3) -- +(-60:0.5);
        \foreach \i in {-1,...,1} {
                \fill[black] (P1)+(-150+\i *15: 0.40) circle (0.15ex);
                \fill[black] (P2)+(90+\i *15: 0.40) circle (0.15ex);
                \fill[black] (P3)+(-30+\i *15: 0.40) circle (0.15ex);
            }
        \foreach \i in {1,2,3} {
             \node[circle,fill=gray!70,draw,minimum size=0.40cm,inner sep=0pt] at (P\i) {};
        }
        \node at(-90:0.7) {$x_{i}$};
        \node at(30:0.7) {$x_{k}$};
        \node at(150:0.7) {$x_{j}$};
        \node at(0,0) {$\ell_{\ast}^{\pm}$};
        \node[right] at(0:1.5) {$\displaystyle \Rightarrow \quad \mathcal{C}_{i,j,k}^{\pm}
        =\int \prod_{I=1}^{3}\dif\eta_{\ell_{i}}^{I}\dif\eta_{\ell_{j}}^{I}\dif\eta_{\ell_{k}}^{I}\:
        A_{(ij)}^{\text{tree}}A_{(jk)}^{\text{tree}}A_{(ki)}^{\text{tree}}\Big\vert_{\ell=\ell^{\pm}_{\ast}} 
        $};
    \end{scope}
    \end{tikzpicture}
    \caption{Gluing tree amplitudes into one-loop leading singularities}
    \label{fig: MaxCut}
\end{figure}

Similarly, we need to put 6 propagators on-shell to fully localize the loop momenta at two loops, 
giving the so-called kissing-triangle and box-triangle cuts. 
Since the kissing-triangle cut is only contained in the double-box integral, we can use it to determine the 
coefficients of double-box integrals in the expansion \eqref{eq:amplitude_expansion}, which gives
\begin{equation} \label{eq:dbcoeff_1}
    \begin{tikzpicture}[baseline={([yshift=-0.55ex]current bounding box.center)}]
          \path (30:0.8) -- +(0:0.8) coordinate(R1);
            \path (-30:0.8) -- +(0:0.8) coordinate(R2);
          \path (30:0.8) -- +(180:0.8) coordinate(L1);
           \path (-30:0.8) --+(180:0.8) coordinate(L2);
               \path (L1)--+(0:0.4) coordinate (Lum);
              \path (L1)--+(-90:0.4) coordinate (Lm);
            \path (R1)--+(-90:0.4) coordinate (Rm);
           \path (L2)--+(0:0.4) coordinate (Ldm);
           \path (R1)--+(180:0.4) coordinate (Rum);
            \path (R2)--+(180:0.4) coordinate (Rdm);
          \draw[line width=0.2mm] (L1) -- (30:0.8) -- (R1) -- (R2)--(-30:0.8)--(L2)--cycle;
                    \draw[line width=0.2mm] (30:0.8) --(-30:0.8);
          \draw[line width=0.2mm]  (L1) -- +(135+30:0.4) (L1) -- +(135-30:0.4);
          \draw[line width=0.2mm]  (L2) -- +(-135-30:0.4) (L2) -- +(-135+30:0.4);
          \foreach \i in {-1,...,1} {
                \fill[black] (L1)+(135+\i *15: 0.3) circle (0.12ex);
                \fill[black] (L2)+(-135+\i *15: 0.3) circle (0.12ex);
                \fill[black] (R1)+(45+\i *15: 0.3) circle (0.12ex);
                \fill[black] (R2)+(-45+\i *15: 0.3) circle (0.12ex);
            }
          \draw[line width=0.2mm]  (R1) -- +(45-30:0.4)  (R1) -- +(45+30:0.4);
          \draw[line width=0.2mm]  (R2) -- +(-45-30:0.4) (R2) -- +(-45+30:0.4);
  
        \draw[line width=0.2mm,dash=on 0.4ex off 0.2ex phase 0pt]  (30:0.8) -- +(90-30:0.4);
        \draw[line width=0.2mm,dash=on 0.4ex off 0.2ex phase 0pt]  (30:0.8) -- +(90+30:0.4);
        \draw[line width=0.2mm,dash=on 0.4ex off 0.2ex phase 0pt]  (-30:0.8) -- +(-90-30:0.4);
        \draw[line width=0.2mm,dash=on 0.4ex off 0.2ex phase 0pt]  (-30:0.8) --+(-90+30:0.4);
             \foreach \i in {-1,...,0} {
                \fill[black] (30:0.8)+(102+\i *24: 0.3) circle (0.12ex);
                \fill[black] (-30:0.8)+(-102-\i *24: 0.3) circle (0.12ex);
            }
             \draw[line width=0.2mm,color=red!80!black] (Lum)+(90:0.15)--+(-90:0.15);
        \draw[line width=0.2mm,color=red!80!black] (Rum)+(90:0.15)--+(-90:0.15);
        \draw[line width=0.2mm,color=red!80!black] (Ldm)+(90:0.15)--+(-90:0.15);
         \draw[line width=0.2mm,color=red!80!black] (Rdm)+(90:0.15)--+(-90:0.15);
         \draw[line width=0.2mm,color=red!80!black] (Lm)+(180:0.15)--+(0:0.15);
        \draw[line width=0.2mm,color=red!80!black] (Rm)+(180:0.15)--+(0:0.15);
        \node[below, scale=0.8,xshift=0.55cm,yshift=-0.1cm] at (-30:0.8) {$x_{t}$};
        \node[right,scale=0.8,xshift=4pt] at (Rm) {$x_{s}$};
        \node[above,scale=0.8,xshift=0.55cm,yshift=0.1cm] at (30:0.8) {$x_r$};
         \node[below, scale=0.8,xshift=-0.55cm,yshift=-0.1cm] at (-30:0.8) {$x_{i}$};
        \node[left,scale=0.8,xshift=-4pt] at (Lm) {$x_{j}$};
        \node[above,scale=0.8,xshift=-0.55cm,yshift=0.1cm] at (30:0.8) {$x_k$};
     \end{tikzpicture} 
     \to
     \begin{tikzpicture}[baseline={([yshift=-0.55ex]current bounding box.center)},shift={(5,0.2)}]
         \coordinate (O) at (0,0);
        \path (O)+(150:1.0) coordinate(P1);
        \path (O)+(-150:1.0) coordinate(P2);
        \path (O)+(30:1.0) coordinate(Q1);
        \path (O)+(-30:1.0) coordinate(Q2);
        \draw[line width=0.2mm]  (P1)--node[scale=0.8,left,xshift=2pt]{$x_{j}$}(P2)--node[scale=0.8,yshift=-6pt]{$x_{i}$}(O)--node[scale=0.8,yshift=6pt]{$x_{k}$}cycle ;
           \draw[line width=0.2mm]  (Q1)--node[scale=0.8,right,xshift=-2pt]{$x_{s}$}(Q2)--node[scale=0.8,yshift=-6pt]{$x_{t}$}(O)--node[scale=0.8,yshift=6pt]{$x_{r}$}cycle;
        \draw[line width=0.2mm] (P1)+(150:0.5)--(P1)--+(90:0.5);
        \draw[line width=0.2mm] (P2)+(-150:0.5)--(P2)--+(-90:0.5); 
        \draw[line width=0.2mm] (Q1)+(30:0.5)--(Q1)--+(90:0.5);
        \draw[line width=0.2mm] (Q2)+(-30:0.5)--(Q2)--+(-90:0.5);
        \foreach \i in {-1,...,1} {
                \fill[black] (P1)+(120+\i *15: 0.40) circle (0.15ex);
                \fill[black] (P2)+(-120+\i *15: 0.40) circle (0.15ex);
                \fill[black] (Q1)+(60+\i *15: 0.40) circle (0.15ex);
                \fill[black] (Q2)+(-60+\i *15: 0.40) circle (0.15ex);
             }
        \draw[line width=0.2mm,dash=on 0.4ex off 0.2ex phase 0pt] (O)--(70:0.65) (O)--(110:0.65) (O)--(-70:0.65) (O)--(-110:0.65);
        \foreach \i in {-1,0} {
                \fill[black] (O)+(97+\i *14: 0.50) circle (0.15ex);
                \fill[black] (O)+(-97-\i *14: 0.50) circle (0.15ex);
             }
        \foreach \v in {O,P1,P2,Q1,Q2}{ 
             \node[circle,fill=gray!70,draw,minimum size=0.40cm,inner sep=0pt] at (\v) {};
            }
    \end{tikzpicture}=\mathcal{D}^{i,j,k}_{r,s,t}=\sum_{(\sigma,\sigma')\in(\pm,\pm)}\frac{(-1)^{ir}\sigma \sigma^\prime \mathcal{C}_{i,j,k;r,s,t}^{\sigma,\sigma'}}{4\sqrt{ \mathcal{G}_{\{i,j,k\}}\mathcal{G}_{\{r,s,t\}}}} \:.
\end{equation}
Here $\mathcal{C}_{i,j,k;r,s,t}^{\sigma,\sigma'}$ in the numerator denote the 4 different products of the tree amplitudes for the corresponding on-shell loop momenta, and the denominator again arises from the Jacobian.

It should be noted that the divergent double-boxes always contain two consecutive massless corners as shown in eq.~\eqref{eq:divergent_db}. The corresponding kissing-triangle cut therefore contains a \emph{soft} cut $x_{b,s-1}^2=x_{b,s}^2=x_{b,s+1}^2=0$ whose effect is simply requiring the loop momentum $x_{b,s}=0$ and hence reducing the loop order by one. The coefficients of these double-boxes in the expansion \eqref{eq:amplitude_expansion} are thus given by one-loop leading singularities,
\begin{equation}\label{eq:soft-box}
       \begin{tikzpicture}[baseline={([yshift=-0.7ex]current bounding box.center)}]
          \path (30:0.8) -- +(0:0.8) coordinate(R1);
        \path (-30:0.8) -- +(0:0.8) coordinate(R2);
          \path (30:0.8) -- +(180:0.8) coordinate(L1);
           \path (-30:0.8) --+(180:0.8) coordinate(L2);
           \path (L1)--+(0:0.4) coordinate (Lum);
              \path (L1)--+(-90:0.4) coordinate (Lm);
            \path (R1)--+(-90:0.4) coordinate (Rm);
           \path (L2)--+(0:0.4) coordinate (Ldm);
           \path (R1)--+(180:0.4) coordinate (Rum);
            \path (R2)--+(180:0.4) coordinate (Rdm);
          \draw[line width=0.2mm] (L1) -- (30:0.8) -- (R1) -- (R2)--(-30:0.8)--(L2)--cycle;
                    \draw[line width=0.2mm] (30:0.8) --(-30:0.8);
          \draw[line width=0.2mm]  (L1) -- +(135:0.35) (L2) -- +(-135:0.35);
          \draw[line width=0.2mm]  (R1) -- +(45+30:0.35) (R1) -- +(45-30:0.35) (R2) -- +(-45+30:0.35) (R2) -- +(-45-30:0.35);
          \foreach \i in {0,...,1} {
           \draw[line width=0.2mm,dash=on 0.4ex off 0.2ex phase 0pt]  (30:0.8) -- +(60+60*\i:0.35);
            \draw[line width=0.2mm,dash=on 0.4ex off 0.2ex phase 0pt]  (-30:0.8) -- +(-120+60*\i:0.35);
              \fill[black] (30:0.8)+(78+\i *24: 0.26) circle (0.13ex);
              \fill[black] (-30:0.8)+(-78-\i *24: 0.26) circle (0.13ex);
          }
          \foreach \i in {-1,0,1}{
            \fill[black] (R1)+(45+\i*15:0.27) circle (0.12ex);
            \fill[black] (R2)+(-45+\i*15:0.27) circle (0.12ex);
          }
        \draw[line width=0.2mm,color=red!80!black] (Lum)+(90:0.15)--+(-90:0.15) (Rum)+(90:0.15)--+(-90:0.15) (Ldm)+(90:0.15)--+(-90:0.15) (Rdm)+(90:0.15)--+(-90:0.15) (Lm)+(180:0.15)--+(0:0.15) (Rm)+(180:0.15)--+(0:0.15);
        \node[below, scale=0.8,xshift=0.55cm,yshift=-0.1cm] at (-30:0.8) {$x_{k}$};
        \node[right,scale=0.8,xshift=4pt] at (Rm) {$x_{j}$};
        \node[above,scale=0.8,xshift=0.55cm,yshift=0.1cm] at (30:0.8) {$x_i$};
     \end{tikzpicture} \quad  \to \quad 
        \begin{tikzpicture}[baseline={([yshift=-0.7ex]current bounding box.center)},shift={(5,0.2)}]
        \coordinate (O) at (0,0);
        \path (O)+(30:1.0) coordinate(Q1);
        \path (O)+(-30:1.0) coordinate(Q2);  
        \draw[line width=0.2mm] (Q1)+(30:0.5)--(Q1)--+(90:0.5);
        \draw[line width=0.2mm]  (Q1)--node[scale=0.8,right,xshift=-2pt]{$x_{j}$}(Q2)--node[scale=0.8,yshift=-6pt]{$x_{k}$}(O)--node[scale=0.8,yshift=6pt]{$x_{i}$}cycle (O)--(-150:0.7) (O)--(150:0.7);
        \foreach \i in {-1,...,1} {
                \fill[black] (Q1)+(60+\i *15: 0.40) circle (0.15ex);
             }
        \draw[line width=0.2mm] (Q2)+(-30:0.5)--(Q2)--+(-90:0.5);
        \foreach \i in {-1,...,1} {
                \fill[black] (Q2)+(-60+\i *15: 0.40) circle (0.15ex);
             }
        \draw[line width=0.2mm,dash=on 0.4ex off 0.2ex phase 0pt] (O)--(70:0.65);
        \draw[line width=0.2mm,dash=on 0.4ex off 0.2ex phase 0pt] (O)--(110:0.65);
        \draw[line width=0.2mm,dash=on 0.4ex off 0.2ex phase 0pt] (O)--(-70:0.65);
        \draw[line width=0.2mm,dash=on 0.4ex off 0.2ex phase 0pt] (O)--(-110:0.65);
        \foreach \i in {-1,0} {
                \fill[black] (O)+(97+\i *14: 0.50) circle (0.15ex);
                 \fill[black] (O)+(-97-\i *14: 0.50) circle (0.15ex);
             }
        \foreach \v in {O,Q1,Q2}{ 
             \node[circle,fill=gray!70,draw,minimum size=0.40cm,inner sep=0pt] at (\v) {};
            }
    \end{tikzpicture} \quad = (-1)^{i}\,\mathcal{D}_{i,j,k} \:,
\end{equation}
if only one side consists of two consecutive massless corners, or simply tree amplitudes,
\begin{equation}\label{eq:soft-soft}
       \begin{tikzpicture}[baseline={([yshift=-0.7ex]current bounding box.center)}]
          \path (30:0.8) -- +(0:0.8) coordinate(R1);
        \path (-30:0.8) -- +(0:0.8) coordinate(R2);
          \path (30:0.8) -- +(180:0.8) coordinate(L1);
           \path (-30:0.8) --+(180:0.8) coordinate(L2);
           \path (L1)--+(0:0.4) coordinate (Lum);
              \path (L1)--+(-90:0.4) coordinate (Lm);
            \path (R1)--+(-90:0.4) coordinate (Rm);
           \path (L2)--+(0:0.4) coordinate (Ldm);
           \path (R1)--+(180:0.4) coordinate (Rum);
            \path (R2)--+(180:0.4) coordinate (Rdm);
          \draw[line width=0.2mm] (L1) -- (30:0.8) -- (R1) -- (R2)--(-30:0.8)--(L2)--cycle;
                    \draw[line width=0.2mm] (30:0.8) --(-30:0.8);
          \draw[line width=0.2mm]  (L1) -- +(135:0.35) (L2) -- +(-135:0.35);
          \draw[line width=0.2mm]  (R1) -- +(45:0.35) (R2) -- +(-45:0.35);
          \foreach \i in {0,...,1} {
           \draw[line width=0.2mm,dash=on 0.4ex off 0.2ex phase 0pt]  (30:0.8) -- +(60+60*\i:0.35);
            \draw[line width=0.2mm,dash=on 0.4ex off 0.2ex phase 0pt]  (-30:0.8) -- +(-120+60*\i:0.35);
              \fill[black] (30:0.8)+(78+\i *24: 0.26) circle (0.13ex);
              \fill[black] (-30:0.8)+(-78-\i *24: 0.26) circle (0.13ex);
          }
        \draw[line width=0.2mm,color=red!80!black] (Lum)+(90:0.15)--+(-90:0.15);
        \draw[line width=0.2mm,color=red!80!black] (Rum)+(90:0.15)--+(-90:0.15);
        \draw[line width=0.2mm,color=red!80!black] (Ldm)+(90:0.15)--+(-90:0.15);
         \draw[line width=0.2mm,color=red!80!black] (Rdm)+(90:0.15)--+(-90:0.15);
         \draw[line width=0.2mm,color=red!80!black] (Lm)+(180:0.15)--+(0:0.15);
        \draw[line width=0.2mm,color=red!80!black] (Rm)+(180:0.15)--+(0:0.15);
     \end{tikzpicture} \quad  \to \quad 
        \begin{tikzpicture}[baseline={([yshift=-0.7ex]current bounding box.center)},shift={(5,0.2)}]
         \coordinate (O) at (0,0);
        \draw[line width=0.2mm,dash=on 0.4ex off 0.2ex phase 0pt] (O)--(70:0.65) (O)--(110:0.65) (O)--(-70:0.65) (O)--(-110:0.65);
        \draw[line width=0.2mm] (O)--(30:1) (O)--(150:1) (O)--(-30:1) (O)--(-150:1);
        \foreach \i in {-1,0} {
                \fill[black] (O)+(97+\i *14: 0.50) circle (0.15ex);
                 \fill[black] (O)+(-97-\i *14: 0.50) circle (0.15ex);
             }
   \node[circle,fill=gray!70,draw,minimum size=0.40cm,inner sep=0pt] at (O) {};
    \end{tikzpicture} \quad = A_{n}^{\text{tree}}
\end{equation}
if both sides consist of two consecutive massless corners.

Although the double-box integrals completely capture the kissing-triangle cuts of the amplitudes, they also introduce some unphysical cuts that should be balanced by double-triangles and double-boxes themselves.
The BDS integrand is constructed precisely such that all unphysical cuts are absent, as we will elaborate in the next section.

\section{Local integrands for BDS$^{\text{3D}}_n$}\label{sec: bds_3d}

It was shown that for six- and eight-point two-loop amplitudes~\cite{Caron-Huot:2012sos,He:2022lfz}, a subset of local integrands, which integrates to BDS$^{\text{3D}}_n$ (i.e., eq.~\eqref{eq: BDS3}) multiplied by tree amplitudes, can be separated from the entire two-loop integrands.
This subset comprises all infrared divergent integrals, along with their counterparts whose role is to cancel unphysical cuts. For convenience, we refer to this as the BDS integrand. To emphasize: 
\begin{align}\label{eq: BDSDef2}
    &\textbf{BDS Integrand}\coloneq \nonumber\\
    &\textit{The set of divergent local integrands, accompanied by their counterparts such that}\nonumber\\
    &\textit{unphysical cuts are absent.}     
\end{align}
In this section, we construct the BDS integrand for all multiplicities. The main tasks are to analyze the \emph{unphysical} singularities present in the divergent double-boxes in \eqref{eq:divergent_db} and to identify the companion integrals required to cancel them. As these integrals serve solely to eliminate unphysical cuts in the divergent sector, their prefactors are necessarily determined by those of divergent double-box integrals. This structure is naturally encoded in terms of scaffolding triangulations, which represent leading singularities.

\subsection{Divergent two-loop Integrands and cancellation of unphysical cuts} \label{sec:3.1}

\begin{figure}
    \begin{subfigure}[b]{0.5\textwidth}
        \centering
        \begin{tikzpicture}
            \begin{scope}
                \draw[line width=0.2mm] (0,0)--(0,1) -- (1,1) -- (1,0) --(0,0);
                \draw[line width=0.2mm] (0,1) -- +(135:0.5);
                \draw[line width=0.2mm] (0,0) -- +(-135:0.5);
                \draw[line width=0.2mm] (1,1) -- (2,1) -- (2,0) -- (1,0);
                \draw[line width=0.2mm] (2,1) --  +(75:0.45);
                \draw[line width=0.2mm,dash=on 0.4ex off 0.2ex phase 0pt] (2,1) --  +(15:0.45);
                \draw[line width=0.2mm,dash=on 0.4ex off 0.2ex phase 0pt] (2,0) --  +(-75:0.45);
                \draw[line width=0.2mm] (2,0) --  +(-15 :0.45);
                \draw[line width=0.2mm,color=red!80!black] (0.5,0.8)--(0.5,1.2) (1.5,0.8)--(1.5,1.2) (0.8,0.5)--(1.2,0.5) 
                                             (0.5,0.2)--(0.5,-0.2) (1.5,-0.2)--(1.5,0.2);
                \foreach \i in {-1,...,1} {
                \fill[black] (2,1)+(45+\i *16: 0.35) circle (0.14ex);
                 \fill[black] (2,0)+(-45+\i *16: 0.35) circle (0.14ex);
                    }
            \end{scope}
            \begin{scope}[shift={(3.5,0)}]
                \draw[line width=0.2mm] (0,0)--(0,1) -- (1,1) -- (1,0) --(0,0);
                \draw[line width=0.2mm] (0,1) -- +(135:0.5);
                \draw[line width=0.2mm] (0,0) -- +(-135:0.5);
                \draw[line width=0.2mm] (1,1) -- (2,1) -- (2,0) -- (1,0);
                \draw[line width=0.2mm] (2,1) --  +(75:0.45);
                \draw[line width=0.2mm,dash=on 0.4ex off 0.2ex phase 0pt] (2,1) --  +(15:0.45);
                \draw[line width=0.2mm,dash=on 0.4ex off 0.2ex phase 0pt] (2,0) --  +(-75:0.45);
                \draw[line width=0.2mm] (2,0) --  +(-15 :0.45);
                \draw[line width=0.2mm,color=red!80!black] (0.5,0.8)--(0.5,1.2) (1.5,0.8)--(1.5,1.2) (0.8,0.5)--(1.2,0.5);
                \foreach \i in {-1,...,1} {
                    \fill[black] (2,1)+(45+\i *16: 0.35) circle (0.14ex);
                    \fill[black] (2,0)+(-45+\i *16: 0.35) circle (0.14ex);
                   \fill[black] (1,0)+(-90+\i *16: 0.35) circle (0.14ex);
                    }
                 \draw[line width=0.2mm] (1,0) --  +(-60 :0.45) (1,0)-- +(-120 :0.45);
            \end{scope}
        \end{tikzpicture}
        \vspace{0.3em}
\caption{Unphysical internal cubic cuts.}
\label{subfig:UnphyCuta}
    \end{subfigure}
    \begin{subfigure}[b]{0.5\textwidth}
        \centering
        \begin{tikzpicture}
            \begin{scope}
            \draw[line width=0.2mm] (0,0)--(0,1) -- (1,1) -- (1,0) --(0,0);
            \draw[line width=0.2mm] (0,1) -- +(135:0.5);
            \draw[line width=0.2mm] (0,0) -- +(-135:0.5);
            \draw[line width=0.2mm] (1,1) -- (2,1) -- (2,0) -- (1,0);
            \draw[line width=0.2mm] (2,1) --  +(75:0.45) (2,1) --  +(15:0.45);
            \draw[line width=0.2mm] (2,0) --  +(-15 :0.45) (2,0) --  +(-75:0.45);
            \draw[line width=0.2mm,color=red!80!black] (0.5,0.8)--(0.5,1.2) (0.5,0.2)--(0.5,-0.2)     
                                             (1.5,-0.2)--(1.5,0.2) (0.8,0.5)--(1.2,0.5) (2.2,0.5)--(1.8,0.5);
    \node[scale=0.8] at (1,1.35) {$x_{i}$};
    \node[scale=0.8] at (0.4,-0.35) {$x_{i-2}$};
    \node[scale=0.8] at (2.4,0.5) {$x_{j}$};
    \node[scale=0.8] at (1.5,-0.35) {$x_{k}$};
    \foreach \i in {-1,...,1} {
                \fill[black] (2,1)+(45+\i *16: 0.35) circle (0.14ex);
                \fill[black] (2,0)+(-45+\i *16: 0.35) circle (0.14ex);
                \fill[black] (1,0)+(-90+\i *16: 0.35) circle (0.14ex);
            }
  \draw[line width=0.2mm] (1,0) --  +(-60 :0.45) (1,0)-- +(-120 :0.45);
  \node at (3.0,0.5) {$\Rightarrow$};
\end{scope}   
\begin{scope}[shift={(4.7,1)}]
         \coordinate (U) at (0,0);
         \coordinate (D) at (0,-1);
         \path (U) -- +(-150:1) coordinate(L);
        \path (U) -- +(-30:1) coordinate(R);
        \draw[line width=0.2mm] (U)--(D) (U)--node[scale=0.8,xshift=-1ex,yshift=1.5ex]{$x_{i}$}(L)--node[scale=0.8,xshift=-1ex,yshift=-1.5ex]{$x_{i-2}$}(D)--node[scale=0.8,xshift=1ex,yshift=-1.5ex]{$x_{k}$}(R)--node[scale=0.8,xshift=1ex,yshift=1.5ex]{$x_{j}$}cycle;
        \draw[line width=0.2mm] (L)--+(150:0.5) (L)--+(-150:0.5) (U)--+(120:0.5) (U)--+(60:0.5) 
                                (R)--+(30:0.5) (R)--+(-30:0.5) (D)--+(-120:0.5) (D)--+(-60:0.5);
        \foreach \i in {-1,...,1} {
                \fill[black] (U)+(90+\i *16: 0.35) circle (0.14ex);
                \fill[black] (D)+(-90+\i *16: 0.35) circle (0.14ex);
                \fill[black] (R)+(0+\i *16: 0.35) circle (0.14ex);
            }
        \foreach \i in {U,D,L,R} {
        \node[circle,fill=gray!70,draw,minimum size=0.40cm,inner sep=0pt] at (\i) {};}
    \end{scope} 
\end{tikzpicture}
\caption{Unphysical elliptic cut.} \label{subfig:UnphyCutb}
    \end{subfigure}
    \caption{Unphysical cuts of divergent double-boxes} 
    \label{fig:UnphyCut}
\end{figure}
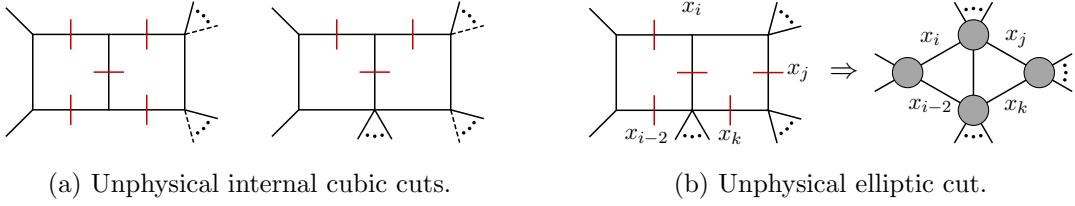

The individual divergent double-boxes possess two kinds of unphysical cuts in addition to the kissing-triangle cuts discussed previously. These two cuts are illustrated in figure~\ref{fig:UnphyCut}. In the following, we describe how these cuts cancel in the amplitudes and how this cancellation leads to a construction of the BDS integrand~\eqref{eq: BDSDef2}.

We start with the easier case: the internal cubic cuts shown in figure \ref{subfig:UnphyCuta}. 
Such cuts can be canceled by the following double-triangles,
\begin{equation*}
    I^{\text{dt}}(i,j;j,k):=\begin{tikzpicture}[baseline={([yshift=-0.75ex]current bounding box.center)}]
          \path (-30:1) -- +(30:1) coordinate(P1);
          \path (-30:1) -- +(90:0.5) coordinate(O);
          \draw[line width=0.2mm] (0,0) -- (30:1) -- (P1)--(-30:1)--cycle;
          \draw[line width=0.2mm] (-30:1) -- (30:1);
          \draw[line width=0.2mm]  (150:0.4)--(0,0)--(-150:0.4);
                 \foreach \i in {-1,...,1} {
                \fill[black] (0,0)+(-180+\i *15: 0.3) circle (0.12ex);
            }
          \draw[line width=0.2mm]  (P1)+(30:0.4)--(P1)--+(-30:0.4);
           \foreach \i in {-1,...,1} {
                \fill[black] (P1)+(0+\i *15: 0.3) circle (0.12ex);
            }
        \path (30:1)--+(90:0.2) coordinate (xj);
        \path (O)+(180:0.2) coordinate (xa);
        \path (O)+(0:0.2) coordinate (xb);
       \foreach \i in {xj,xa,xb} {
         \fill[mathblue] (\i) circle (0.3ex);}
        \draw[line width=0.4mm,color=mathblue] (xa)node[left,scale=0.8,color=black]{$x_{a}$}-- (xb)node[right,scale=0.8,color=black]{$x_{b}$} (xa)--(xj) (xb)--(xj)node[right,scale=0.8,color=black]{$x_{j}$};
        \draw[line width=0.4mm,color=mathblue] (xa) -- +(-120:0.55)node[scale=0.8,color=black,shift=(-120:0.2)]{$x_{i}$} ; 
        \draw[line width=0.4mm,color=mathblue] (xb) -- +(-60:0.55)node[scale=0.8,color=black,shift=(-60:0.2)]{$x_{k}$} ; 
        \draw[line width=0.2mm,dash=on 0.4ex off 0.2ex phase 0pt] (-30:1)--+(-60:0.4);
        \draw[line width=0.2mm,dash=on 0.4ex off 0.2ex phase 0pt] (-30:1)--+(-120:0.4);
          \foreach \i in {-1,...,0} {
                \fill[black] (-30:1)+(-102-\i *24: 0.3) circle (0.12ex);
            }
     \end{tikzpicture}=\int \frac{x_{i,j}^2 x_{j,k}^2 \: \dif^{3}x_{a}\dif^{3}x_{b}}{x_{a,i}^2 x_{a,j}^2  x_{a,b}^2 x_{b,j}^2 x_{b,k}^2} \:.
\end{equation*}

For the general double-boxes (include finite ones) with a single 3-vertex, one can easily verify that the following combination, which we denote as $\hat{I}^{\text{db}}(i_{1},i_{2},j;j,k_{1},k_{2})$,
\begin{align} \label{eq:mod_db1}
        \hat{I}^{\text{db}}(i_{1},i_{2},j;j,k_{1},k_{2}):=I^{\text{db}}(i_{1},i_{2},j;j,k_{1},k_{2}) 
        -\sum_{\alpha,\beta} (-1)^{\alpha+\beta} I^{\text{dt}} (i_{\alpha},j;j,k_{\beta})
\end{align}
vanishes on the cut $x_{a,b}^2=x_{a,j}^2=x_{b,j}^2=0$.

Similarly, for the double-boxes with two internal 3-vertices, we can define
\begin{align} \label{eq:mod_db2}
   \Hat{I}^{\text{db}}(i,j,k;k,\ell,i)&:=  I^{\text{db}}(i,j,k;k,\ell,i) + I^{\text{dt}}(i,k;k,i) \nonumber \\
         &\quad-I^{\text{dt}}(i,k;\ell,i)-I^{\text{dt}}(i,j;k,i)+I^{\text{dt}}(i,j;\ell,i) \nonumber \\
         &\quad-I^{\text{dt}}(i,k;k,\ell)+I^{\text{dt}}(j,k;k,\ell)-I^{\text{dt}}(i,k;k,j)  \:,
\end{align}
which is free of the unphysical cuts shown in figure \ref{subfig:UnphyCuta}. The degenerate cases with massless corners 
can be easily obtained by dropping terms such as $I^{\text{dt}}(j{-}1,j;j,k)$ since they are only constants in mass regularization (see appendix~\ref{sec: EdgeF}).

The cut shown in figure \ref{subfig:UnphyCutb} is more tricky and first occurs at eight points. This cut is elliptic, containing a non-rationalizable Jacobian factor, and unphysical, as it involves odd-particle sub-amplitudes. 
We denote such a cut in general as 
\begin{equation} \label{eq:def_Ecut}
    \Ecut{i,j}{k,\ell}\colon \{x_{a,i}^2=0,x_{a,j}^2=0,x_{a,b}^2=0,x_{b,k}^2=0,x_{b,\ell}^2=0\} \:.
\end{equation}
Moreover, such a cut (with $i,j,k,\ell$ all having the same parity) can only be cancelled among the double-boxes themselves. As the coefficients of double-boxes in the expansion~\eqref{eq:amplitude_expansion} have already been fixed by the kissing-triangle cuts, which are eqs.~\eqref{eq:dbcoeff_1}--\eqref{eq:soft-soft}, this cancellation requires identities among these leading singularities.

To see this, let us use the double-box $I^{\text{db}}(9,10,1;1,3,5)$ as an example (corresponding to $i=1$, $j=3$, $k=5$ and $i-2=9$ in figure~\ref{subfig:UnphyCutb}). The corresponding unphysical elliptic cut is $\Ecut{9,1}{3,5}$. It is straightforward to identify the other double-boxes possessing this cut, which are simply those whose arguments include $\{9,1,3,5\}$:
\begin{equation*}
    I^{\text{db}}(7,9,1;1,3,5), I^{\text{db}}(5,9,1;1,3,5),\text{and 8 more}. 
\end{equation*}
One can check that the residues of all these double-boxes on the cut $\Ecut{9,1}{3,5}$ are at most differ by a sign.  
Now, it's not hard to verify that the cut $\Ecut{9,1}{3,5}$ introduced by the divergent double-box 
$I^{\text{db}}(9,10,1;1,3,5)$ cancels in the combination
\begin{align*}
     \mathcal{D}_{1,3,5}I^{\text{db}}(9,10,1;1,3,5)+  \mathcal{D}_{1,3,5}^{7,9,1}I^{\text{db}}(7,9,1;1,3,5)
    +\mathcal{D}_{1,3,5}^{5,9,1}I^{\text{db}}(5,9,1;1,3,5) \:,
\end{align*}
due to the relations $\mathcal{D}_{1,3,5}=\mathcal{D}_{1,3,5}^{7,9,1}+\mathcal{D}_{1,3,5}^{5,9,1}$ and 
\begin{align*}
    I^{\text{db}}(9,10,1;1,3,5)\vert_{\Ecut{9,1}{3,5}}&=-I^{\text{db}}(7,9,1;1,3,5)\vert_{\Ecut{9,1}{3,5}} \\
    &=-I^{\text{db}}(5,9,1;1,3,5)\vert_{\Ecut{9,1}{3,5}} \:.
\end{align*}

For higher-point amplitudes, the above procedure becomes more and more complicated for two issues. 
First, the identities among the leading singularities $\mathcal{D}_{i,j,k}$ and $\mathcal{D}_{i,j,k}^{r,s,t}$ 
become more complicated, as their ingredients now involve higher-point amplitudes. 
Second, as was already the case above, the finite boxes that serve to balance the unphysical elliptic cuts of the divergent double-boxes also generate new unphysical elliptic cuts of their own. For instance, the double-box $I^{\text{db}}(7,9,1;1,3,5)$ contains an unphysical elliptic cut $\Ecut{7,9}{3,5}$ which requires other double-boxes to cancel it.
On the other hand, the collection of all double-boxes containing such unphysical elliptic cuts in the two-loop $n$-point amplitude is free of any such cuts. This special combination of double-boxes can be written explicitly as
\begin{align} \label{eq:col_db1}
    &\sum_{\substack{i\in [n]^{\sf{(o)}}, \\ \{\triangle\}\subset[n]^{(\sf{e})}}}
    \hspace{-1em}\mathcal{D}_{\triangle} I^{\text{db}}(i{-}1,i,i{+}1;\triangle)
    \hspace{1em}+ \hspace{-1em}\sum_{\substack{\{\triangle_{a}\}\nmid \{\triangle_{b}\},\\ \{\triangle_{a},\triangle_{b}\}\subset [n]^{(\sf{e})}}} \hspace{-0.5em}\mathcal{D}_{\triangle_{a}}^{\triangle_{b}}
    I^{\text{db}}(\triangle_{a};\triangle_{b}) \nonumber \\ 
    &+ \sum_{\{i,j\}\subset [n]^{\sf{(o)}}}  A_{n} I^{\text{db}}(i{-}1,i,i{+}1;j{-}1,j,j{+}1)
   +(\sf{e}\leftrightarrow\sf{o})\:,
\end{align}
where $[n]^{(\sf{e})}$ and $[n]^{(\sf{o})}$ denote the sets of even and odd numbers from 1 to $n$, respectively, 
$\{\triangle\}$ represents a triplet $\{r,s,t\}$, 
and $\{\triangle_{a}\}\nmid \{\triangle_{b}\} $ means that the two triplets do not intersect when embedded as triangles in an $n$-gon.
The second issue is thus solved, and the eq.~\eqref{eq:col_db1} provides a better starting point for the construction of the 
$n$-point BDS integrand.

To complete the construction, we still need to address the first issue, i.e., 
to find a basis that accounts for the identities among the $\mathcal{D}_{i,j,k}$ and $\mathcal{D}_{i,j,k}^{r,s,t}$. 
Surprisingly, the scaffolding triangulation representation of the on-shell diagrams described in the last section provides 
a beautiful solution. 
By replacing the amplitude factors in $\mathcal{D}_{i,j,k}$ and $\mathcal{D}_{i,j,k}^{r,s,t}$ (defined by eqs.~\eqref{eq: 1LD} and \eqref{eq:dbcoeff_1}) through the BCFW recursion relations, we find
\begin{equation} \label{eq:relofD}
    \mathcal{D}_{\triangle} = \sum_{\{\triangle\}\in \tau} \operatorname{OSD}_{\tau} \:, \qquad 
    \mathcal{D}_{\triangle_{a}}^{\triangle_{b}} = \sum_{\{\{\triangle_{a}\},\{\triangle_{b}\}\}\subset \tau} \operatorname{OSD}_{\tau} \:,
\end{equation}
where $\tau$ is a set of $(n-4)/2$ triplets representing a scaffolding triangulation of an $n$-gon. 
The previous example $\mathcal{D}_{1,3,5}=\mathcal{D}_{1,3,5}^{7,9,1}+\mathcal{D}_{1,3,5}^{5,9,1}$ for 10 points now is the consequence of 
\begin{equation*}
    \mathcal{D}_{1,3,5}=\underbrace{\operatorname{OSD}_{\{\{1,3,5\},\{7,9,1\},\{5,7,1\}\}}}_{=\mathcal{D}_{1,3,5}^{7,9,1}}+
    \underbrace{\operatorname{OSD}_{\{\{1,3,5\},\{5,9,1\},\{5,7,9\}\}}}_{=\mathcal{D}_{1,3,5}^{5,9,1}} \:.
\end{equation*}

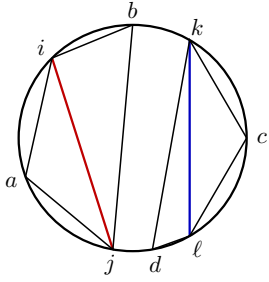
\begin{figure}
    \centering
    \begin{tikzpicture}
        \coordinate (Pi) at (135:1.5); 
        \coordinate (Pj) at (-100:1.5);
        \coordinate (Pk) at (60:1.5); 
        \coordinate (Pl) at (-60:1.5);
        \coordinate (i1) at (200:1.5);
        \coordinate (j1) at (90:1.5);
        \coordinate (k1) at (0:1.5);
        \coordinate (l1) at (-80:1.5);
        \draw[line width=0.3mm,color=red!75!black] (Pi) -- (Pj);
        \draw[line width=0.3mm,color=blue!75!black] (Pk) -- (Pl);
        \draw[line width=0.2mm] (i1)--(Pi)--(j1) (i1)--(Pj)--(j1) (k1)--(Pk)--(l1) (k1)--(Pl)--(l1); 
        \node[scale=0.8] at (135:1.7) {$i$};
        \node[scale=0.8] at (-100:1.7) {$j$};
        \node[scale=0.8] at (60:1.7) {$k$};
        \node[scale=0.8] at (-60:1.7) {$\ell$};
        \node[scale=0.8] at (200:1.7) {$a$};
        \node[scale=0.8] at (90:1.7) {$b$};
        \node[scale=0.8] at (0:1.7) {$c$};
        \node[scale=0.8] at (-80:1.7) {$d$};
        \draw[line width=0.3mm] (0,0) circle (1.5);
    \end{tikzpicture}
    \caption{Cancellation of the elliptic cut $\Ecut{i,j}{k,\ell}$: four 2STs arising from four triangles containing the corresponding edge-pair} 
    \label{fig:cancelofUEcut}
\end{figure}

We are now in a position to state the construction of the $n$-point BDS integrands \eqref{eq: BDSDef2}. First, we insert eq.~\eqref{eq:relofD} and the BCFW recursion relation for the tree amplitude $A_{n}^{\text{tree}}$ into eq.~\eqref{eq:col_db1} and then collect terms according to $\operatorname{ODS}_{\tau}$. This gives
\begin{equation}
    \sum_{\tau\in \mathcal{T}_{n}^{(\sf{e})}} \operatorname{ODS}_{\tau}
    \sum_{\{\{\triangle_{a}\},\{\triangle_{b}\}\}\subset \tau_{c}}I^{\text{db}}(\triangle_{a};\triangle_{b})+
    (\sf{e}\leftrightarrow \sf{o}) \:,
\end{equation}
where $\mathcal{T}^{(\sf{e})}_{n}$ ($\mathcal{T}^{(\sf{o})}_{n}$) denotes the set of all even (odd) scaffolding triangulations of an $n$-gon; the set $\tau_{c}\supset\tau$ also contains all external triangles such as $\{i-1,i,i+1\}$ of a scaffolding triangulation to include the double boxes such as $I^{\text{db}}(i{-}1,i,i{+}1;j{-}1,j,j{+}1)$ and $I^{\text{db}}(i{-}1,i,i{+}1;\triangle)$ in eq.~\eqref{eq:col_db1}. 
In this step, given a scaffolding triangulation $\tau$, we find a patricular combination of double boxes, $\sum_{\{\{\triangle_{a}\},\{\triangle_{b}\}\}\subset \tau_{c}}I^{\text{db}}(\triangle_{a};\triangle_{b})$, which avoids any unphysical elliptic cuts $\Ecut{i,j}{k,\ell}$. 
Indeed, any such cut can only occur when $\{i,j\}$ and $\{k,\ell\}$ are two chords of the triangulation $\tau_{c}$; we can thus localize it to the following combination (cf. figure \ref{fig:cancelofUEcut})
\begin{equation} \label{eq: cancel_Epcut}
    I^{\text{db}}(i,a,j;d,\ell,k) + I^{\text{db}}(i,a,j;\ell,c,k) 
    + I^{\text{db}}(i,j,b;d,\ell,k) + I^{\text{db}}(i,j,b;\ell,c,k)\:,
\end{equation} 
which vanishes on the cut $\Ecut{i,j}{k,\ell}$. This fact will be justified from a different perspective in the next section.

Then, we introduce the double-triangles as in \eqref{eq:mod_db1} and \eqref{eq:mod_db2} to cancel the unphysical internal cubic cuts. This completes our construction and gives
    \begin{equation}\label{eq:tri rep}
    \sum_{\tau\in \mathcal{T}_{n}^{(\sf{e})}} \operatorname{ODS}_{\tau}
    \underbrace{\sum_{\{\{\triangle_{a}\},\{\triangle_{b}\}\}\subset \tau_{c}}\Big(I^{\text{db}}(\triangle_{a};\triangle_{b})+
    I^{\text{dt}}\text{'s via eqs.~\eqref{eq:mod_db1} and \eqref{eq:mod_db2}}\Big)}_{:=\operatorname{BDS}_{n,\tau}^{\text{3D}}}+
    (\sf{e}\leftrightarrow \sf{o}) \:.
\end{equation}
As we know, the IR divergence of two-loop amplitudes is proportional to tree amplitudes; this requires that $\operatorname{BDS}_{n,\tau}^{\text{3D}}$ share the same divergent piece. It turns out that the situation is much better: all $\operatorname{BDS}_{n,\tau}^{\text{3D}}$ are actually the same, as we will show in the next section.

In conclusion, the double-box integrals contributing to the BDS integrand can be identified as two shaded triangles (2STs) in an $n$-gon, from which we can straightforwardly read off their on-shell prefactors: tree amplitudes, $\mathcal{D}_{i,j,k}$, and $\mathcal{D}^{r,s,t}_{i,j,k}$. In particular: 
\begin{itemize}
    \item These on-shell prefactors are simply the sum over on-shell diagrams that can be identified as scaffolding triangulations that contain the corresponding 2ST,
    \item In the basis of on-shell diagrams, i.e., a given scaffolding triangulation, each is attached to the sum over all 2STs compatible with that triangulation. This guarantees that all IR-divergences are captured and all unphysical cuts are canceled. 
\end{itemize}
Therefore, the $n$-point BDS integrand can be constructed in a graphical manner. One simply starts with all possible scaffolding triangulations, where for each triangulation we sum over all 2STs. Each 2ST represents  double-box integrals defined by the vertices of the two triangles. Summing over all triangulations then gives the BDS integrand. We will use the last property to give the explicit BDS integrand in the next subsection.

\subsection{Examples: the six- and eight-point BDS Integrands}
Having established the general construction of the $n$-point BDS integrand from eq.~\eqref{eq:col_db1} and its final realization through scaffolding triangulations in eq.~\eqref{eq:tri rep}, let us now illustrate how this mechanism operates in explicit cases.

These examples make transparent how the combinatorial structure of scaffolding triangulations and their compatible 2STs reproduces the known BDS integrands at six and eight points~\cite{Caron-Huot:2012sos,He:2022lfz}.

\paragraph{Six-point example.} At six points, the situation is particularly simple: there are only two distinct scaffolding triangulations, corresponding to connecting either the even or the odd vertices of the hexagon,
\begin{equation} 
   \tau_{\sf{o}} = \{\{1,3,5\}\}=\begin{tikzpicture}[baseline={([yshift=-0.7ex]current bounding box.center)}]
        \begin{scope}[baseline={([yshift=-0.6ex]current bounding box.center)},shift={(8,-0.8)}]
            \foreach \i  in {1,...,6} {
        \coordinate (P\i) at (-60-60*\i:0.7);
    }
    \foreach \i  in {1,...,6} {
        \coordinate (P\i) at (-60-60*\i:0.7);
        \node[scale=0.7] at (P\i) [shift=(-60-60*\i:0.25)] {\i};
             }
    \draw[line width=0.15mm] (P1) \foreach \i in {1,3,5} { -- (P\i) } -- cycle;
         \draw[line width=0.15mm] (P1)--(P3);
             \draw[line width=0.15mm] (P1)--(P5);
                 \draw[line width=0.15mm] (P3)--(P5);
    \draw[line width=0.3mm] (P1) \foreach \i in {2,...,6} { -- (P\i) } -- cycle;
        \end{scope}
     \end{tikzpicture},\ 
     \tau_{\sf{e}} = \{\{2,4,6\}\}=\begin{tikzpicture}[baseline={([yshift=-0.7ex]current bounding box.center)}]
        \begin{scope}[baseline={([yshift=-0.6ex]current bounding box.center)},shift={(11,-0.8)}]
            \foreach \i  in {1,...,6} {
        \coordinate (P\i) at (-60-60*\i:0.7);
    }
    \foreach \i  in {1,...,6} {
        \coordinate (P\i) at (-60-60*\i:0.7);
        \node[scale=0.7] at (P\i) [shift=(-60-60*\i:0.25)] {\i};
             }
         \draw[line width=0.15mm] (P2)--(P4);
             \draw[line width=0.15mm] (P2)--(P6);
                 \draw[line width=0.15mm] (P4)--(P6);
    \draw[line width=0.3mm] (P1) \foreach \i in {2,...,6} { -- (P\i) } -- cycle;
        \end{scope}
     \end{tikzpicture}.
\end{equation}
Each triangulation is associated with an on-shell data,
$\operatorname{OSD}_{\{\{1,3,5\}\}}$ and $\operatorname{OSD}_{\{\{2,4,6\}\}}$, 
which in this case coincide with the one-loop maximal cuts 
$\mathcal{D}_{1,3,5}$ and $\mathcal{D}_{2,4,6}$, respectively.  

For each $\tau$, the algorithmic construction proceeds as follows:
\begin{enumerate}
    \item Start with the triangulation $\tau_{\sf{o}}$ (or $\tau_{\sf{e}}$).
    \item List all 2STs compatible with it, and summing over all compatible 2STs yields the corresponding BDS integrand.   
    For $\tau_{\sf{o}}$, it can be represented as
    \begin{equation} 
    \operatorname{BDS}^{\text{3D}}_{6,\tau_{\sf{o}}}
    =\left.\begin{tikzpicture}[baseline={([yshift=-0.7ex]current bounding box.center)}]
        \foreach \i  in {1,...,6} {
        \coordinate (P\i) at (-60-60*\i:0.7);
    }
    \foreach \i  in {1,...,6} {
        \coordinate (P\i) at (-60-60*\i:0.7);
        \node[scale=0.7] at (P\i) [shift=(-60-60*\i:0.25)] {\i};
             }
    \draw[line width=0.15mm] (P1) \foreach \i in {1,3,5} { -- (P\i) } -- cycle;
        \fill[color=gray!70] (P1)--(P2)--(P3)--cycle;
     \fill[color=gray!70] (P3)--(P4)--(P5)--cycle;
         \draw[line width=0.15mm] (P1)--(P3);
             \draw[line width=0.15mm] (P1)--(P5);
                 \draw[line width=0.15mm] (P3)--(P5);
    \draw[line width=0.3mm] (P1) \foreach \i in {2,...,6} { -- (P\i) } -- cycle;
     \end{tikzpicture}
     +\begin{tikzpicture}[baseline={([yshift=-0.7ex]current bounding box.center)}]
        \foreach \i  in {1,...,6} {
        \coordinate (P\i) at (-60-60*\i:0.7);
    }
    \foreach \i  in {1,...,6} {
        \coordinate (P\i) at (-60-60*\i:0.7);
        \node[scale=0.7] at (P\i) [shift=(-60-60*\i:0.25)] {\i};
             }
    \draw[line width=0.15mm] (P1) \foreach \i in {1,3,5} { -- (P\i) } -- cycle;
        \fill[color=gray!70] (P1)--(P2)--(P3)--cycle;
     \fill[color=gray!70] (P1)--(P3)--(P5)--cycle;
         \draw[line width=0.15mm] (P1)--(P3);
             \draw[line width=0.15mm] (P3)--(P5);
                 \draw[line width=0.15mm] (P1)--(P5);
    \draw[line width=0.3mm] (P1) \foreach \i in {2,...,6} { -- (P\i) } -- cycle;
     \end{tikzpicture}
     +\text{4 more}\:.\right.
\end{equation}
 Similarly for $\tau_{\sf{e}}$, its representation is followed by a one-site cyclic shift.  
    \item For each 2ST, assign its corresponding double box $I^{\text{db}}(\triangle_a;\triangle_b)$ and the double-triangle modification $I^{\text{dt}}$ according to eqs.~\eqref{eq:mod_db1}--\eqref{eq:mod_db2}. For example,
    \begin{align}
        &\begin{tikzpicture}[baseline={([yshift=-0.7ex]current bounding box.center)}]
        \foreach \i  in {1,...,6} {
        \coordinate (P\i) at (-60-60*\i:0.7);
        \node[scale=0.7] at (P\i) [shift=(-60-60*\i:0.25)] {\i};
         }
    \draw[line width=0.15mm] (P1) \foreach \i in {1,3,5} { -- (P\i) } -- cycle;
    \fill[color=gray!70] (P1)--(P2)--(P3)--cycle (P3)--(P4)--(P5)--cycle;
    \draw[line width=0.15mm] (P1)--(P3) (P1)--(P5) (P3)--(P5);
    \draw[line width=0.3mm] (P1) \foreach \i in {2,...,6} { -- (P\i) } -- cycle;
     \end{tikzpicture}
     \to \hat{I}^{\text{db}}(1,2,3;3,4,5)=I^\text{db}(1,2,3;3,4,5)+I^\text{dt}(1,3;3,5)\:.\\
&     \begin{tikzpicture}[baseline={([yshift=-0.7ex]current bounding box.center)}]
        \foreach \i  in {1,...,6} {
        \coordinate (P\i) at (-60-60*\i:0.7);
        \node[scale=0.7] at (P\i) [shift=(-60-60*\i:0.25)] {\i};
    }
    \draw[line width=0.15mm] (P1) \foreach \i in {1,3,5} { -- (P\i) } -- cycle;
    \fill[color=gray!70] (P1)--(P2)--(P3)--cycle (P1)--(P3)--(P5)--cycle;
    \draw[line width=0.15mm] (P1)--(P3) (P3)--(P5) (P1)--(P5);
    \draw[line width=0.3mm] (P1) \foreach \i in {2,...,6} { -- (P\i) } -- cycle;
    \end{tikzpicture}
    \to  \hat{I}^{\text{db}}(1,2,3;3,5,1)= 
    \raisebox{-1.9ex}{$
    \begin{aligned}
    &I^\text{db}(1,2,3;3,5,1)+I^\text{dt}(1,3;3,1)\\
     &\quad    -I^\text{dt}(1,3;3,5)-I^\text{dt}(1,3;5,1)\:.
    \end{aligned}$}
    \end{align}
    These two 2ST configurations, corresponding precisely to the “crab” and “2-mass hard” integrals of~\cite{Caron-Huot:2012sos}.
\end{enumerate}
Combining both triangulations with their respective on-shell data yields the complete six-point BDS integrand:
\begin{equation}
    \begin{split}
       A_6^{\text{tree}} \operatorname{BDS}^{\text{3D}}_{6}
    &= \operatorname{OSD}_{\tau_{\sf{e}}}\operatorname{BDS}^{\text{3D}}_{6,\tau_{\sf{e}}}
    - \operatorname{OSD}_{\tau_{\sf{o}}} \operatorname{BDS}^{\text{3D}}_{6,\tau_{\sf{o}}}\\
    &=\sum_{\{i,j\}\subset\{2,4,6\}}A_6 \,\hat{I}^\text{db}(i{-}1,i,i{+}1;j{-}1,j,j{+}1)
       \\
       &\quad +\sum_{i=1,3,5}(-1)^{i-1}\mathcal{D}_{2,4,6}\, \hat{I}^\text{db}(i{-}1,i,i{+}1;\Delta_{2,4,6})+
    (\sf{e}\leftrightarrow \sf{o}) 
    \end{split}
\end{equation}
in perfect agreement with the explicit construction of~\cite{Caron-Huot:2012sos}.  
Thus, the sum over compatible 2STs within each scaffolding reproduces exactly the BDS integrand at six points.

\paragraph{Eight-point example.}
We now move to eight-points where the BDS integrand was given in~\cite{He:2022lfz}. At eight points the structure becomes richer. There are four distinct on-shell data: 
\begin{equation} 
   \begin{split}
       \text{Odd scaffolding: }&\begin{tikzpicture}[baseline={([yshift=-0.7ex]current bounding box.center)}]
        \foreach \i  in {1,...,8} {
        \coordinate (P\i) at (-45-45*\i:0.7);
    }
    \foreach \i  in {1,...,8} {
        \coordinate (P\i) at (-45-45*\i:0.7);
        \node[scale=0.7] at (P\i) [shift=(-45-45*\i:0.25)] {\i};
             }
    \draw[line width=0.15mm] (P1) \foreach \i in {3,5,7} { -- (P\i) } -- cycle;
    \draw[line width=0.3mm] (P1) \foreach \i in {2,...,8} { -- (P\i) } -- cycle;
     \end{tikzpicture}
     \supset \tau_{1}=\begin{tikzpicture}[baseline={([yshift=-0.7ex]current bounding box.center)}]
        \foreach \i  in {1,...,8} {
        \coordinate (P\i) at (-45-45*\i:0.7);
    }
    \foreach \i  in {1,...,8} {
        \coordinate (P\i) at (-45-45*\i:0.7);
        \node[scale=0.7] at (P\i) [shift=(-45-45*\i:0.25)] {\i};
             }
    \draw[line width=0.15mm] (P1) \foreach \i in {3,5,7} { -- (P\i) } -- cycle;
    \draw[line width=0.15mm] (P1)--(P5);
    \draw[line width=0.3mm] (P1) \foreach \i in {2,...,8} { -- (P\i) } -- cycle;
     \end{tikzpicture},\quad
      \tau_{3}=\begin{tikzpicture}[baseline={([yshift=-0.7ex]current bounding box.center)}]
        \foreach \i  in {1,...,8} {
        \coordinate (P\i) at (-45-45*\i:0.7);
    }
    \foreach \i  in {1,...,8} {
        \coordinate (P\i) at (-45-45*\i:0.7);
        \node[scale=0.7] at (P\i) [shift=(-45-45*\i:0.25)] {\i};
             }
    \draw[line width=0.15mm] (P1) \foreach \i in {3,5,7} { -- (P\i) } -- cycle;
    \draw[line width=0.15mm] (P3)--(P7);
    \draw[line width=0.3mm] (P1) \foreach \i in {2,...,8} { -- (P\i) } -- cycle;
     \end{tikzpicture},\\
    \text{Even scaffolding: } &\begin{tikzpicture}[baseline={([yshift=-0.7ex]current bounding box.center)}]
        \foreach \i  in {1,...,8} {
        \coordinate (P\i) at (-45-45*\i:0.7);
    }
    \foreach \i  in {1,...,8} {
        \coordinate (P\i) at (-45-45*\i:0.7);
        \node[scale=0.7] at (P\i) [shift=(-45-45*\i:0.25)] {\i};
             }
    \draw[line width=0.15mm] (P2) \foreach \i in {4,6,8} { -- (P\i) } -- cycle;
    \draw[line width=0.3mm] (P1) \foreach \i in {2,...,8} { -- (P\i) } -- cycle;
     \end{tikzpicture}
     \supset \tau_{2}=\begin{tikzpicture}[baseline={([yshift=-0.7ex]current bounding box.center)}]
        \foreach \i  in {1,...,8} {
        \coordinate (P\i) at (-45-45*\i:0.7);
    }
    \foreach \i  in {1,...,8} {
        \coordinate (P\i) at (-45-45*\i:0.7);
        \node[scale=0.7] at (P\i) [shift=(-45-45*\i:0.25)] {\i};
             }
    \draw[line width=0.15mm] (P2) \foreach \i in {4,6,8} { -- (P\i) } -- cycle;
    \draw[line width=0.15mm] (P2)--(P6);
    \draw[line width=0.3mm] (P1) \foreach \i in {2,...,8} { -- (P\i) } -- cycle;
     \end{tikzpicture}
     ,\quad 
    \tau_{4}= \begin{tikzpicture}[baseline={([yshift=-0.7ex]current bounding box.center)}]
        \foreach \i  in {1,...,8} {
        \coordinate (P\i) at (-45-45*\i:0.7);
    }
    \foreach \i  in {1,...,8} {
        \coordinate (P\i) at (-45-45*\i:0.7);
        \node[scale=0.7] at (P\i) [shift=(-45-45*\i:0.25)] {\i};
             }
    \draw[line width=0.15mm] (P2) \foreach \i in {4,6,8} { -- (P\i) } -- cycle;
    \draw[line width=0.15mm] (P4)--(P8);
    \draw[line width=0.3mm] (P1) \foreach \i in {2,...,8} { -- (P\i) } -- cycle;
     \end{tikzpicture},
   \end{split}
\end{equation}
which satisfy eq.~\eqref{eq:8pt relation}. Since each triangulation is related by clyclic shift, let's focus on the triangulation $\tau_{1} = \{\{1,3,5\},\{5,7,1\}\}$.

According to our procedure, list all 2STs compatible with it, and summing over all compatible ones and  yields 
\begin{equation} 
   \begin{split}
     \operatorname{BDS}^{\text{3D}}_{8,\tau_{1}}=&\begin{tikzpicture}[baseline={([yshift=-0.7ex]current bounding box.center)}]
        \foreach \i  in {1,...,8} {
        \coordinate (P\i) at (-45-45*\i:0.6);
        \node[scale=0.7] at (P\i) [shift=(-45-45*\i:0.25)] {\i};
    }
    \draw[line width=0.15mm] (P1) \foreach \i in {3,5,7} { -- (P\i) } -- cycle (P1)--(P5);
        \fill[color=gray!70] (P1)--(P2)--(P3)--cycle (P3)--(P4)--(P5)--cycle;
    \draw[line width=0.3mm] (P1) \foreach \i in {2,...,8} { -- (P\i) } -- cycle;
     \end{tikzpicture}
     +
    \begin{tikzpicture}[baseline={([yshift=-0.7ex]current bounding box.center)}]
        \foreach \i  in {1,...,8} {
        \coordinate (P\i) at (-45-45*\i:0.6);
        \node[scale=0.7] at (P\i) [shift=(-45-45*\i:0.25)] {\i};
    }
    \draw[line width=0.15mm] (P1) \foreach \i in {3,5,7} { -- (P\i) } -- cycle;
    \draw[line width=0.15mm] (P1)--(P5);
        \fill[color=gray!70] (P1)--(P2)--(P3)--cycle;
     \fill[color=gray!70] (P5)--(P6)--(P7)--cycle;
    \draw[line width=0.3mm] (P1) \foreach \i in {2,...,8} { -- (P\i) } -- cycle;
     \end{tikzpicture}
     + \begin{tikzpicture}[baseline={([yshift=-0.7ex]current bounding box.center)}]
        \foreach \i  in {1,...,8} {
        \coordinate (P\i) at (-45-45*\i:0.6);
                \node[scale=0.7] at (P\i) [shift=(-45-45*\i:0.25)] {\i};
    }
    \fill[color=gray!70] (P1)--(P2)--(P3)--cycle;
     \fill[color=gray!70] (P1)--(P3)--(P5)--cycle;
    \draw[line width=0.15mm] (P1) \foreach \i in {3,5,7} { -- (P\i) } -- cycle;
    \draw[line width=0.15mm] (P1)--(P5);
    \draw[line width=0.3mm] (P1) \foreach \i in {2,...,8} { -- (P\i) } -- cycle;
     \end{tikzpicture}+
     \begin{tikzpicture}[baseline={([yshift=-0.7ex]current bounding box.center)}]
        \foreach \i  in {1,...,8} {
        \coordinate (P\i) at (-45-45*\i:0.7);
    }
    \foreach \i  in {1,...,8} {
        \coordinate (P\i) at (-45-45*\i:0.6);
        \node[scale=0.7] at (P\i) [shift=(-45-45*\i:0.25)] {\i};
             }
    \fill[color=gray!70] (P1)--(P2)--(P3)--cycle;
     \fill[color=gray!70] (P1)--(P5)--(P7)--cycle;
    \draw[line width=0.15mm] (P1) \foreach \i in {3,5,7} { -- (P\i) } -- cycle;
    \draw[line width=0.15mm] (P1)--(P5);
    \draw[line width=0.3mm] (P1) \foreach \i in {2,...,8} { -- (P\i) } -- cycle;
     \end{tikzpicture}
      +
     \begin{tikzpicture}[baseline={([yshift=-0.7ex]current bounding box.center)}]
        \foreach \i  in {1,...,8} {
        \coordinate (P\i) at (-45-45*\i:0.6);
    }
    \foreach \i  in {1,...,8} {
        \coordinate (P\i) at (-45-45*\i:0.6);
        \node[scale=0.7] at (P\i) [shift=(-45-45*\i:0.25)] {\i};
             }
     \fill[color=gray!70] (P1)--(P3)--(P5)--(P7)--cycle;
    \draw[line width=0.15mm] (P1) \foreach \i in {3,5,7} { -- (P\i) } -- cycle;
    \draw[line width=0.15mm] (P1)--(P5);
    \draw[line width=0.3mm] (P1) \foreach \i in {2,...,8} { -- (P\i) } -- cycle;
     \end{tikzpicture}
     +\text{10 more}\:.
   \end{split}
\end{equation}
Each 2ST then map to the double box with acommpanying triangles as following. For example, 
\begin{align}
     & 
     \begin{tikzpicture}[baseline={([yshift=-0.7ex]current bounding box.center)}]
        \foreach \i  in {1,...,8} {
        \coordinate (P\i) at (-45-45*\i:0.7);
        \node[scale=0.7] at (P\i) [shift=(-45-45*\i:0.25)] {\i};
    }
    \draw[line width=0.15mm] (P1) \foreach \i in {3,5,7} { -- (P\i) } -- cycle (P1)--(P5);
    \fill[color=gray!70] (P1)--(P2)--(P3)--cycle (P5)--(P6)--(P7)--cycle;
    \draw[line width=0.3mm] (P1) \foreach \i in {2,...,8} { -- (P\i) } -- cycle;
     \end{tikzpicture}
     \to {I}^\text{db}(1,2,3;5,6,7)\\
   &  \begin{tikzpicture}[baseline={([yshift=-0.7ex]current bounding box.center)}]
        \foreach \i  in {1,...,8} {
        \coordinate (P\i) at (-45-45*\i:0.7);
    }
    \foreach \i  in {1,...,8} {
        \coordinate (P\i) at (-45-45*\i:0.7);
        \node[scale=0.7] at (P\i) [shift=(-45-45*\i:0.25)] {\i};
             }
    \fill[color=gray!70] (P1)--(P2)--(P3)--cycle;
     \fill[color=gray!70] (P1)--(P5)--(P7)--cycle;
    \draw[line width=0.15mm] (P1) \foreach \i in {3,5,7} { -- (P\i) } -- cycle;
    \draw[line width=0.15mm] (P1)--(P5);
    \draw[line width=0.3mm] (P1) \foreach \i in {2,...,8} { -- (P\i) } -- cycle;
     \end{tikzpicture}\to
     \raisebox{-0.9em}{$
     \begin{aligned}
         \hat{I}^\text{db}(1,2,3;5,7,1)&={I}^\text{db}(1,2,3;5,7,1)-I^\text{dt}(1,3;5,1)\\
         &\quad+I^\text{dt}(1,3;5,7)
     \end{aligned}$}\\
    & \begin{tikzpicture}[baseline={([yshift=-0.7ex]current bounding box.center)}]
        \foreach \i  in {1,...,8} {
        \coordinate (P\i) at (-45-45*\i:0.7);
    }
    \foreach \i  in {1,...,8} {
        \coordinate (P\i) at (-45-45*\i:0.7);
        \node[scale=0.7] at (P\i) [shift=(-45-45*\i:0.25)] {\i};
             }
     \fill[color=gray!70] (P1)--(P3)--(P5)--(P7)--cycle;
    \draw[line width=0.15mm] (P1) \foreach \i in {3,5,7} { -- (P\i) } -- cycle;
    \draw[line width=0.15mm] (P1)--(P5);
    \draw[line width=0.3mm] (P1) \foreach \i in {2,...,8} { -- (P\i) } -- cycle;
     \end{tikzpicture}\to
     \raisebox{-4ex}{
     $\begin{aligned}
         &\hat{I}^\text{db}(1,3,5;5,7,1)=
         {I}^\text{db}(1,3,5;5,7,1)+{I}^\text{dt}(1,5;5,1)\\
       & \hspace{3em} -{I}^\text{dt}(1,5;7,1) - {I}^\text{dt}(1,5;5,7)- {I}^\text{dt}(1,3;5,1)\\
       & \hspace{3em} -{I}^\text{dt}(3,5;5,1)+ {I}^\text{dt}(1,3;7,1)+ {I}^\text{dt}(3,5;5,7)
     \end{aligned}$}
\end{align}
These configurations reproduce exactly the $I_{B}^{\text{db}}$, $I_{E}^{\text{db}}$, and $I_{G}^{\text{db}}$ integrals identified in~\cite{He:2022lfz}.
(The first and third, $\hat{I}^{\text{db}}(1,2,3;3,4,5)$ and $\hat{I}^{\text{db}}(1,2,3;3,5,1)$, have already appeared in the six-point case and will not be repeated here.)

Combining the four triangulations $\tau_{i}$ with their associated on-shell data gives the full eight-point BDS integrand:
\begin{align}
    &\quad A_8^{\text{tree}} \operatorname{BDS}^{\text{3D}}_{8}
    =\sum_{i=1}^4 \operatorname{OSD}_{\tau_{i}}\operatorname{BDS}^{\text{3D}}_{8,\tau_{i}} \nonumber \\
    &=A_8^{\text{tree}}\sum_{i=1}^8 \hat{I}^\text{db}(i{-}1,i,i{+}1;i{-}3,i{-}2,i{-}1){+}A_8 \sum_{i=1}^4 I^\text{db}(i{-}1,i,i{+}1;i{+}3,i{+}4,i{+}5) \nonumber \\
    &\quad+\sum_{i=1}^8 (-1)^{i-1}\Bigg[ \sum_{j=1}^2 \mathcal{D}_{i{-}1,i{+}1,i{+}2j{+}1} \hat{I}^{\text{db}} (i{-}1,i, i{+}1;i{+}1,i{+}2j{+}1,i{-}1) \nonumber \\
        &\quad{+} \mathcal{D}_{i{-}1,i{+}3,i{+}5 }\hat{I}^{\text{db}}(i{-}1,i,i{+}1;i{+}3,i{+}5,i{-}1){+}  \mathcal{D}_{i{+}1,i{+}3,i{+}5} \hat{I}^{\text{db}}(i{-}1,i,i{+}1;i{+}1,i{+}3,i{+}5)\Bigg] \nonumber \\
        &\quad +\sum_{i=1}^4 (-1)^i  \mathcal{D}^{i,i+2,i+4}_{i{+}4,i{+}6,i}\, \hat{I}^{\text{db}}(i,i{+}2,i{+}4;i{+}4,i{+6},i)
\end{align}
again in perfect agreement with~\cite{He:2022lfz}. 

\medskip
These examples make manifest that the $n$-point BDS integrand is nothing but the canonical object obtained by summing over all 2ST-compatible triangulations of the $n$-gon scaffolding, thereby geometrizing the exponentiation structure of infrared divergences.

\section{``Triangulations" of the BDS function} \label{sec: tri_func}

So far, we have introduced a graphical method for constructing the BDS integrand based on the combinatorics of scaffolding triangulations of a polygon. However, why integrands associated with distinct triangulations yield the same \emph{integrated result} remains somewhat mysterious.

From a graphical perspective, when two different triangulations produce the same result, this suggests that only the external edges carry physical significance, while the internal edges encode certain cancellations. Remarkably, we will see that this is precisely the case. The key observation is that the double-box can be expressed as a sum of nine one-fold integrals over the four-dimensional one-loop box. For those double-boxes that can be represented by a 2ST in a scaffolding triangulation, each such function is associated with an edge-pairing, that is, one edge from each of the two triangles comprising this 2ST. We therefore refer to these nine terms as edge-pairing functions. Crucially, it is exactly those edge-pairing functions involving internal edges that cancel when all 2STs are summed over within a given triangulation.

\subsection{Two-loop integrals as edge-pairing functions} \label{sec: 4.1}

In this subsection, we give a derivation of eq.~\eqref{eq: IntroEdge}. For convenience, we recall the equation here:
\begin{align}\label{eq: EdgePairRep}
        I^{\text{db}}(i,j,k;r,s,t)
        &=   \Ep{i,j}{r,s} - \Ep{i,j}{r,t} + \Ep{i,j}{s,t} - \Ep{i,k}{r,s} +\Ep{i,k}{r,t} \nonumber \\
        &\quad -\Ep{i,k}{s,t}+\Ep{j,k}{r,s} -\Ep{j,k}{r,t}+\Ep{j,k}{s,t}  \:,
\end{align}
where we have introduced the \emph{edge-pairing function}
\begin{equation}\label{eq: EDef}
    \Ep{i , j}{ k , \ell}=\int \frac{\dif c\, [\dif \vec{a}_{L}]\, \dif^2 \vec{b}_{R}}{ \pi \sqrt{c} \, (1+c)} 
    \frac{ X_{i\ell}X_{jk}- X_{ik}X_{j\ell}}{4\left((c{+}1)X_{L}^2+ 2X_{L}\cdot X_{R}+X_{R}^2\right)^2}\:,
\end{equation} 
with $X_{L}:=a_i\, X_i+a_j\, X_j$ and $X_{R}:=b_k\, X_k+ b_\ell \, X_\ell$. Here, we introduce the index sets $L=\{i,j\}$ and $R=\{k,\ell\}$, and define $[\dif^{|I|-1}\vec{a}_{I}]:=\dif^{|I|}\vec{a}_{I}\,\delta(a_{i_0}-1)$ for any $i_0 \in I$ and an index set $I$.
Note that when $i,j,k,\ell$ are all distinct, the edge-pairing function is essentially a one-fold integral of a $c$-deformed one-loop box function in four dimensions.

The derivation simply follows the procedure described in \cite{Paulos:2012nu,Caron-Huot:2012sos,Spiering:2024sea}. First, by introducing the Feynman parameters and performing loop-by-loop integration~(c.f.~\cite{Bourjaily:2019jrk}), we obtain 
\begin{equation}
      I^{\text{db}}(i,j,k;r,s,t)=\int\frac{[\dif^{2}\vec{a}_{L}]\dif^{3}\vec{b}_{R}}{(-X_{L}^{2})^{1/2}}
    \frac{3}{16}\Biggl(\frac{5}{2}\frac{\mathcal{G}_{\{i,j,k,Y_{[6]}\}}^{\{r,s,t,Y_{[6]}\}}}{(-Y_{[6]}\cdot Y_{[6]})^{7/2}}
     -\frac{\mathcal{G}_{\{i,j,k\}}^{\{r,s,t\}}}{(-Y_{[6]}\cdot Y_{[6]})^{5/2}}\Biggr)  \:,
\end{equation}
where now $L:=\{i,j,k\}$, $R:=\{r,s,t\}$, and $Y_{[6]}=X_{L}+X_{R}$. Here, the Gram determinant is computed in the embedding space. Applying the Feynman trick again,
\begin{equation}
    \frac{1}{A^{(2n-1)/2}B^{1/2}}=\frac{\Gamma(n)}{\Gamma\big(n-\tfrac{1}{2}\big)\sqrt{\pi}} \int_{0}^{\infty}\frac{\dif c}{\sqrt{c}\,(A+c B)^{n}} \quad \text{for $n\in \mathbb{Z}_{\geq 1}$,}
\end{equation}
we arrive at
\begin{equation}
    I^{\text{db}}(i,j,k;r,s,t)=\int\frac{[\dif^{2}\vec{a}_{L}]\dif^{3}\vec{b}_{R}\,\dif c}{4\pi\sqrt{c}}
    \Biggl(\frac{6\,\mathcal{G}_{\{i,j,k,Y_{[6]}\}}^{\{r,s,t,Y_{[6]}\}}}{\bigl(-Y_{[6]}\cdot Y_{[6]}-c X_{L}^{2}\bigr)^{4}}
      -\frac{2\,\mathcal{G}_{\{i,j,k\}}^{\{r,s,t\}}}{\bigl(-Y_{[6]}\cdot Y_{[6]}-c X_{L}^{2}\bigr)^{3}}\Biggr) \:,
\end{equation}
which is simply a one-fold integral of a $c$-deformed one-loop \emph{hexagon} in four dimensions. 
The eq.~\eqref{eq: EdgePairRep} then can be easily obtained through the familiar one-loop reduction (c.f.~\cite{Arkani-Hamed:2017ahv}).

Note that eq.~\eqref{eq: EdgePairRep} can be naturally encoded in the 2ST representation of the double-box integrals. Starting from a given 2ST, we regard the edges of each shaded triangle as a set. Pairing one edge from each set produces nine distinct combinations, which correspond precisely to the nine edge-pairing functions given in eq.~\eqref{eq: EdgePairRep}.
Next, we assign weights to the edges of the 2ST by giving a minus sign to the innermost edge-pair --- defined as the pair with no other edges of the 2ST lying between them --- and a plus sign to the outer edges. The overall sign of an edge-pairing function is then simply the product of the weights of its constituent edges, i.e., 
\begin{equation}
w_{i,j}w_{p,q}\Ep{i,j}{p,q}
\end{equation}
where $w_{i,j}=\pm1$ and $w_{p,q}=\pm1$ is the weight for the two edges. This can be represented graphically as follows: 
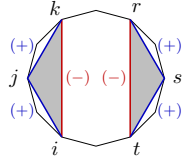
\begin{equation*}
      \centering
      \begin{tikzpicture}
    \begin{scope}
          \foreach \i  in {1,...,12} {
        \coordinate (P\i) at (-30*\i:0.9);
             }
        \draw[line width =0.1mm] (P1) \foreach \i in {2,...,12} {--(P\i)} -- cycle;
         \fill[color=gray!50] (P6)--(P8)--(P4)--cycle;
          \fill[color=gray!50] (P10)--(P12)--(P2)--cycle;
          \draw[line width=0.2mm,color=red!75!black] (P4)--node[scale=0.55,xshift=1em]{$(-)$}(P8)
                                                     (P2)--node[scale=0.55,xshift=-1em]{$(-)$}(P10);
         \draw[line width=0.2mm,color=blue!75!black]  (P4)--node[scale=0.55,xshift=-1.4em,yshift=-0.2em]{$(+)$}(P6)--node[scale=0.55,xshift=-1.4em,yshift=0.2em]{$(+)$}(P8);    
         \draw[line width=0.2mm,color=blue!75!black]  (P2)--node[scale=0.55,xshift=1.4em,yshift=-0.2em]{$(+)$}(P12)--node[scale=0.55,xshift=1.4em,yshift=0.2em]{$(+)$}(P10);                                  
            \node[scale=0.7] at (P4) [shift=(-30*4:0.25)] {$i$};
              \node[scale=0.7] at (P6) [shift=(-30*6:0.25)] {$j$};
              \node[scale=0.7] at (P8) [shift=(-30*8:0.25)] {$k$};
                \node[scale=0.7] at (P10) [shift=(-30*10:0.25)] {$r$};
              \node[scale=0.7] at (P12) [shift=(-30*12:0.25)] {$s$};
              \node[scale=0.7] at (P2) [shift=(-30*2:0.25)] {$t$};
         \end{scope}
    \end{tikzpicture}
\end{equation*}

We remark on several useful aspects of the edge-pairing function:

\paragraph{IR divergence and degenerate cases.} The IR divergence of the double-box is inherited by those edge-pairing functions in which either two edges join at the same point or at least one edge becomes lightlike.
In such cases, the edge-pairing functions become simple polylogarithms under mass regularization. For convenience, we collect the results below:
\begingroup
\allowdisplaybreaks
        \begin{align}
        \Ep{i , j}{j , k} &= -1+\frac{\pi^2}{24}+\frac{3 \log^2 2}{4}+\frac{1}{2} \log \frac{4 \mu^2_{\mathrm{IR}} x_{i,k}^2}{x_{i,j}^2x_{j,k}^2}-\frac{1}{2}\log 2 \log \frac{4 \mu^2_{\mathrm{IR}} x_{i,k}^2}{x_{i,j}^2x_{j,k}^2} \:, \label{eq: Ep_deg1}\\
        \Ep{i,j}{j,i}&=-\frac{\pi^2}{12}+\frac{3 \log^2 2}{2}+\log \frac{4 \mu^2_{\mathrm{IR}}}{x_{i,j}^2}-\log 2 \log\frac{4 \mu^2_{\mathrm{IR}}}{x_{i,j}^2} \:, \label{eq: Ep_deg2} \\
        \Ep{i,j}{j,j{+}1}&=-\frac{\pi^2}{24}+\frac{ \log^2 2}{4}
        -\frac{1}{2} \text{Li}_2\ab(1{-} \frac{x_{i,j+1}^2}{x_{i,j}^2})
        +\frac{1}{2} \text{Li}_2\ab(1{-} \frac{x_{i,j+1}^2}{2x_{i,j}^2}) \:,  \label{eq: Ep_deg3} \\
        \Ep{i{-}1,i}{i,j}&=-\frac{\pi^2}{24}+\frac{ \log^2 2}{4}
        -\frac{1}{2} \text{Li}_2\ab(1{-} \frac{x_{i-1,j}^2}{x_{i,j}^2})
        +\frac{1}{2} \text{Li}_2\ab(1{-} \frac{x_{i-1,j}^2}{2x_{i,j}^2}) \:,\label{eq: Ep_deg4} \\
        \Ep{i{-}1,i}{i,i{+}1}&=-\frac{\pi^2}{24}+\frac{ \log^2 2}{4}-\frac{1}{2} \log 2 \log \frac{4 \mu^2_{\mathrm{IR}}}{x_{i-1,i+1}^2} \:, \label{eq: Ep_deg5}\\
        \Ep{i{+}2 , i}{i , i{+}1}&=\Ep{i{-}1 , i}{i , i{-}2}=-\frac{\pi^2}{24}+\frac{ \log^2 2}{4} \:,\label{eq: Ep_deg6} \\
        \Ep{i,i{+}1}{k,\ell} &= -\frac{1}{8} \log^2 \ab(\frac{x_{i,\ell}^2 x_{i+1,k}^2}{x_{i,k}^2 x_{i+1,\ell}^2 })-\frac{1}{4} \log \ab(\frac{x_{i,\ell}^2 x_{i+1,k}^2}{x_{i,k}^2 x_{i+1,\ell}^2 }) 
        \log \ab(\frac{4 \mu^2_{\mathrm{IR}} x_{k,\ell}^2}{x_{i+1,\ell}^2 x_{i+1,k}^2})  \nonumber \\
        &\quad +\frac{1}{2} \text{Li}_2 \ab(1{-}\frac{x_{i+1,k}^2}{x_{i,k}^2})
        -\frac{1}{2} \text{Li}_2 \ab(1{-}\frac{x_{i+1,\ell}^2}{x_{i,\ell}^2})
        -\frac{1}{2} \text{Li}_2 \ab( 1{-}\frac{x_{i,\ell}^2 x_{i+1,k}^2}{x_{i,k}^2 x_{i+1,\ell}^2})  \:, \label{eq: Ep_deg7}  \\
        \Ep{i,i{+}1}{j,j{+}1}&= \frac{1}{2} \log \frac{2 \mu^2_{\mathrm{IR}}}{x_{i+1,j+1}^2} 
        \log \ab(\frac{x_{i,j}^2 x_{i+1,j+1}^2}{x_{i,j+1}^2 x_{i+1,j}^2 })
        - \frac{1}{2} \text{Li}_2 \ab(1{-}\frac{x_{i,j+1}^2 x_{i+1,j}^2}{x_{i,j}^2 x_{i+1,j+1}^2 })\nonumber \\
        &\quad +\frac{1}{2} \text{Li}_2 \ab(1{-}\frac{x_{i,j+1}^2}{x_{i,j}^2}) 
        +\frac{1}{2} \text{Li}_2 \ab(1{-}\frac{x_{i+1,j}^2}{x_{i,j}^2})\nonumber \\
        &\quad -\frac{1}{2} \text{Li}_2 \ab(1{-}\frac{x_{i+1,j+1}^2}{x_{i,j+1}^2})
        -\frac{1}{2} \text{Li}_2 \ab(1{-}\frac{x_{i+1,j+1}^2}{x_{i+1,j}^2})\:, \label{eq: Ep_deg8}\\
        \Ep{i{-}1,i}{i{+}1,j}&= \Ep{j,i{-}1}{i,i{+}1} = - \frac{\pi^2}{24} 
        - \frac{1}{8} \log^2 \ab(\frac{4 \mu^2_{\mathrm{IR}} x_{i-1,j}^2 x_{i+1,j}^2}{x_{i-1,i+1}^2 x_{i,j}^4})  \label{eq: Ep_deg9}\\
        & \phantom{=\Ep{j,i{-}1}{i,i{+}1} =} 
        - \frac{1}{2} \text{Li}_2 \ab(1-\frac{x_{i,j}^2}{x_{i-1,j}^2}) 
        -\frac{1}{2} \text{Li}_2 \ab(1{-}\frac{x_{i,j}^2}{2 x_{i+1,j}^2}) \:, \nonumber \\
        \Ep{i,i{+}1}{i{+}2,i{+}3}&=\frac{\pi^2}{24} - \frac{1}{4}\log^2 \ab( \frac{2 \mu^2_{\mathrm{IR}}\, x_{i,i+3}^2}{x_{i,i+2}^2 x_{i+1,i+3}^2})   \nonumber \\ 
        &\quad -\frac{1}{2} \text{Li}_2\ab(1{-}\frac{x_{i,i+2}^2}{x_{i,i+3}^2})- \frac{1}{2} 
        \text{Li}_2\ab(1{-}\frac{x_{i+1,i+3}^2}{x_{i,i+3}^2})\label{eq: Ep_deg10} \:,
        \end{align}
    \endgroup
These integrals can be obtained using similar techniques to those developed in~\cite{Caron-Huot:2012sos,He:2022lfz}. 
However, a special treatment is needed for eqs.~\eqref{eq: Ep_deg3}, \eqref{eq: Ep_deg4}, and \eqref{eq: Ep_deg6} (i.e., $\Ep{i,j}{j,j{+}1}$, $\Ep{i{-}1,i}{i,j}$, $\Ep{i{+2},i{+}1}{i,i{+}1}$, and $\Ep{i{-1},i}{i,i{-}2}$): the numerators in eq.~\eqref{eq: EDef} for these cases are $O(\mu_{\text{IR}})$. More details are provided in appendix~\ref{sec: EdgeF}.

\paragraph{Combining with double triangles.}  
As mentioned in section~\ref{sec:3.1}, the double-boxes with internal cubic vertices possess unphysical cuts. Correspondingly, the edge-pairing functions $\Ep{i,j}{j,k}$ and $\Ep{i,j}{j,i}$, with $i,j,k$ sharing the same parity, are \emph{not} polylogarithms of uniform transcendental weight (see eqs.~\eqref{eq: Ep_deg1} and \eqref{eq: Ep_deg2}). This can be correct by introducing double-triangles as in eqs.~\eqref{eq:mod_db1} and \eqref{eq:mod_db2}. Therefore, the edge-pairing functions are modified to
\begin{align}
        \Ephat{i,j}{j,k}&: = \Ep{i,j}{j,k}+\hat{I}^{\text{dt}}(i,j;j,k)   \nonumber \\
        &=\int \frac{\dif c\, [\dif \vec{a}_{L}]\, \dif^2 \vec{b}_{R}}{ \pi \sqrt{c}} 
\ab(\frac{ -1}{1+c}+1) \frac{ X_{ij} X_{jk}}{4\left((c{+}1)A^2+ 2A\cdot B+ B^2\right)^2}\:,
\end{align}
and likewise for $\Ephat{i,j}{j,i}$. 

Once combined with the double triangles, these edge-pairing functions gives polylogarithms of uniform transcendental weight:
\begin{align}
        \Ephat{i,j}{j,k}&=\frac{\pi^2}{24}+\frac{3 \log^2 2}{2}-\frac{1}{2}\log 2 \log\frac{4 \mu^2_{\mathrm{IR}}x_{i,k}^2 }{x_{i,j}^2 x_{j,k}^2} \:, \label{eq: Ephatresult1}\\
        \Ephat{i,j}{j,i}&=-\frac{\pi^2}{12}+\frac{3 \log^2 2}{2}-\log 2 \log\frac{4 \mu^2_{\mathrm{IR}}}{x_{i,j}^2}\,.
        \label{eq: Ephatresult2}
\end{align}
One can now easily reconstruct the integrated result of various double-box integrals using these edge-pairing functions. A collection is presented in appendix~\ref{sec: double box in literature}.

\paragraph{Cancellation of unphysical elliptic cuts.}  

The other unphysical cut discussed in Section~\ref{sec:3.1}, the elliptic cut $\Ecut{i,j}{k,\ell}$ defined by eq.~\eqref{eq:def_Ecut}, is naturally associated with the edge-pairing function $\Ep{i,j}{k,\ell}$. First, only $\Ep{i,j}{k,\ell}$ matches $\Ecut{i,j}{k,\ell}$ at the kinematic level in the decomposition \eqref{eq: EdgePairRep}.
Second, the elliptic nature of $\Ep{i,j}{k,\ell}$ becomes manifest upon performing the $a_i, b_i$ integrations in eq.~\eqref{eq: EDef}. As promised, this yields a one-fold integral of the Bloch--Wigner dilogarithm,
\begin{equation}
    \Ep{i , j}{ k , \ell}
    =\frac{1}{2\pi} \int_0^\infty \frac{(v-1)\,\dif c}{(1+c)\,\sqrt{c}\,\Delta(c)}
    \left[ \Li_2(z) -  \Li_2(\bar{z})
    + \tfrac{1}{2} \log(z\bar{z}) \,\log\frac{1-z}{1-\bar{z}} \right]\:,
\end{equation}
where we have introduced
\begin{align}
        u &=\frac{x^2_{i,j}x^2_{k,\ell}}{x^2_{i,k}x^2_{j,\ell}}\:, &
\qquad  v &=\frac{x^2_{i,\ell}x^2_{j,k}}{x^2_{i,k}x^2_{j,\ell}}\:, \\
    z,\bar{z} &= \frac{1-(1{+}c)u+v \pm \Delta}{2}\:,  &
\Delta(c) &= \sqrt{\big(1-(1{+}c)u - v\big)^2 - 4(1{+}c)uv} \:.
\end{align}
The factor $\sqrt{c}\,\Delta$ is a square root of a cubic polynomial in $c$ and cannot be rationalized; the $c$–integral therefore generically yields elliptic functions.

The cancellation of $\Ep{i,j}{k,\ell}$ in the BDS integrand $\operatorname{BDS}^{\text{3D}}_{n,\tau}$ is also straightforward, since it only arises from the combination \eqref{eq: cancel_Epcut} in $\operatorname{BDS}^{\text{3D}}_{n,\tau}$. Substituting eq.~\eqref{eq: EdgePairRep} into this combination then gives the desired result.

\subsection{BDS$^{\text{3D}}_n$ for all $n$}

Armed with the knowledge of edge-pairing functions given in the previous subsection, we can now straightforwardly convert the BDS integrands into explicit functions. For a given scaffolding triangulation $\tau$, one sums over all compatible 2STs to obtain $\operatorname{BDS}_{n,\tau}^{\text{3D}}$, where each 2ST is expressed in terms of nine edge-pairing functions.

From the above discussion of elliptic cuts, we have seen that the majority of edge-pairing functions, namely those involving two disjoint internal edges, cancel out in $\operatorname{BDS}_{n,\tau}^{\text{3D}}$. 
More generally, if an edge-pairing $\Ep{i,j}{k,\ell}$ appears in an even number of 2STs for the triangulation $\tau$, 
it cancels out in $\operatorname{BDS}_{n,\tau}^{\text{3D}}$.
In this way, we can easily identify the uncanceled ones, which fall into three categories:
\begin{enumerate}[leftmargin=5em]
    \item[Type-A:] 
    Both edges are external edges and do not belong to the same triangle in a given triangulation, 
    namely $\Ep{i,i{+}1}{j,j{+}1}$ with $2\leq j-i\leq n-2$, and $\Ep{i{-}1,i}{i,i{+}1}$ with even (odd) $i$ for even (odd) scaffolding triangulations. Graphically, they are
    
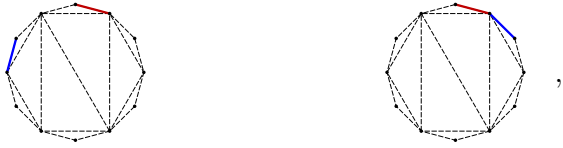
\begin{equation*}
    \begin{tikzpicture}[baseline={([yshift=-0.8ex]current bounding box.center)}]
        \begin{scope}
          \foreach \i  in {1,...,12} {
        \coordinate (P\i) at (-30*\i:0.9);
             }
        \draw[dash=on 2pt off 0.6pt phase 0pt,line width =0.1mm] (P1) \foreach \i in {2,...,12} {--(P\i)} -- cycle;
        \draw[dash=on 2pt off 0.6pt phase 0pt,line width =0.1mm] (P6)--(P8)--(P10)--(P12)--(P2)--(P4)--cycle (P4)--(P8) (P2)--(P10) (P2)--(P8);
        \draw[color=blue!100!black,line width=0.3mm] (P6)--(P7);
        \draw[color=red!75!black,line width=0.3mm] (P9)--(P10);
             \foreach \i  in {1,...,12} {
        \fill[black] (P\i) circle (0.14ex);
             }
         \end{scope}
         \begin{scope}[shift={(5,0)}]
          \foreach \i  in {1,...,12} {
        \coordinate (P\i) at (-30*\i:0.9);
             }
        \draw[dash=on 2pt off 0.6pt phase 0pt,line width =0.1mm] (P1) \foreach \i in {2,...,12} {--(P\i)} -- cycle;
        \draw[dash=on 2pt off 0.6pt phase 0pt,line width =0.1mm] (P6)--(P8)--(P10)--(P12)--(P2)--(P4)--cycle (P4)--(P8) (P2)--(P10) (P2)--(P8);
        \draw[color=blue!100!black,line width=0.3mm] (P10)--(P11);
        \draw[color=red!75!black,line width=0.3mm] (P9)--(P10);
           \foreach \i  in {1,...,12} {
        \fill[black] (P\i) circle (0.14ex);
             }
         \end{scope}
    \end{tikzpicture} \quad ,
\end{equation*}
from which it is clear that they arise from only one 2ST, and appear in the final sum with coefficient $+1$.
     \item[Type-B:] Two edges belong to the same triangle and hence share a common vertex;
these include $\Ephat{i,j}{j,k}$ together with its cyclic rotations for $\{i,j,k\}\in \tau$, and $\Ep{i{-}1,i}{i,i{-}2}$ together with its reflection. Graphically, they are 
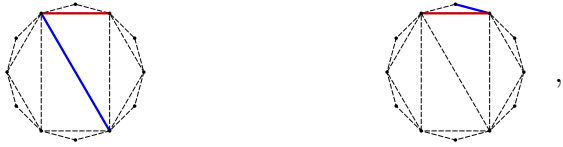
\begin{equation*}
    \begin{tikzpicture}[baseline={([yshift=-0.8ex]current bounding box.center)}]
        \begin{scope}
          \foreach \i  in {1,...,12} {
        \coordinate (P\i) at (-30*\i:0.9);
             }
        \draw[dash=on 2pt off 0.6pt phase 0pt,line width =0.1mm] (P1) \foreach \i in {2,...,12} {--(P\i)} -- cycle;
        \draw[dash=on 2pt off 0.6pt phase 0pt,line width =0.1mm] (P6)--(P8)--(P10)--(P12)--(P2)--(P4)--cycle (P4)--(P8) (P2)--(P10) (P2)--(P8);
        \draw[color=blue!100!black,line width=0.3mm] (P2)--(P8);
        \draw[color=red!75!black,line width=0.3mm] (P8)--(P10);
             \foreach \i  in {1,...,12} {
        \fill[black] (P\i) circle (0.14ex);
             }
         \end{scope}
         \begin{scope}[shift={(5,0)}]
          \foreach \i  in {1,...,12} {
        \coordinate (P\i) at (-30*\i:0.9);
             }
        \draw[dash=on 2pt off 0.6pt phase 0pt,line width =0.1mm] (P1) \foreach \i in {2,...,12} {--(P\i)} -- cycle;
        \draw[dash=on 2pt off 0.6pt phase 0pt,line width =0.1mm] (P6)--(P8)--(P10)--(P12)--(P2)--(P4)--cycle (P4)--(P8) (P2)--(P10) (P2)--(P8);
        \draw[color=blue!100!black,line width=0.3mm] (P10)--(P9);
        \draw[color=red!75!black,line width=0.3mm] (P8)--(P10);
           \foreach \i  in {1,...,12} {
        \fill[black] (P\i) circle (0.14ex);
             }
         \end{scope}
    \end{tikzpicture} \quad ,
\end{equation*}
from which it is clear that $\Ephat{i,j}{j,k}$ arises from three 2STs and $\Ep{i{-}1,i}{i,i{-}2}$ arises from only one. 
Both appear in the final sum with coefficient $-1$.
    \item[Type-C:] Two edges are the same internal edge, namely $\Ephat{i,j}{j,i}$ with $i$ and $j$ of the same parity.
 Graphically, they are
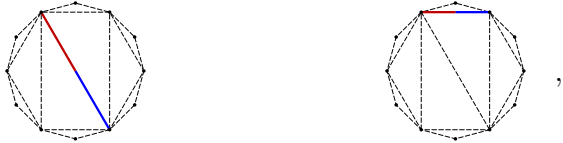
\begin{equation*}
    \begin{tikzpicture}[baseline={([yshift=-0.8ex]current bounding box.center)}]
        \begin{scope}
          \foreach \i  in {1,...,12} {
        \coordinate (P\i) at (-30*\i:0.9);
             }
        \draw[dash=on 2pt off 0.6pt phase 0pt,line width =0.1mm] (P1) \foreach \i in {2,...,12} {--(P\i)} -- cycle;
        \draw[dash=on 2pt off 0.6pt phase 0pt,line width =0.1mm] (P6)--(P8)--(P10)--(P12)--(P2)--(P4)--cycle (P4)--(P8) (P2)--(P10);
        \draw[color=blue!100!black,line width=0.3mm] (P2)--(0,0);
        \draw[color=red!75!black,line width=0.3mm] (P8)--(0,0);
             \foreach \i  in {1,...,12} {
        \fill[black] (P\i) circle (0.14ex);
             }
         \end{scope}
         \begin{scope}[shift={(5,0)}]
          \foreach \i  in {1,...,12} {
        \coordinate (P\i) at (-30*\i:0.9);
             }
        \draw[dash=on 2pt off 0.6pt phase 0pt,line width =0.1mm] (P1) \foreach \i in {2,...,12} {--(P\i)} -- cycle;
        \draw[dash=on 2pt off 0.6pt phase 0pt,line width =0.1mm] (P6)--(P8)--(P10)--(P12)--(P2)--(P4)--cycle (P4)--(P8) (P2)--(P10) (P2)--(P8);
        \draw[color=blue!100!black,line width=0.3mm] (P10)--+(180:0.45);
        \draw[color=red!75!black,line width=0.3mm] (P8)--+(0:0.45);
           \foreach \i  in {1,...,12} {
        \fill[black] (P\i) circle (0.14ex);
             }
         \end{scope}
    \end{tikzpicture} \quad,
\end{equation*}
from which it is clear that they arise from only one 2ST, and appear in the final sum with coefficient $+1$.
\end{enumerate}

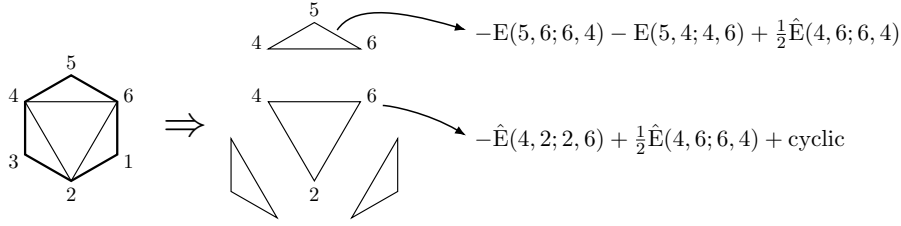
\begin{figure}
    \centering
    \begin{tikzpicture}[baseline={([yshift=-0.7ex]current bounding box.center)}]
        \begin{scope}[shift={(0.8,0)}]
            \foreach \i  in {1,...,6} {
            \coordinate (P\i) at (-60/2+60-60*\i:0.7);
            \node[scale=0.7] at (P\i) [shift=(-60/2+60-60*\i:0.25)] {\i};
                }
            \draw[line width=0.15mm] (P2)--(P4) (P2)--(P6) (P4)--(P6);
            \draw[line width=0.3mm] (P1) \foreach \i in {2,...,6} { -- (P\i) } -- cycle;
            \node[scale=1.5] at (1.5,0) {$\Rightarrow$};
        \end{scope}
        \begin{scope}[baseline={([yshift=-0.6ex]current bounding box.center)},shift={(4,0)}]
            \foreach \i  in {2,4,6} {
                \coordinate (P\i) at (-60/2+60-60*\i:0.7);
                \node[scale=0.7] at (P\i) [shift=(-60/2+60-60*\i:0.25)] {\i};
                }
            \draw[line width=0.15mm] (P2)--(P4)--(P6)--cycle; 
            \draw[-{Latex[length=1.5mm]},line width=0.2mm] (0.9,0.3).. controls (1,0.3) and (1.5,0.2) .. (2,-0.1);
            \node[below right, scale=0.8,yshift=2ex] at(2,-0.1) {$-\Ephat{4,2}{2,6}+\tfrac{1}{2}\Ephat{4,6}{6,4}+\text{cyclic}$};
        \end{scope}
        \begin{scope}[baseline={([yshift=-0.6ex]current bounding box.center)},shift={(4,0.7)}]
            \foreach \i  in {4,5,6} {
                \coordinate (P\i) at (-60/2+60-60*\i:0.7);
                \node[scale=0.7] at (P\i) [shift=(-60/2+60-60*\i:0.25)] {\i};
            }
            \draw[line width=0.15mm] (P4)--(P5)--(P6)--cycle;
            \draw[-{Latex[length=1.5mm]},line width=0.2mm] (0.3,0.6).. controls (0.5,0.9) and (1,0.9) .. (2,0.6);
            \node[below right, scale=0.8,yshift=2ex] at(2,0.6) {$
           -\Ep{5,6}{6,4}-\Ep{5,4}{4,6}+\tfrac{1}{2}\Ephat{4,6}{6,4}
            $};
        \end{scope}
        \begin{scope}[baseline={([yshift=-0.6ex]current bounding box.center)},shift={(4+0.49,-0.49)}]
            \foreach \i  in {1,2,6} {
                \coordinate (P\i) at (-60/2+60-60*\i:0.7);
                }  
            \draw[line width=0.15mm] (P1)--(P2)--(P6)--cycle;
        \end{scope}
        \begin{scope}[baseline={([yshift=-0.6ex]current bounding box.center)},shift={(4-0.49,-0.49)}]
            \foreach \i  in {2,3,4} {
                \coordinate (P\i) at (-60/2+60-60*\i:0.7);
            }
            \draw[line width=0.15mm] (P2)--(P3)--(P4)--cycle;
        \end{scope}
     \end{tikzpicture}
    \caption{Edge-pairing functions grouped by triangles: the six-point case}
    \label{fig: demo}
\end{figure}

It is evident that the contribution arising from Type-B and Type-C depends on the details of the triangulation $\tau$; moreover, each edge-pairing function of Type-B and Type-C is closely tied to the triangles in the triangulation. Indeed, each edge-pairing function of Type-B is associated with one triangle, while each edge-pairing function of Type-C is associated with two. 
Therefore, we can group them according to triangles in the triangulation, which gives, for even cases,
\begin{align} \label{eq: BDS_tri}
    \operatorname{BDS}_{n,\tau}^{\text{3D}}&=
    \sum_{2\leq j-i\leq n-2}\Ep{i,i{+}1}{j,j{+}1}+\sum_{i\in [n]_{\sf{e}}}\Ep{i{-}1,i}{i,i{+}1} \nonumber \\
    &\quad+\sum_{i\in [n]_{\sf{e}}}\ab(-\Ep{i{-}1,i}{i,i{-}2}-\Ep{i{-}1,i{-}2}{i{-}2,i}+\tfrac{1}{2}\Ephat{i{-}2,i}{i,i{-}2}) \nonumber \\
    &\quad+\sum_{\{i,j,k\}\in\tau}\ab(-\Ephat{i,j}{j,k}+\tfrac{1}{2}\Ephat{i,k}{k,i}+\text{cyclic in } i,j,k) \:,
\end{align}
and a similar expression for odd cases.
Here, the second and last lines of eq.~\eqref{eq: BDS_tri} arise from the external and internal triangles, respectively. The coefficient $\frac{1}{2}$ for $\Ephat{i,j}{j,i}$ accounts for the fact that such an edge-pairing is shared by two triangles. See figure~\ref{fig: demo} for an illustration of the six-point case.

Very nicely, for the combination in the last line of eq.~\eqref{eq: BDS_tri}, using the results \eqref{eq: Ephatresult1} and \eqref{eq: Ephatresult2} gives
\begin{equation} \label{eq: Ep_Identity1}
    -\Ephat{i,j}{j,k}+\tfrac{1}{2}\Ephat{i,k}{k,i}+\text{cyclic in } i,j,k = -\frac{\pi^2}{4} \:.
\end{equation}
Thus we have proved that $\operatorname{BDS}_{n,\tau}^{\text{3D}}$ is the same for \emph{all even (odd) scaffolding triangulations}.
For the combination in the second line of eq.~\eqref{eq: BDS_tri}, by using eqs.~\eqref{eq: Ep_deg5}, \eqref{eq: Ep_deg6} and \eqref{eq: Ephatresult2}, we find
\begin{align} \label{eq: Ep_Identity2}
   &\quad -\Ep{i{-}1,i}{i,i{-}2}-\Ep{i{-}1,i{-}2}{i{-}2,i}+\tfrac{1}{2}\Ephat{i{-}2,i}{i,i{-}2} \nonumber \\
    &=\frac{\pi^2}{24}+\frac{\log^2 2}{4}-\frac{1}{2}\log 2 \log \frac{4 \mu^2_\mathrm{IR}}{x_{i-2,i}^2}=\Ep{i{-}2,i{-}1}{i{-}1,i}+\frac{\pi^2}{12} \:.
\end{align}
Then, substituting eqs.~\eqref{eq: Ep_Identity1} and \eqref{eq: Ep_Identity2} into \eqref{eq: BDS_tri} leads to 
\begin{align} \label{eq: BDS_nice}
    \operatorname{BDS}_{n,\tau}^{\text{3D}} &=  \sum_{2\leq j-i \leq n-2}\Ep{i,i{+}1}{j,j{+}1}+\sum_{i\in [n]}\Ep{i{-}1,i}{i,i{+}1} 
    -\frac{\pi^2}{12}(n-6) \:,
\end{align}
where we have used the fact that there are $n/2$ external triangles and $(n-4)/2$ internal ones in a scaffolding triangulation to determine the constant term. This establishes the triangulation independence of $\operatorname{BDS}_{n,\tau}^{\text{3D}}$, and we can now drop the subscript $\tau$!

Finally, let us build the connection between the expression~\eqref{eq: BDS_nice} and the first two equations we started with in the introduction. First, the integrated result \eqref{eq:BDS} can be easily confirmed by applying eqs.~\eqref{eq: Ep_deg5} and \eqref{eq: Ep_deg8} to \eqref{eq: BDS_nice}. Second, to obtain the connection with the four-dimensional BDS ansatz, we note that $\operatorname{BDS}_{n}^{\text{4D}}$ is the sum of all one-mass boxes $I_\text{1mb}(i,i{+}1,i{+}2,i{+}3)$ and two-mass-easy boxes $I_\text{2me}(i,i{+}1,j,j{+}1)$ with $2\leq j-i\leq n-2$. The four-dimensional box integrals under mass regularization are related to the edge-pairing functions via
\begin{align}
    &\Ep{i,i{+}1}{i{+}2,i{+}3}+ \frac{1}{2}\Ep{i,i{+}1}{i{+1},i{+}2}+\frac{1}{2}\Ep{i{+1},i{+2}}{i{+}2,i{+}3} \nonumber \\
    &\qquad =\frac{1}{2}I_\text{1mb}(i,i{+}1,i{+}2,i{+}3)-\frac{\pi^2}{12}+\frac{1}{4}\log 2 \log \ab(\frac{ x_{i,i+3}^4}{x_{i,i+2}^2 x_{i+1,i+3}^2}) \:,  \label{eq: box_ep1}\\
    &\Ep{i,i{+}1}{j,j{+}1}=\frac{1}{2}I_\text{2me}(i,i{+}1,j,j{+}1){-}\frac{1}{2} \log 2 \log \left(\frac{x_{i,j}^2 x_{i+1,j+1}^2}{x_{i,j+1}^2 x_{i+1,j}^2 }\right) \:. \label{eq: box_ep2}
\end{align}
Using the above two equations in the expression \eqref{eq: BDS_nice} then gives
\begin{align}
     \operatorname{BDS}_{n}^{\text{3D}} &=\frac{1}{2}\ab(\sum_{i\in [n]}I_\text{1mb}(i,i{+}1,i{+}2,i{+}3)
      + \sum_{2<j-i<n-2}I_\text{2me}(i,i{+}1,j,j{+}1)) -\frac{\pi^2}{12}(2n-6) \nonumber \\
      &\quad +  \sum_{i\in [n]}\frac{1}{4}\log 2 \log \ab(\frac{ x_{i,i+3}^4}{x_{i,i+2}^2 x_{i+1,i+3}^2}) 
      -\sum_{2<j-i<n-2}\frac{1}{2} \log 2 \log \left(\frac{x_{i,j}^2 x_{i+1,j+1}^2}{x_{i,j+1}^2 x_{i+1,j}^2 }\right) \nonumber \\
      &=\frac{1}{2} \operatorname{BDS}_{n}^{\text{4D}}-\frac{\pi^2}{6}(n-3) \:,
\end{align}
where we used the fact that the second line vanishes to reach the second equality.

\section{Conclusions and outlook}
In this work, we investigate the planar two-loop $n=2k$-point amplitude in ABJM theory, focusing on the component determined by the infrared behavior of the theory. More precisely, we analyze the so-called BDS integrand --- the collection of IR-divergent integrands supplemented by counterparts whose role is to cancel unphysical cuts. We show that the full answer admits a natural organization in terms of scaffolding triangulations of $n$-gons, where each triangulation defines a basis for the expansion of the result. Remarkably, this graphical representation persists after integration, allowing for a straightforward demonstration that the IR divergence of the $n$-point two-loop amplitude in ABJM theory is entirely captured by the BDS expression of $\mathcal{N}=4$ sYM.

The fact that distinct triangulations yield different integrands whose integrated results coincide provides the first explicit example of a triangulation of a function in loop amplitudes of quantum field theory. This strongly hints at a geometric interpretation of the BDS function itself (for related discussions, see~\cite{Mason:2010pg}). Our framework thus offers an ideal setting to examine whether the geometric description extends from rational functions to the integrated level.

Perhaps the  next step is to explore whether the construction of scaffolding triangulations can be generalized to higher-loop infrared divergences. If such an extension is possible, it would be fascinating to investigate how, upon resumming the loop expansion, these triangulations furnish a non-perturbative geometric or combinatorial description. To test this idea concretely, we clearly need higher-loop data, which remain very limited~---~so far, our understanding essentially stops at the four-point case~\cite{Bianchi:2014iia,Bianchi:2011aa}. In this regard, computing the three-loop data~\cite{3L6-pt} would represent a crucial step toward extending this picture to the higher-loop regime.

Beyond the BDS integrand, we expect more general (non-scaffolding) triangulations to appear, corresponding to the richer on-shell diagrams that arise from unitarity cuts. While we do not expect something similar to triangulation independence, it would be intriguing to explore how the remaining local integrands map onto the combinatorics of these triangulations. Furthermore, beginning at six points~\cite{Caron-Huot:2012sos, He:2022lfz}, the integrated amplitude involves sign functions of kinematic invariants, whose structure may play an analogous role to the IR divergences in determining the BDS integrand. Uncovering their potential connection to generalized triangulations presents an interesting direction for future investigation.

\section*{Acknowledgement}

We thank Ham-Woo Puinn and Kai Yan for helpful discussions and collaboration on related projects. We also thank Qu Cao, Jin Dong, Song He,  Zhenjie Li, and Qinglin Yang for stimulating discussions. Y.-t. Huang is supported by the Taiwan Ministry of Science and Technology Grant No. 112-2628-M-002-003-MY3 and
114-2923-M-002-011-MY. C.-K. Kuo is funded by the European Union (ERC, 101118787). 
C. Zhang is supported by the European Research Council (ERC) under the European Union’s research and innovation programme grant agreement 101043686 (ERC Consolidator Grant LoCoMotive). Views and opinions expressed are
however those of the author(s) only and do not necessarily reflect those of the European Union or the European
Research Council. Neither the European Union nor the
granting authority can be held responsible for them.

\appendix

\section{Result of double-boxes in literature}\label{sec: double box in literature}
In this appendix, we collect the known results for the double-boxes from~\cite{Caron-Huot:2012sos,He:2022lfz} for comparison. Below, $\hat{I}$ means that the double triangle has been included.
\begin{align}
    \Hat{I}^{\text{db}}(1,2,3;3,4,1)&:=
    \begin{tikzpicture}[baseline={([yshift=-0.8ex]current bounding box.center)}]
        \draw[line width=0.2mm] (0,0)--(0,1) -- (1,1) -- (1,0) --(0,0);
        \draw[line width=0.2mm] (0,1) -- +(135:0.5);
        \draw[line width=0.2mm] (0,0) -- +(-135:0.5);
        \draw[line width=0.2mm] (1,1) -- (2,1) -- (2,0) -- (1,0);
        \draw[line width=0.2mm] (2,1) --  +(45:0.45);
            \draw[line width=0.2mm] (2,0) --  +(-45:0.45);
            \node[scale=0.9] at (1,1.35) {$3$};
            \node[scale=0.9] at (1,-0.35) {$1$};
            \node[scale=0.9] at (2.4,0.5) {$4$};
            \node[scale=0.9] at (-0.3,0.5) {$2$};
    \end{tikzpicture} 
    =\frac{\pi^2}{3}-\log{\frac{4\mu^2_{\mathrm{IR}}}{ x_{1,3}^2}} \log{\frac{4\mu^2_{\mathrm{IR}}}{x_{2,4}^2}}.
\end{align}
The above integral corresponds to the four-point double-box in \cite{Caron-Huot:2012sos}.

\begin{align}
   & \Hat{I}^{\text{db}}(i{-}1,i,i{+}1;i{+}1,j,i{-}1):=
        \begin{tikzpicture}[baseline={([yshift=-0.8ex]current bounding box.center)}]
            \begin{scope}
                \draw[line width=0.2mm] (0,0)--(0,1) -- (1,1) -- (1,0) --(0,0);
                \draw[line width=0.2mm] (0,1) -- +(135:0.5);
                \draw[line width=0.2mm] (0,0) -- +(-135:0.5);
                \draw[line width=0.2mm] (1,1) -- (2,1) -- (2,0) -- (1,0);
                \draw[line width=0.2mm] (2,1) --  +(75:0.45);
                \draw[line width=0.2mm] (2,1) --  +(15:0.45);
                \draw[line width=0.2mm] (2,0) --  +(-75:0.45);
                \draw[line width=0.2mm] (2,0) --  +(-15 :0.45);
                \foreach \i in {-1,...,1} {
                \fill[black] (2,1)+(45+\i *16: 0.35) circle (0.14ex);
                 \fill[black] (2,0)+(-45+\i *16: 0.35) circle (0.14ex);
                    }
            \node[scale=0.9] at (1,1.35) {$i{+}1$};
            \node[scale=0.9] at (1,-0.35) {$i{-}1$};
            \node[scale=0.9] at (2.4,0.5) {$j$};
            \node[scale=0.9] at (-0.3,0.5) {$i$};
            \end{scope}
        \end{tikzpicture} \nonumber\\
    &=-\frac{\pi^2}{4}-\frac{1}{4}\log^2\bigg({\frac{4 \mu_{\mathrm{IR}}^2x_{i-1,j}^2 x_{i+1,j}^2}{x_{i-1,i+1}^2 x_{i,j}^4}}\bigg)-\mathrm{Li}_2\bigg(1{-}\frac{x_{i,j}^2}{x_{i-1,j}^2}\bigg)-\mathrm{Li}_2\bigg(1{-}\frac{x_{i,j}^2}{x_{i+1,j}^2}\bigg).
\end{align}
The above corresponds to the ``2-mass hard'' topology in \cite{Caron-Huot:2012sos}.

\begin{align}
    &\Hat{I}^{\text{db}}(i{-}1,i,i{+}1;i{-}3,i{-}2,i{-}1):=
    \begin{tikzpicture}[baseline={([yshift=-0.8ex]current bounding box.center)}]
        \draw[line width=0.2mm] (0,0)--(0,1) -- (1,1) -- (1,0) --(0,0);
        \draw[line width=0.2mm] (0,1) -- +(135:0.5);
        \draw[line width=0.2mm] (0,0) -- +(-135:0.5);
        \draw[line width=0.2mm] (1,1) -- (2,1) -- (2,0) -- (1,0);
        \draw[line width=0.2mm] (2,1) --  +(45:0.45);
        \draw[line width=0.2mm]  (1,1) -- +(90-30:0.45);
           \draw[line width=0.2mm]  (1,1) -- +(90+30:0.45);
           \foreach \i in {-1,...,1} {
                   \fill[black] (1,1)+(90+\i *16: 0.35) circle (0.14ex);
                    }
            \draw[line width=0.2mm] (2,0) --  +(-45:0.45);
            \node[scale=0.9] at (0.4,1.35) {$i{+}1$};
            \node[scale=0.9] at (1.7,1.35) {$i{-}3$};
            \node[scale=0.9] at (1,-0.35) {$i{-}1$};
            \node[scale=0.9] at (2.4,0.5) {$i{-}2$};
            \node[scale=0.9] at (-0.3,0.5) {$i$};
    \end{tikzpicture} \nonumber\\
    &=\frac{\pi^2}{4}+\frac{1}{2}\log\left(\frac{x_{i,i-2}^2 x_{i+1,i-3}^2 }{x_{i,i-3}^2 x_{i+1,i-2}^2  }\right)\log \left({\frac{4 \mu^2_{\mathrm{IR}}x_{i+1,i-3}^2}{x_{i-1,i+1}^2 x_{i-1,i-3}^2}}\right)-\frac{1}{4}\log^2\left({\frac{4 \mu^2_{\mathrm{IR}}x_{i+1,i-3}^2}{x_{i-1,i+1}^2x_{i-1,i-3}^2}}\right)\nonumber\\
 &\quad -\mathrm{Li}_2\bigg(1{-}\frac{x_{i-1,i+1}^2}{x_{i+1,i-2}^2}\bigg)-\mathrm{Li}_2\bigg(1{-}\frac{x_{i-1,i-3}^2}{x_{i,i-3}^2}\bigg)+\frac{1}{2}\mathrm{Li}_2\left(1{-}\frac{x_{i,i-3}^2 x_{i+1,i-2}^2}{x_{i,i-2}^2 x_{i+1,i-3}^2 }\right).
\end{align}
The above integral corresponds to the ``crab'' topology in \cite{Caron-Huot:2012sos}.

\begin{align}
    &I^{\text{db}}(i{-}1,i,i{+}1;j{-}1,j,j{+}1):=
    \begin{tikzpicture}[baseline={([yshift=-0.8ex]current bounding box.center)}]
        \draw[line width=0.2mm] (0,0)--(0,1) -- (1,1) -- (1,0) --(0,0);
        \draw[line width=0.2mm] (0,1) -- +(135:0.5);
        \draw[line width=0.2mm] (0,0) -- +(-135:0.5);
        \draw[line width=0.2mm] (1,1) -- (2,1) -- (2,0) -- (1,0);
        \draw[line width=0.2mm] (2,1) --  +(45:0.45);
        \draw[line width=0.2mm]  (1,1) -- +(90-30:0.45);
           \draw[line width=0.2mm]  (1,1) -- +(90+30:0.45);
           \foreach \i in {-1,...,1} {
                   \fill[black] (1,1)+(90+\i *16: 0.35) circle (0.14ex);
                    }
            \foreach \i in {-1,...,1} {
                   \fill[black] (1,0)+(-90+\i *16: 0.35) circle (0.14ex);
                    }
            \draw[line width=0.2mm] (1,0) --  +(-60 :0.45) (1,0)-- +(-120 :0.45);
            \draw[line width=0.2mm] (2,0) --  +(-45:0.45);
            \node[scale=0.9] at (0.4,1.35) {$i{+}1$};
            \node[scale=0.9] at (1.7,1.35) {$j{-}1$};
            \node[scale=0.9] at (0.4,-0.35) {$i{-}1$};
            \node[scale=0.9] at (1.7,-0.35) {$j{+}1$};
            \node[scale=0.9] at (2.4,0.5) {$j$};
            \node[scale=0.9] at (-0.3,0.5) {$i$};
    \end{tikzpicture} \\
    &=\frac{1}{2}\log{2}\log\left(\frac{x_{i-1,j+1}^2 x_{i+1,j-1}^2}{x_{i-1,j-1}^2 x_{i+1,j+1}^2 }\right) + 
    \frac{1}{2} \log\left(\frac{x_{i-1,j-1}^2 x_{i,j}^2}{x_{i-1,j}^2 x_{i,j-1}^2 }\right) \log\left(\frac{x_{i-1,j+1}^2 x_{i+1,j-1}^2}{x_{i-1,j-1}^2 x_{i+1,j+1}^2 }\right) \nonumber \\
 &\quad +\frac{1}{4}\log^2\left(\frac{x_{i-1,j-1}^2 x_{i,j}^2}{x_{i-1,j}^2 x_{i,j-1}^2 }\right)+\frac{1}{4}\log\left(\frac{x_{i-1,j+1}^2 x_{i+1,j-1}^2}{x_{i-1,j-1}^2 x_{i+1,j+1}^2 }\right)  \log\left(\frac{x_{i-1,i+1}^2 x_{j-1,j+1}^2}{x_{i-1,j-1}^2 x_{i+1,j+1}^2}\right) \nonumber \\
 &\quad + \frac{1}{2} \mathrm{Li}_2\left(1{-}\frac{x_{i-1,j-1}^2 x_{i,j}^2}{x_{i-1,j}^2 x_{i,j-1}^2 } \right)- \frac{1}{2} \mathrm{Li}_2\left(1{-}\frac{x_{i-1,j-1}^2 x_{i,j+1}^2 }{x_{i-1,j+1}^2 x_{i,j-1}^2 } \right) + \frac{1}{2} \mathrm{Li}_2\left(1{-}\frac{x_{i-1,j}^2 x_{i,j+1}^2 }{x_{i-1,j+1}^2 x_{i,j}^2 }\right) \nonumber \\
&\quad -\frac{1}{2} \mathrm{Li}_2\left(1{-}\frac{x_{i-1,j-1}^2 x_{i+1,j}^2 }{x_{i-1,j}^2 x_{i+1,j-1}^2 }\right)+ \frac{1}{2} \mathrm{Li}_2\left(1{-}\frac{x_{i,j-1}^2 x_{i+1,j}^2}{x_{i,j}^2 x_{i+1,j-1}^2 } \right)+\frac{1}{2} \mathrm{Li}_2\left(1{-}\frac{x_{i,j+1}^2 x_{i+1,j-1}^2 }{x_{i,j-1}^2 x_{i+1,j+1}^2 }\right) \nonumber\\
&\quad +\frac{1}{2} \mathrm{Li}_2\left(1{-}\frac{x_{i-1,j+1}^2 x_{i+1,j}^2}{x_{i-1,j}^2 x_{i+1,j+1}^2}\right)-\frac{1}{2}  \mathrm{Li}_2\left(1{-}\frac{x_{i,j+1}^2 x_{i+1,j}^2 }{x_{i,j}^2 x_{i+1,j+1}^2 }\right)+\Ep{i{-}1,i{+}1}{j{-}1,j{+}1} \:. \nonumber
\end{align}
The above integral is denoted as $I_B^{\text{db}}$ in \cite{He:2022lfz}.

\begin{align} \label{eq: IG1}
   & \Hat{I}^{\text{db}}(i{-}1,i,i{+}1;j,k,i{-}1):=
        \begin{tikzpicture}[baseline={([yshift=-0.8ex]current bounding box.center)}]
            \begin{scope}[shift={(3.5,0)}]
                \draw[line width=0.2mm] (0,0)--(0,1) -- (1,1) -- (1,0) --(0,0);
                \draw[line width=0.2mm] (0,1) -- +(135:0.5);
                \draw[line width=0.2mm] (0,0) -- +(-135:0.5);
                \draw[line width=0.2mm] (1,1) -- (2,1) -- (2,0) -- (1,0);
                \draw[line width=0.2mm] (2,1) --  +(75:0.45);
                \draw[line width=0.2mm] (2,1) --  +(15:0.45);
                \draw[line width=0.2mm] (2,0) --  +(-75:0.45);
                \draw[line width=0.2mm] (2,0) --  +(-15 :0.45);
                \foreach \i in {-1,...,1} {
                    \fill[black] (2,1)+(45+\i *16: 0.35) circle (0.14ex);
                    \fill[black] (2,0)+(-45+\i *16: 0.35) circle (0.14ex);
                    }
                 \draw[line width=0.2mm]  (1,1) -- +(90-30:0.45);
           \draw[line width=0.2mm]  (1,1) -- +(90+30:0.45);
           \foreach \i in {-1,...,1} {
                   \fill[black] (1,1)+(90+\i *16: 0.35) circle (0.14ex);
                    }
                 \node[scale=0.9] at (0.3,1.35) {$i{+}1$};
                 \node[scale=0.9] at (1.6,1.35) {$j$};
            \node[scale=0.9] at (1,-0.35) {$i{-}1$};
            \node[scale=0.9] at (2.4,0.5) {$k$};
            \node[scale=0.9] at (-0.3,0.5) {$i$};
            \end{scope}
        \end{tikzpicture} \\
    &=\frac{1}{2}\log{2}\log\left(\frac{x_{i-1,j}^2 x_{i+1,k}^2}{x_{i-1,k}^2 x_{i+1,j}^2}\right)+\frac{1}{2}\log\left(\frac{x_{i-1,j}^2 x_{i,k}^2 }{x_{i-1,k}^2 x_{i,j}^2}\right)\log\left({\frac{4 \mu^2_{\mathrm{IR}}x_{j,k}^2}{x_{i-1,j}^2x_{i-1,k}^2}}\right) \nonumber \\
 &\quad +\frac{1}{4}\log{\left(\frac{x_{i-1,j}^2 x_{i+1,k}^2}{x_{i-1,k}^2 x_{i+1,j}^2}\right)}\log\left(\frac{x_{i-1,i+1}^2 x_{j,k}^2}{x_{i-1,j}^2 x_{i+1,k}^2 }\right)-\frac{1}{2}\log{\left(\frac{x_{i-1,j}^2 x_{i,k}^2}{x_{i-1,k}^2 x_{i,j}^2 }\right)}\log\left(\frac{x_{i-1,i+1}^2 x_{j,k}^2 }{x_{i-1,j}^2 x_{i+1,k}^2 }\right)\nonumber \\
 &\quad -\mathrm{Li}_2\bigg(1{-}\frac{x_{i-1,j}^2}{x_{i,j}^2}\bigg)+\mathrm{Li}_2\bigg(1{-}\frac{x_{i-1,k}^2}{x_{i,k}^2}\bigg)+\frac{1}{2}\mathrm{Li}_2\left(1{-}\frac{x_{i-1,j}^2 x_{i,k}^2}{x_{i-1,k}^2 x_{i,j}^2}\right)\nonumber \\
 &\quad -\frac{1}{2}\mathrm{Li}_2\left(1{-}\frac{x_{i,k}^2 x_{i+1,j}^2}{x_{i,j}^2 x_{i+1,k}^2}\right)-\Ep{i{-}1,i{+}1}{j,k}\:. \nonumber
\end{align}
The above integral is denoted as $I_E^{\text{db}}$ in \cite{He:2022lfz}.
\begin{align}
   & \Hat{I}^{\text{db}}(i{-}1,i,i{+}1;i{+}1,j,k):=
        \begin{tikzpicture}[baseline={([yshift=-0.8ex]current bounding box.center)}]
            \begin{scope}[shift={(3.5,0)}]
                \draw[line width=0.2mm] (0,0)--(0,1) -- (1,1) -- (1,0) --(0,0);
                \draw[line width=0.2mm] (0,1) -- +(135:0.5);
                \draw[line width=0.2mm] (0,0) -- +(-135:0.5);
                \draw[line width=0.2mm] (1,1) -- (2,1) -- (2,0) -- (1,0);
                \draw[line width=0.2mm] (2,1) --  +(75:0.45);
                \draw[line width=0.2mm] (2,1) --  +(15:0.45);
                \draw[line width=0.2mm] (2,0) --  +(-75:0.45);
                \draw[line width=0.2mm] (2,0) --  +(-15 :0.45);
                \foreach \i in {-1,...,1} {
                    \fill[black] (2,1)+(45+\i *16: 0.35) circle (0.14ex);
                    \fill[black] (2,0)+(-45+\i *16: 0.35) circle (0.14ex);
                   \fill[black] (1,0)+(-90+\i *16: 0.35) circle (0.14ex);
                    }
                 \draw[line width=0.2mm] (1,0) --  +(-60 :0.45) (1,0)-- +(-120 :0.45);
                 \node[scale=0.9] at (1,1.35) {$i{+}1$};
            \node[scale=0.9] at (0.3,-0.35) {$i{-}1$};
            \node[scale=0.9] at (1.6,-0.35) {$k$};
            \node[scale=0.9] at (2.4,0.5) {$j$};
            \node[scale=0.9] at (-0.3,0.5) {$i$};
            \end{scope}
        \end{tikzpicture} \\
    &=\frac{1}{2}\log{2}\log\left(\frac{x_{i-1,j}^2 x_{i+1,k}^2}{x_{i-1,k}^2 x_{i+1,j}^2 }\right)+\frac{1}{2}\log\left(\frac{x_{i,j}^2 x_{i+1,k}^2}{x_{i,k}^2 x_{i+1,j}^2}\right)\log\left({\frac{4 \mu^2_{\mathrm{IR}}x_{j,k}^2}{x_{j,i+1}^2x_{k,i+1}^2}}\right) \nonumber\\
 &\quad +\frac{1}{4}\log{\left(\frac{x_{i-1,j}^2 x_{i+1,k}^2 }{x_{i-1,k}^2 x_{i+1,j}^2}\right)}\log\left(\frac{x_{i-1,i+1}^2 x_{j,k}^2}{x_{i-1,j}^2 x_{i+1,k}^2}\right)
 -\frac{1}{2} \log{\left(\frac{x_{i,j}^2 x_{i+1,k}^2}{x_{i,k}^2 x_{i+1,j}^2}\right)}\log\left(\frac{x_{i-1,i+1}^2 x_{j,k}^2}{x_{i-1,j}^2 x_{i+1,k}^2}\right)  \nonumber \\
 &\quad +\mathrm{Li}_2\bigg(1{-}\frac{x_{i+1,j}^2}{x_{i,j}^2}\bigg) -\mathrm{Li}_2\bigg(1{-}\frac{x_{i+1,k}^2}{x_{i,k}^2}\bigg) -\frac{1}{2}\mathrm{Li}_2\left(1{-}\frac{ x_{i-1,k}^2 x_{i,j}^2}{x_{i-1,j}^2 x_{i,k}^2 }\right) \nonumber \\
 &\quad +\frac{1}{2}\mathrm{Li}_2\left(1{-}\frac{x_{i,j}^2 x_{i+1,k}^2}{x_{i,k}^2 x_{i+1,j}^2}\right)-\Ep{i{-}1,i{+}1}{j,k}\:. \nonumber
\end{align}
The above integral is related to eq.~\eqref{eq: IG1} through the horizontal reflection.

\begin{align}
   & \Hat{I}^{\text{db}}(i,j,k;k,\ell,i):=
        \begin{tikzpicture}[baseline={([yshift=-0.8ex]current bounding box.center)}]
            \begin{scope}
                \draw[line width=0.2mm] (0,0)--(0,1) -- (1,1) -- (1,0) --(0,0);
                \draw[line width=0.2mm] (1,1) -- (2,1) -- (2,0) -- (1,0);
                \draw[line width=0.2mm] (2,1) --  +(75:0.45);
                \draw[line width=0.2mm] (2,1) --  +(15:0.45);
                \draw[line width=0.2mm] (2,0) --  +(-75:0.45);
                \draw[line width=0.2mm] (2,0) --  +(-15 :0.45);
                \draw[line width=0.2mm] (0,1) --  +(180-75:0.45);
                \draw[line width=0.2mm] (0,1) --  +(180-15:0.45);
                \draw[line width=0.2mm] (0,0) --  +(180+75:0.45);
                \draw[line width=0.2mm] (0,0) --  +(180+15 :0.45);
                \foreach \i in {-1,...,1} {
                \fill[black] (2,1)+(45+\i *16: 0.35) circle (0.14ex);
                 \fill[black] (2,0)+(-45+\i *16: 0.35) circle (0.14ex);
                 \fill[black] (0,0)+(-135+\i *16: 0.35) circle (0.14ex);
                 \fill[black] (0,1)+(135+\i *16: 0.35) circle (0.14ex);
                    }
            \node[scale=0.9] at (1,1.35) {$k$};
            \node[scale=0.9] at (1,-0.35) {$i$};
            \node[scale=0.9] at (2.4,0.5) {$\ell$};
            \node[scale=0.9] at (-0.3,0.5) {$j$};
            \end{scope}
        \end{tikzpicture} \\
    &=-\frac{\pi^2}{6}+\frac{1}{2}\log{2}\log\left(\frac{x_{i,\ell}^2 x_{j,k}^2}{x_{i,k}^2 x_{j,\ell}^2 }\right)+\frac{1}{2}\log{2}\log\left(\frac{x_{i,j}^2 x_{k,\ell}^2}{x_{i,k}^2 x_{j,\ell}^2}\right)+\Ep{i,\ell}{k,j}+\Ep{i,j}{k,\ell} \:. \nonumber
\end{align}
The above integral is denote as $I_G^{\text{db}}$ in \cite{He:2022lfz}.

\section{Mass regularization and degenerate edge-pairing function} \label{sec: EdgeF}

\paragraph{Mass regularization.}
Some degenerate cases of the integrals discussed in this paper exhibit IR divergences when some of the $x_{i,j}^2$ vanish. 
We regulate these divergences via mass regularization (cf.~\cite{Caron-Huot:2012sos}): we simply replace all $x_{i,j}^2$ by $x_{i,j}^2 + 2\mu_{\text{IR}}^2$ in integration, and drop terms of $O(\mu_{\text{IR}}^2)$ and beyond at the end of computation.

\paragraph{Degenerate edge-pairing functions.}
We observe that the numerator in eq.~\eqref{eq: EDef} may appear to vanish 
for some degenerate cases, such as $\Ep{i,j}{j,j{+}1}$. For briefness,
we focus on a special case $\Ep{1,2}{3,1}$. Within mass regularization, it can be expressed as
\begin{equation}
        \Ep{1,2}{3,1}=\int \frac{\dif c\: \dif a_1 \dif b_3 \dif b_1}{ \pi (1+c)\sqrt{c}}   \frac{ -2 \mu^2_{\mathrm{IR} } x_{1,3}^2}{\left(2 (a_1{+}b_1) b_3 x_{1,3}^2 {+} 2 \mu^2_{\mathrm{IR} } \left(  (a_1{+}1 {+} b_1{+}b_3)^2 {+} c (a_1{+}1)^2  \right)  \right)^2} \:,
\end{equation}
where we have set $a_{2}=1$.

Next, we make the substitution 
$b_{3}\to 2\mu_{\text{IR}}^2 b_{3}$, the integral becomes
\begin{equation}
        \Ep{1,2}{3,1}=\int \frac{\dif c\: \dif a_1 \dif b_3 \dif b_1}{ \pi (1+c)\sqrt{c}}   \frac{ - x_{1,3}^2}{\left(2 (a_1{+}b_1) b_3 x_{1,3}^2 {+}  \left(  (a_1{+}1 {+} b_1{+}2 \mu^2_{\mathrm{IR} }b_3)^2 {+} c (a_1{+}1)^2  \right)  \right)^2} \:.
\end{equation}
Now the integral is manifestly convergent, and we can therefore safely drop the term $2\mu^2_{\mathrm{IR}}b_3$ in the denominator. The integration over $b_{3}$ can then be easily performed, yielding
\begin{equation}
        \Ep{1,2}{3,1}=\int \frac{\dif c\: \dif a_1 \dif b_1}{ \pi (1+c)\sqrt{c}}    
         \frac{-1}{  2 (a_1{+}b_1)  \left( (a_1{+}1 {+} b_1)^2 {+} c (a_1{+}1)^2 \right) } \:,
\end{equation}
which is nothing but a pure number. One can easily check that
\begin{equation}
        \Ep{1,2}{3,1}
        = -\frac{\pi^2}{24} + \frac{\log^2 2}{4} \:.
\end{equation}
The results for the other cases, eqs.~\eqref{eq: Ep_deg3} and \eqref{eq: Ep_deg4}, can be verified in a similar manner. 
Moreover, one can show that 
\begin{equation}
    I^{\text{dt}}(j{-}1,j;j,k)= \frac{\pi^2}{12} \:.
\end{equation}

\bibliographystyle{utphys}
\bibliography{bib}

\end{document}